\documentclass[12pt]{article}
\usepackage{psfrag,amsmath,euscript,amssymb,graphicx}

\allowdisplaybreaks

\newcommand{\be}{\begin{equation}}
\newcommand{\ee}{\end{equation}}
\newcommand{\bea}{\begin{eqnarray}}
\newcommand{\eea}{\end{eqnarray}}
\newcommand{\nl}{\nonumber\\}
\newcommand{\order}{{\cal O}}

\newcommand{\sla}[1]{\rlap{\hspace{0.02cm}/}{#1}}
\newcommand{\ov}[1]{\overleftarrow{#1} }
\def\nb{\bar{n}}
\def\calAslash{\rlap{\hspace{0.08cm}/}{{\EuScript A}}}
\def\Dslash{\rlap{\hspace{0.07cm}/}{D}}

\def\bm#1{\mbox{\boldmath$#1$\unboldmath}}

\def\A{{\EuScript A}}
\def\H{{\EuScript H}}
\def\J{{\EuScript J}}
\def\Q{{\EuScript Q}}
\def\X{{\EuScript X}}

\begin{document}

\begin{titlepage}

\begin{flushright}
CLNS~04/1865\\
SLAC-PUB-10412\\
{\tt hep-ph/0404217}\\[0.2cm]
June 25, 2004
\end{flushright}

\vspace{0.7cm}
\begin{center}
\Large\bf
Sudakov Resummation for Subleading SCET Currents and Heavy-to-Light Form 
Factors
\end{center}

\vspace{0.8cm}
\begin{center}
{\sc R.J.~Hill$^{(a)}$, T.~Becher$^{(a)}$, S.J.~Lee$^{(b)}$, and
M.~Neubert$^{(b)}$}\\
\vspace{0.7cm}
$^{(a)}${\sl Stanford Linear Accelerator Center, Stanford University\\
Stanford, CA 94309, U.S.A.} \\
\vspace{0.3cm}
$^{(b)}${\sl Institute for High-Energy Phenomenology\\
Newman Laboratory for Elementary-Particle Physics, Cornell University\\
Ithaca, NY 14853, U.S.A.}
\end{center}

\vspace{0.5cm}
\begin{abstract}
\vspace{0.2cm}\noindent
The hard-scattering contributions to heavy-to-light form factors at large 
recoil are studied systematically in soft-collinear effective theory (SCET). 
Large logarithms arising from multiple energy scales are resummed by 
matching QCD onto SCET in two stages via an intermediate effective theory. 
Anomalous dimensions in the intermediate theory are computed, and their form 
is shown to be constrained by conformal symmetry. Renormalization-group 
evolution equations are solved to give a complete leading-order analysis of 
the hard-scattering contributions, in which all single and double logarithms 
are resummed. In two cases, spin-symmetry relations for the soft-overlap 
contributions to form factors are shown not to be broken at any order in 
perturbation theory by hard-scattering corrections. One-loop matching 
calculations in the two effective theories are performed in sample cases, 
for which the relative importance of renormalization-group evolution and 
matching corrections is investigated. The asymptotic behavior of Sudakov 
logarithms appearing in the coefficient functions of the soft-overlap and 
hard-scattering contributions to form factors is analyzed. 
\end{abstract}
\vfil

\end{titlepage}

\section{Introduction}

Weak-interaction form factors for exclusive heavy-to-light transitions at 
large recoil energy, such as $B\to\pi\,l\,\nu$ with $E_\pi\sim m_B/2$, are 
an important input to measurements of the parameters of the unitarity 
triangle. The QCD description in this energy regime is complicated by the 
competition between different scattering mechanisms and the resulting 
proliferation of relevant energy scales. The tools of effective field theory 
provide an efficient means of separating the contributions from different 
scales and setting up a controlled expansion in the small ratios 
$\Lambda_{\rm QCD}/E$ and $\Lambda_{\rm QCD}/m_b$.

The appropriate theory in the present case is soft-collinear effective 
theory (SCET), which is constructed to describe processes with both soft and 
collinear partons 
\cite{Bauer:2000yr,Bauer:2001yt,Chay:2002vy,Beneke:2002ph,Hill:2002vw}.
Using SCET, it has been argued that there are two competing contributions to 
large-recoil heavy-to-light form factors at leading power in 
$\Lambda_{\rm QCD}/E$ (ignoring scaling violations from Sudakov logarithms), 
referred to as the soft-overlap (or Feynman) mechanism and the 
hard-scattering (or hard spectator-scattering) mechanism 
\cite{Bauer:2002aj,Beneke:2003pa,Lange:2003pk}. In the first of these, the 
spectator quark in the $B$-meson is absorbed into the light final-state meson 
with no large momentum transfer. For this to happen, both the initial- and 
final-state partons must be arranged in an endpoint configuration with 
atypically small values of certain momentum components. In the second 
mechanism, a large momentum is transferred to the spectator quark via hard 
gluon exchange. The suppression due to the wave-function fall-off in the 
first case, and the suppression due to hard momentum transfer in the second 
case, are of the same order in power counting.

In this paper, we present a renormalization-group (RG) analysis of the 
hard-scattering contributions. In SCET this mechanism is described by 
non-local four-quark operators, whose matrix elements factorize into products 
of leading-twist light-cone distribution amplitudes (LCDAs) for the $B$-meson 
and the light final-state meson. The matrix elements are multiplied by 
calculable coefficient functions, and the resulting convolution integrals are 
convergent to all orders in perturbation theory 
\cite{Beneke:2003pa,Lange:2003pk}. The coefficient functions at an 
appropriate low-energy hadronic scale may be computed to any order in 
$\alpha_s$ by perturbative matching of QCD onto the effective theory and 
subsequent RG evolution down to the low-energy scale. The analysis applies also 
to more complicated decay processes such as $B\to\pi\pi$ and 
$B\to K^*\gamma$, for which QCD factorization formulae relate the decay 
amplitudes to the $B\to\pi$ or $B\to K^*$ form factors plus a residual 
hard-scattering term \cite{Beneke:1999br,Beneke:2001at,Bosch:2001gv}.

The soft-overlap contributions to heavy-to-light form factors can be defined 
in SCET in terms of matrix elements of effective-theory operators obeying 
spin-symmetry relations appropriate for a heavy-collinear transition current 
\cite{Lange:2003pk}. The relevant operators are rather complicated and lead 
to ``non-factorizable'' matrix elements sensitive to endpoint momentum 
configurations, transverse momentum components, and non-valence Fock states.
However, these soft overlap contributions can be described in terms of 
universal functions $\zeta_M(E)$ that only depend on the light final-state 
meson but not on the Lorentz structure of the currents whose matrix elements 
define the various form factors. For instance, there is one function 
$\zeta_P(E)$ for decays into pseudoscalar mesons, and similarly only one 
function each, $\zeta_{V_\parallel}(E)$ and $\zeta_{V_\perp}(E)$, for decays 
into longitudinally and transversely polarized vector mesons. This implies 
spin-symmetry relations between different soft form factors, which were 
first derived in \cite{Charles:1998dr} by considering the large-energy limit 
of QCD.  

The presence of both a soft-overlap and a hard-scattering contribution is 
summarized by the factorization formula \cite{Beneke:2000wa}
\be\label{eq:soft_plus_hard}
   F_i^{B\to M}(E) = C_i(E)\,\zeta_M(E)
   + \int_0^\infty\!{d\omega\over\omega}\,\phi_B(\omega) 
   \int_0^1\!du\,f_M\,\phi_M(u)\,T_i(E,\omega,u) \,,  
\ee
which is valid at leading power in $\Lambda_{\rm QCD}/E$. Here $\phi_B$ and 
$\phi_M$ are the LCDAs of the $B$-meson and light final-state meson, and $f_M$ 
is the decay constant of the light meson. The Wilson coefficients of the 
effective-theory operators, $C_i(E)$ and $T_i(E,\omega,u)$, may be calculated 
in perturbation theory to any order in $\alpha_s$, and a RG analysis can be 
used to relate the coefficients at different renormalization scales. However, 
the large-$E$ behavior of the soft-overlap contribution cannot be addressed 
satisfactorily in perturbation theory, since the long-distance matrix 
elements, $\zeta_M(E)$, depend on the energy $E$ in a non-perturbative way 
\cite{Lange:2003pk}. Thus the issue of whether one of the soft-overlap or the 
hard-scattering contributions is enhanced relative to the other in the formal 
limit $E\to\infty$ cannot be addressed using short-distance methods. 
Phenomenologically, it appears that the soft-overlap terms are dominant for 
physical values of the coupling and mass parameters. Although not a complete 
answer to the question, the relative suppression of the coefficients 
multiplying the long-distance matrix elements may be computed using 
perturbative methods. Studying the resummation of Sudakov logarithms for these 
coefficients, we find that the soft-overlap contribution is suppressed in the 
formal asymptotic limit $E\to\infty$, but that this suppression is mild for 
realistic values of $E$.

Because the form factors involve three different physical scales, namely
$\mu^2\sim m_b^2$ (hard), $\mu^2\sim m_b\Lambda_{\rm QCD}$ (hard-collinear), 
and $\mu^2\sim\Lambda_{\rm QCD}^2$ (soft), integrating out modes of 
progressively smaller virtuality results in a sequence of effective theories. 
At the high scale $\mu^2\sim m_b^2$, the effective theory is described by the 
usual QCD Lagrangian (with five quark flavors) plus an effective 
weak-interaction Lagrangian obtained by integrating out virtual $W$ and $Z$ 
bosons and top quarks. Integrating out modes of virtuality $m_b^2$ we arrive 
at an intermediate effective theory containing soft modes and hard-collinear 
modes of virtuality $p_{hc}^2\sim m_b\Lambda_{\rm QCD}$. In this paper we are 
mainly concerned with this intermediate theory, called SCET$_{\rm I}$ 
\cite{Bauer:2001yt,Chay:2002vy,Beneke:2002ph}. Integrating out the 
hard-collinear modes of virtuality $m_b\Lambda_{\rm QCD}$ yields the final 
low-energy theory, denoted SCET$_{\rm II}$, consisting of soft and collinear 
modes of virtuality $p_s^2,\,p_c^2\sim\Lambda_{\rm QCD}^2$. In this case, 
soft-collinear messenger modes are also required 
\cite{Becher:2003qh,Becher:2003kh}.

In the following section we briefly review some relevant elements of SCET.
Section~\ref{sec:currents} lists the leading and subleading SCET$_{\rm I}$ 
current operators, which are required for the discussion of heavy-to-light 
form factors, together with the matching coefficients for these operators 
obtained at tree level. For the example of the scalar current, we also 
calculate the one-loop matching coefficients of the leading and subleading 
operators. Section~\ref{sec:anom} contains the main result of the paper, 
i.e., the anomalous dimensions of the subleading  SCET$_{\rm I}$ current 
operators. These quantities exhibit an interesting symmetry property, which 
can be traced back to a conformal symmetry (at the classical level) in the 
hard-collinear sector. We briefly review the constraints imposed by conformal 
symmetry in Section~\ref{sec:solution}, and use these results to diagonalize 
the non-local part of the evolution operator. We also present a formal 
algebraic solution of the evolution equations for the currents and their 
Wilson coefficient functions. In Section~\ref{sec:SCETII}, we discuss the 
operator representation and renormalization for the hard-scattering 
contributions in SCET$_{\rm II}$. In two cases, namely for the form-factor 
ratios $A_1/V$ and $T_2/T_1$, we find that the spin-symmetry relations 
holding for the soft-overlap contributions are not broken by the 
hard-scattering terms, and therefore, at leading order in $1/E$ and to all 
orders in $\alpha_s$, only eight of the ten form factors describing 
$B\to P,V$ decays are independent. The relevant matching coefficients (jet 
functions) can be related to two universal quantities ${\cal J}_\parallel$ 
and ${\cal J}_\perp$, which we compute including one-loop radiative 
corrections. As an application of our results, we present in 
Section~\ref{sec:applic} a RG-improved analysis of the hard-scattering 
contributions to the large-recoil heavy-to-light form factors, in which all 
single and double logarithms are resummed. Section~\ref{sec:Sudakov} treats 
the asymptotic limit of Sudakov suppression factors in the soft-overlap and 
hard-scattering terms. In Section~\ref{sec:summary} we present our summary 
and conclusions.

\section{Soft-collinear effective theory}

In processes involving energetic light particles, such as the pion emitted at
large recoil in semileptonic $B$ decay, it is convenient to introduce
light-cone coordinates
\be
   p^\mu = n\cdot p\,{\nb^\mu\over 2} + \nb\cdot p\,{n^\mu\over 2}
    + p_{\perp}^\mu
   \equiv p_+^\mu + p_-^\mu + p_\perp^\mu \,,
\ee
the second equality serving to introduce the vectors $p_+$ and $p_-$. The
light-like vectors $n^\mu$, $\nb^\mu$ satisfy $n^2=\nb^2=0$ and 
$n\cdot\nb=2$. As mentioned in the Introduction, depending on the value of 
the renormalization scale the effective theory is described by hard-collinear 
and soft modes (SCET$_{\rm I}$), or by collinear, soft, and soft-collinear 
messenger modes (SCET$_{\rm II}$). It is conventional to quote the scaling
behavior of the components $(p_+,p_-,p_\perp)$ with the energy in terms of
a small parameter $\lambda\sim\Lambda_{\rm QCD}/E$. The collinear and soft 
momenta of the partons inside the external meson states in $B\to M$
transitions scale like $p_c\sim E(\lambda^2,1,\lambda)$ and 
$p_s\sim E(\lambda,\lambda,\lambda)$. Hard-collinear momenta are defined to 
scale as $p_{hc}\sim E(\lambda,1,\lambda^{1/2})$, whereas soft-collinear 
momenta scale as $p_{sc}\sim E(\lambda^2,\lambda,\lambda^{3/2})$. Throughout 
this paper we will identify
\begin{equation}
   E = v\cdot p_- = \frac12\,(n\cdot v) (\nb\cdot p)\gg\Lambda_{\rm QCD}
\end{equation}
with the energy of a collinear or hard-collinear particle in the $B$-meson rest frame. 
Strictly speaking, the energy $v\cdot p$ differs from $E$ by terms which can be 
neglected at leading and first subleading power.

Fields in SCET$_{\rm I}$ and their scalings with the expansion parameter 
$\lambda^{1/2}$ are \cite{Bauer:2001yt,Chay:2002vy,Beneke:2002ph}
\bea
   \xi_{hc} &=& {\sla{n}\sla{\nb}\over 4}\,\psi_{hc}\sim \lambda^{1/2} \,,
    \qquad q_s\sim \lambda^{3/2} \,, \qquad h\sim \lambda^{3/2} \,, \nl
   A_{hc}^\mu &\sim& (\lambda,1,\lambda^{1/2}) \,, \qquad
    A_s^\mu\sim (\lambda,\lambda,\lambda) \,,
\eea
where $h$ is the heavy-quark field in Heavy-Quark Effective Theory (HQET)
\cite{Neubert:1993mb}. The leading-order soft and hard-collinear quark 
Lagrangians are
\bea\label{eq:scetLagrangian}
   {\cal L}_s^{(0)} &=& \bar{q}_s\,i\Dslash_s\,q_s
    + \bar h\,iv\cdot D_s\,h \,, \nl
   {\cal L}_{hc}^{(0)} &=& \bar\xi_{hc}\,{\sla{\nb}\over 2}
    \left( in\cdot D_{hc+s} + i\Dslash_{hc\perp}\,{1\over i\nb\cdot D_{hc}}\,
    i\Dslash_{hc\perp} \right) \xi_{hc} \,.
\eea
Here $iD_s^\mu=i\partial^\mu+g A_s^\mu$ denotes the covariant derivative 
built with soft gluon fields, etc. For the pure gauge sector we may write
${\cal L}^{(0)}_g={\cal L}_s+{\cal L}_{hc}^{(0)}$, where ${\cal L}_s$ is the 
usual gluon Lagrangian restricted to soft fields, and ${\cal L}_{hc}^{(0)}$ 
is obtained by substituting $iD^\mu\to iD_{hc}^\mu+g A_{s +}^\mu$ into the 
Yang-Mills Lagrangian. In interactions involving both soft and hard-collinear 
fields, the soft fields are evaluated at $x_-$. In particular, this rule applies to the 
first term in the hard-collinear Lagrangian, in which 
$in\cdot D_{hc+s}=in\cdot\partial+g n\cdot A_{hc}(x)+g n\cdot A_s(x_-)$. 
Corrections from the multipole expansion 
$\phi_s(x)=\phi_s(x_-)+x_\perp\cdot\partial_\perp\,\phi_s(x_-)+\dots$ appear 
as higher-order terms in the expansion in $\lambda^{1/2}$. We shall also need 
the subleading interaction Lagrangian \cite{Beneke:2002ph}
\be
   {\cal L}_{q\xi}^{(1/2)} = \bar\xi_{hc}\,i\Dslash_{hc\perp} W_{hc}\,q_s
   + \mbox{h.c.} \,,
\ee
which transforms a soft quark into a hard-collinear one (and vice versa). 
Here $W_{hc}$ is a hard-collinear Wilson line in the $\nb$ direction 
necessary to ensure gauge invariance. Additional terms in the Lagrangian are 
required to describe the soft-overlap contribution in SCET$_{\rm I}$ 
\cite{Bauer:2002aj}; however, they will not be of relevance to our discussion 
here.

The Lagrangian interactions are a special case of general gauge-invariant
operators. Imposing homogeneous gauge transformations in the soft and
hard-collinear sectors \cite{Beneke:2002ni}, which strictly respect the SCET
power counting, the ``homogenized'' hard-collinear fields are restricted to 
appear in the combinations
\be\label{BBs}
   \X_{hc} = W_{hc}^\dagger\,\xi_{hc}\sim \lambda^{1/2} \,, \quad
   \A_{hc\perp}^\mu = W_{hc}^\dagger\,(iD_{hc\perp}^\mu W_{hc})\sim
   \lambda^{1/2} \,, \quad
   W_{hc}^\dagger\,in\cdot D_{hc+s}\,W_{hc}\sim \lambda \,,
\ee
and may be acted on by partial derivatives $i\nb\cdot\partial\sim 1$ and 
$i\partial_\perp^\mu\sim\lambda^{1/2}$. 
This result follows from considering the most 
general operator in the gauge where $\nb\cdot A_{hc}=0$, $n\cdot A_s=0$, 
and then returning to an arbitrary gauge by using the transformation laws 
appropriate for the homogenized fields. 
The $\order(1)$ components $i\nb\cdot\partial$ may appear an arbitrary number of times in operators of a 
given order in the power counting. This is accounted for by smearing 
hard-collinear fields along the $\nb$ direction, i.e., $\phi_{hc}(r\nb)$ with 
arbitrary $r$. 
Soft fields may appear as $q_s,\,h\sim\lambda^{3/2}$ and $iD_s^\mu\sim\lambda$, and 
position arguments $x_\perp\sim\lambda^{-1/2}$ and $x_+\sim\lambda^{0}$ arising from 
the multipole expansion may also appear in interactions involving both soft and hard-collinear fields.    
Until Section~\ref{sec:SCETII} we deal exclusively with 
SCET$_{\rm I}$, and from now on will drop the label ``$hc$'' on the 
hard-collinear fields.

\section{Flavor-changing currents in SCET$_{\bf I}$}
\label{sec:currents}

Heavy-to-light form factors describing current-induced $B\to M$ transitions
(with $M$ a light meson) at large recoil are subleading quantities in the 
large-energy limit, in the sense that the transitions they describe cannot be 
mediated by leading-order SCET$_{\rm I}$ currents and Lagrangian 
interactions. For a leading-order analysis of these form factors, it is 
sufficient to include heavy-collinear current operators through the first 
subleading order in SCET$_{\rm I}$ power counting 
\cite{Bauer:2002aj,Beneke:2003pa}. These operators contain a heavy-quark 
field, a hard-collinear quark field, and (in the case of subleading 
operators) transverse derivatives and gluon fields. They provide a 
representation in SCET$_{\rm I}$ of the QCD heavy-light current operators 
$J_{\rm QCD}=\bar q\,\Gamma\,b$, which we renormalize at a scale 
$\mu_{\rm QCD}$. (The vector and axial-vector currents are not renormalized.) 
At a given order in power counting, a minimal basis is determined by first 
writing the most general gauge-invariant operators constructed from the 
available fields and external parameters, and then requiring invariance under 
small variations of the external parameters \cite{Chay:2002vy,Manohar:2002fd}.

\subsection{Determination of the operator basis}

It is convenient to restrict attention to the case $v_\perp^\mu =0$, where 
$v$ denotes the $B$-meson velocity. It follows that
\be\label{eq:nbar}
   \nb^\mu = {1\over n\cdot v}
   \left( 2v^\mu - {n^\mu\over n\cdot v} \right) .
\ee
With this choice there are two remaining reparameterization transformations,
which we consider in their infinitesimal form. The first enforces invariance 
under the rescaling
\bea\label{eq:rescale}
   n^\mu\to(1+\alpha)\,n^\mu \,, \qquad
   \nb^\mu\to(1 - \alpha)\,\nb^\mu
   \qquad (\mbox{with $\alpha \sim 1$}) \,.
\eea
The second allows small changes in the perpendicular components, such that
\be\label{eq:reparam}
   n^\mu\to n^\mu + \epsilon_\perp^\mu \,, \qquad
   \nb^\mu\to\nb^\mu - {\epsilon_\perp^\mu\over(n\cdot v)^2}
   \qquad (\mbox{with $\epsilon_\perp \sim \lambda^{1/2}$}) \,.
\ee
The power counting assigned to $\alpha$ and $\epsilon_\perp$ is the largest
possible (i.e., providing the strongest constraints) such that the scaling of 
hard-collinear momenta is unaltered. In both cases, the variation of 
$\nb^\mu$ is determined by (\ref{eq:nbar}) and the variation of $n^\mu$. An 
alternative approach would be to introduce $\nb$ and $v$ as arbitrary 
vectors, subject only to the conditions $\nb^2=0$, $n\cdot\nb= 2$, and 
$v^2=1$. Then $\nb^\mu$ would not transform under (\ref{eq:reparam}), and 
there would be an additional reparameterization transformation
\be
   n^\mu\to n^\mu \,, \qquad \nb^\mu\to\nb^\mu + e_\perp^\mu
   \qquad (\mbox{with $e_\perp \sim 1$}) \,.
\ee
In this case a consistent power counting requires $v_\perp\sim\order(1)$, and
many more operators appear at a given order than when $v_\perp=0$ 
\cite{Pirjol:2002km}. At higher order it would also be necessary to impose 
invariance under heavy-quark velocity transformations, $v\to v+\delta v$ with
$v\cdot\delta v=0$. However, these transformations enter only at 
$\order({\lambda})$ and so are irrelevant for our leading-order analysis of 
heavy-to-light form factors.

Requiring invariance under these transformations, it is straightforward to
write down the most general operators with given quantum numbers. The 
leading-order currents are $\order(\lambda^2)$, and for the scalar case we 
find
\be
   J_S^{(0)}(s,x) = e^{-im_b v\cdot x}\,\bar\X(x+ s\nb)\,h(x_-) \,,  
\ee
where the phase factor arises from the definition of the HQET field $h$, and 
only the first term, $h(x_-)$, has been retained in the multipole expansion. 
In the form-factor analysis we can use translational invariance to set $x=0$ 
in the weak current operators, and so we restrict attention to 
$J_S^{(0)}(s)\equiv J_S^{(0)}(s,x=0)$. Similarly, for the vector and tensor 
currents we have
\be
   J_{Vi}^{(0)\,\mu}(s) = \bar{\X}(s\nb)\,\Gamma^\mu_i\,h(0) \,, \qquad
   J_{Ti}^{(0)\,\mu\nu}(s) = \bar{\X}(s\nb)\,\Gamma^{\mu\nu}_i\,h(0) \,,
\ee
with the Dirac structures
\bea
   \Gamma_1^\mu &=& \gamma^\mu \,, \qquad \Gamma_2^\mu = v^\mu \,, \qquad
    \Gamma_3^\mu = {n^\mu\over n\cdot v} \,, \nl
   \Gamma_1^{\mu\nu} &=& \gamma^{[\mu}\gamma^{\nu]} \,, \qquad
    \Gamma_2^{\mu\nu} = v^{[\mu}\gamma^{\nu]} \,, \qquad
    \Gamma_3^{\mu\nu} = {n^{[\mu}\gamma^{\nu]}\over n\cdot v} \,, \qquad
    \Gamma_4^{\mu\nu} = {n^{[\mu}v^{\nu]}\over n\cdot v} \,.
\eea
Square brackets around indices denote anti-symmetrization. Note that 
all structures invariant under the rescaling (\ref{eq:rescale}) are allowed 
at leading order, since the reparameterization transformations
(\ref{eq:reparam}) enter only at subleading order.

Before writing the most general set of subleading operators, we first 
consider the
variation of the leading operators under (\ref{eq:reparam}). To first order 
in $\lambda^{1/2}$, the fermion fields transform as
\bea\label{eq:variation_leading}
   \delta\bar\X(x+s\nb) 
   &=& \bar\X(x+s\nb)\left({\sla{\nb}\over 2}{\sla{\epsilon_\perp}\over 2}
    - i{\nb\cdot x\over 2}\,\epsilon_\perp\cdot g A_{s\perp}(x_-) \right)
    , \nl
   \delta h(x_-) &=& {\nb\cdot x\over 2}\,
    \epsilon_\perp\cdot \partial_\perp h(x_-) \,.
\eea
The soft gluon appearing in the variation of $\X(x)$ arises from the field 
redefinition enforcing homogeneous gauge transformations in the 
hard-collinear sector~\cite{Beneke:2002ni}. The combination $\X$ of 
hard-collinear fields may be expressed as 
$\X=R^\dagger\,(\sla{n}\sla{\nb}/4)\,\psi'_{hc}$, where $\psi_{hc}'$ is the 
(unhomogenized) hard-collinear fermion field in the gauge $\nb\cdot A_{hc}=0$, 
and $R(x)$ is a soft Wilson line from $x_-$ to $x$. Under the 
transformation (\ref{eq:reparam}), $\delta\psi'_{hc}=\order(\lambda)$, and the 
remaining variation of $\X$ arises from the projection $\sla{n}\sla{\nb}/4$ 
and the homogenizing factor $R$. Next, we consider the variations of the 
various Dirac structures under the transformation (\ref{eq:reparam}), finding
\bea\label{eq:variation_dirac}
   \delta\Gamma_1^\mu &=& 0 \,, \qquad
    \delta\Gamma_2^\mu = 0 \,, \qquad
    \delta\Gamma_3^\mu = {\epsilon_\perp^\mu\over n\cdot v} \,, \nl
   \delta\Gamma_1^{\mu\nu} &=& 0 \,, \qquad
    \delta\Gamma_2^{\mu\nu} = 0 \,, \qquad
    \delta\Gamma_3^{\mu\nu}
    = {\epsilon_\perp^{[\mu}\gamma^{\nu]}\over n\cdot v} \,, \qquad
    \delta\Gamma_4^{\mu\nu} = {\epsilon_\perp^{[\mu}v^{\nu]}\over n\cdot v}
    \,.
\eea
The variations of the subleading operators must cancel these contributions.

The subleading $\order(\lambda^{1/2})$ operators must contain exactly one 
insertion of $\A_{\perp}^\mu$ or $(-i\ov{\partial_\perp}^\mu)$ acting on the 
hard-collinear field $\bar\X$, or $x_\perp\cdot D_{s\perp}$ with the 
derivative acting on $h$. To determine the most general form, we note the 
following transformation properties, working now to zeroth order in 
$\lambda^{1/2}$:
\bea\label{eq:variation_subleading}
   \delta\A_\perp^\mu = 0 \,, \qquad 
   \delta\bigg(
    {-i\ov{\partial}_{\!\perp}^\mu\over -i\nb\cdot\ov{\partial}} \bigg)
   = -{\epsilon_\perp^\mu\over 2} \,, \qquad
   \delta\big( x_\perp\cdot D_{s\perp} \big)
   = -{\nb\cdot x\over 2}\,\epsilon_\perp\cdot D_{s\perp} \,.
\eea
Thus, $(-i\ov{\partial}_{\!\perp}^\mu)$ and $x_\perp\cdot D_{s\perp}$ are 
restricted to appear in specific reparameterization-invariant combinations 
with the leading-order currents, and there are no constraints on the 
appearance of $\A_\perp^\mu$. Inspection of (\ref{eq:variation_leading}), 
(\ref{eq:variation_dirac}), and (\ref{eq:variation_subleading}) allows us to 
deduce the form of the most general operators through 
$\order({\lambda}^{1/2})$. For the scalar current\\[-0.6cm]
\bea\label{eq:scalar_currents}
   J^A_S(s,x) &=& e^{-im_b v\cdot x}\,\bar\X(x+s\nb) \bigg( 1
    - {i\ov{\sla{\partial}}_{\!\perp}\over i\nb\cdot\ov{\partial}}\,
    {\sla{\nb}\over 2} + x_\perp\cdot D_{s\perp} \bigg) h(x_-) \,, \nl
   J^B_S(s,r,x) &=& e^{-im_b v\cdot x}\,
    \bar\X(x+s\nb)\,\calAslash_\perp(x+r\nb)\,h(x_-) \,.
\eea
Again, using translational invariance we can specialize to $x=0$ and define 
$J^A_S(s)\equiv J^A_S(s,0)$ and $J^B_S(s,r)\equiv J^B_S(s,r,0)$. Similarly, 
for the vector and tensor currents at $x=0$ we obtain
\bea\label{eq:vector_currents}
   J^{A\,\mu}_{V1,2}(s) &=& \bar\X(s\nb) \bigg( 1
    - {i\ov{\sla{\partial}}_{\!\perp}\over i\nb\cdot\ov{\partial}}\,
    {\sla{\nb}\over 2} \bigg) \Gamma_{1,2}^\mu\,h(0) \,, \nl
   J^{A\,\mu}_{V3}(s) &=& \bar\X(s\nb) \bigg( 1
    - {i\ov{\sla{\partial}}_{\!\perp}\over i\nb\cdot\ov{\partial}}\,
    {\sla{\nb}\over 2} \bigg) \Gamma_3^\mu\,h(0)
    + {2\over n\cdot v}\,\bar\X(s\nb)\,
    {i{\ov{\partial}_{\!\perp}^\mu}\over i\nb\cdot\ov{\partial}}\,h(0) \,,
    \nl
   J^{B\,\mu}_{V1,2,3}(s,r) &=& \bar\X(s\nb)\,\calAslash_\perp(r\nb)\,
    \Gamma_{1,2,3}^\mu\,h(0) \,, \nl[0.2cm]
   J^{B\,\mu}_{V4}(s,r) &=& \bar\X(s\nb)\,\A_\perp^\mu(r\nb)\,h(0) \,,
\eea
and
\bea\label{eq:tensor_currents}
   J^{A\,\mu\nu}_{T1,2}(s) &=& \bar\X(s\nb) \bigg( 1
    - {i\ov{\sla{\partial}}_{\!\perp}\over i\nb\cdot\ov{\partial}}\,
    {\sla{\nb}\over 2} \bigg) \Gamma_{1,2}^{\mu\nu}\,h(0) \,, \nl
   J^{A\,\mu\nu}_{T3}(s) &=& \bar\X(s\nb) \bigg( 1
    - {i\ov{\sla{\partial}}_{\!\perp}\over i\nb\cdot\ov{\partial}}\,
    {\sla{\nb}\over 2} \bigg) \Gamma_3^{\mu\nu}\,h(0)
    + {2\over n\cdot v}\,\bar\X(s\nb)\,
    {i\ov{\partial}_{\!\perp}^{[\mu}\gamma^{\nu]}
     \over i\nb\cdot\ov{\partial}}\,h(0) \,, \nl
   J^{A\,\mu\nu}_{T4}(s) &=& \bar\X(s\nb) \bigg( 1
    - {i\ov{\sla{\partial}}_{\!\perp}\over i\nb\cdot\ov{\partial}}\,
    {\sla{\nb}\over 2} \bigg) \Gamma_4^{\mu\nu}\,h(0)
    + {2\over n\cdot v}\,\bar\X(s\nb)\,
    {i\ov{\partial}_{\!\perp}^{[\mu} v^{\nu]}\over i\nb\cdot\ov{\partial}}\,
    h(0) \,, \nl
   J^{B\,\mu\nu}_{T1,2,3,4}(s,r) &=& \bar\X(s\nb)\,\calAslash_\perp(r\nb)\,
    \Gamma^{\mu\nu}_{1,2,3,4}\,h(0) \,, \nl[0.2cm]
   J^{B\,\mu\nu}_{T5,6,7}(s,r) &=& \bar\X(s\nb)\,\A_\perp^{[\mu}(r\nb)\,
    \Gamma^{\nu]}_{1,2,3}\,h(0) \,.
\eea
In the tensor case, one combination of the $J^{B\,\mu\nu}_{Ti}$ is redundant 
in four dimensions, being proportional to the anti-symmetric product 
$\gamma_{\perp}^{[\mu}\gamma_\perp^{\nu}\gamma_\perp^{\rho]}$. This 
combination will be isolated in the next section, when we define a new basis 
of current operators that are renormalized multiplicatively. 

We will refer to the quark-antiquark operators as ``$A$-type currents'', and 
to the quark-antiquark-gluon operators as ``$B$-type currents''. The $A$-type 
currents contain both leading and subleading contributions, which are linked 
by reparameterization invariance. Pseudoscalar and axial-vector currents are 
obtained from the above expressions for scalar and vector currents by 
insertion of $\gamma_5$ next to the light fermion field. (This simple 
prescription holds only in the ``naive dimensional regularization'' (NDR) 
scheme.) The results (\ref{eq:scalar_currents}), (\ref{eq:vector_currents}), 
and (\ref{eq:tensor_currents}) agree with the basis of operators found in
\cite{Pirjol:2002km}, specialized to the case where $v_\perp=0$. However, 
the derivation presented here is much simpler. We also stress that our 
definition of the subleading currents, in which the $A$-type currents contain
$\partial_\perp$ rather than a covariant derivative, will ensure that the 
$A$-type (two-particle) and $B$-type (three-particle) operators do not mix
under renormalization.

The subleading contributions to the $A$-type currents containing 
perpendicular derivatives on the hard-collinear fields do not contribute at 
leading order in the form factor analysis. The extra derivatives yield an 
$\order(\lambda)$ suppression on top of the suppression factors already 
present when the leading currents mediate the decay process by either the 
soft-overlap or hard-scattering mechanisms. A formal demonstration of this 
point can be found in \cite{Beneke:2003pa}. The remaining $A$-type currents 
(at $x=0$) all take the form $\bar{\X}(s\nb)\Gamma\,h(0)$, differing only by 
their Dirac structure $\Gamma$. Spin symmetries arising from the 
constraints $\sla{n}\X=0$, $\sla{v}h=0$ may then be used to relate matrix 
elements involving the same initial- and final-state mesons. Although the 
SCET$_{\rm II}$ representation of the $A$-type currents is rather complicated 
\cite{Lange:2003pk}, owing to the symmetry relations the results can be 
expressed in terms of a small number of non-perturbative
functions. The 
$A$-type currents thus give rise to the first, soft-overlap term in 
(\ref{eq:soft_plus_hard}).  

\begin{figure}
\begin{center}
\newcommand{\jw}{0.22\textwidth}
\newcommand{\jwh}{0.062\textwidth}
\begin{tabular}{ccccc}
QCD && SCET${}_{\rm I}$ && SCET${}_{\rm II}$ \\
\includegraphics[width=\jw]{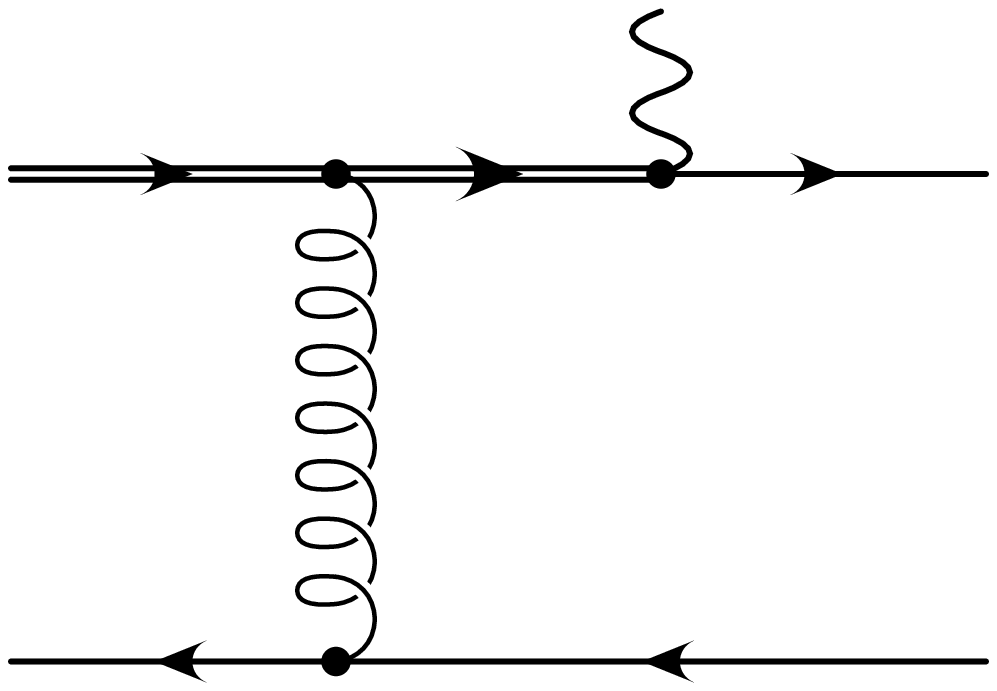} & 
\raisebox{\jwh}{$\longrightarrow$} & 
\includegraphics[width=\jw]{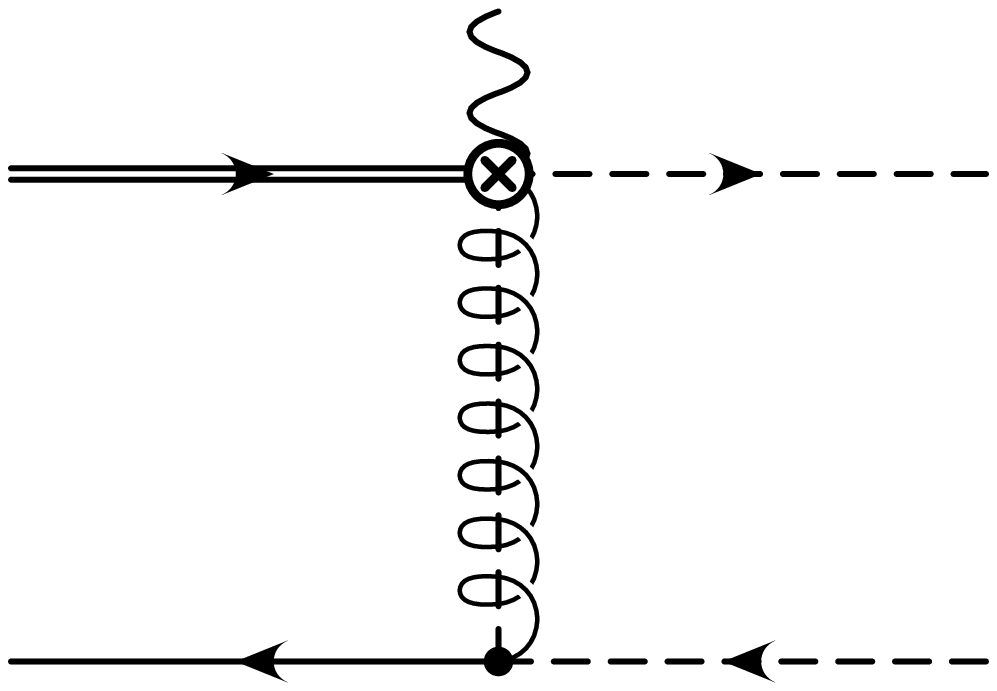} &
\raisebox{\jwh}{$\longrightarrow$} &
\raisebox{0.013\textwidth}{\includegraphics[width=\jw]{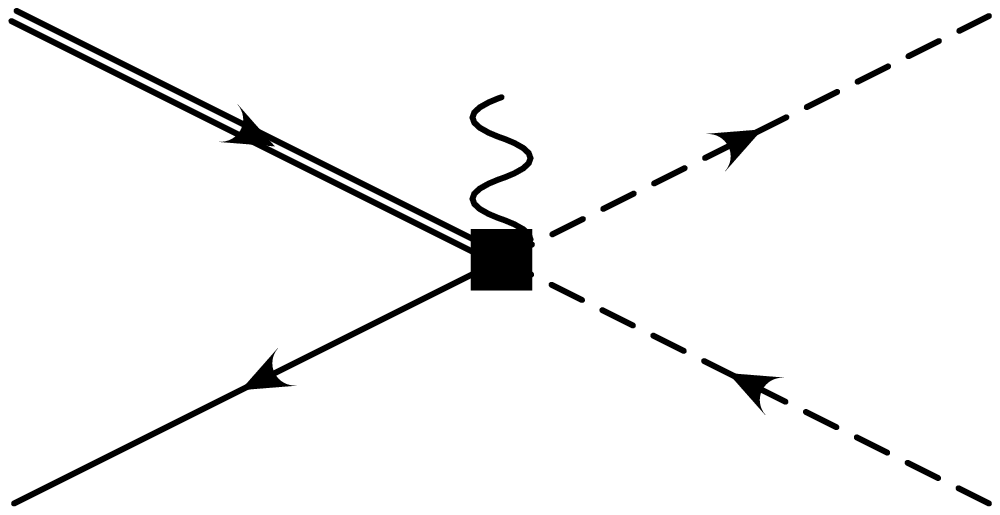}}
\end{tabular}
\caption{Two-step matching procedure for a hard-scattering contribution. The 
dashed lines in the SCET${}_{\rm I}$ diagram represent hard-collinear fields, 
those in the SCET${}_{\rm II}$ four-quark operator denote collinear fields. 
In all cases, the heavy quark is depicted as a double line, and the wavy line 
represents the flavor-changing current.}
\label{fig:hardScattering}
\end{center}
\end{figure}

In general, the $B$-type currents break the spin symmetries. When applied to 
form-factor matrix elements, the hard-collinear gluon emitted from the 
current is absorbed by the spectator quark, allowing the decay to proceed 
via the hard-scattering mechanism. Figure~\ref{fig:hardScattering} 
illustrates the two-step matching of a typical hard-scattering amplitude. In 
the first step, the QCD current is matched onto a $B$-type SCET$_{\rm I}$ 
current. In the second step, the hard-collinear gluon is integrated out, and 
the hard-scattering amplitude is described in the final low-energy theory by 
non-local four-quark operators. Section~\ref{sec:SCETII} examines this 
SCET$_{\rm II}$ representation, where matrix elements take the form of the 
second, symmetry-breaking term in (\ref{eq:soft_plus_hard}).

\subsection{Matching calculations}
\label{sec:match}

We proceed to find the SCET$_{\rm I}$ representations of the QCD scalar, 
vector, and tensor currents
\be\label{eq:currentDefs}
   S = \bar q\,b \,, \qquad
   V^\mu = \bar q\,\gamma^\mu b \,, \qquad
   T^{\mu\nu} = (-i)\,\bar q\,\sigma^{\mu\nu} b
    = \bar q\,\gamma^{[\mu}\gamma^{\nu]}\,b \,.
\ee
The QCD operators $S$ and $T^{\mu\nu}$ require renormalization and are 
defined in the modified minimal subtraction ($\overline{\rm MS}$) 
scheme at a fixed scale 
$\mu_{\rm QCD}=\order(m_b)$. The $\mbox{SCET}_{\rm I}$ representations of 
these operators, evaluated at position $x=0$, are given by an expansion 
(summed over $i,j$)
\bea\label{Jexpansion}
   \bar q\,\Gamma\,b
   &\to& \int ds\,\tilde C_i^A(s)\,J^A_i(s)
    + \frac{1}{2E} \int dr\,ds\,\tilde C^B_j(s,r)\,J^B_j(s,r)
    + \dots \nl
   &=& C^A_i(E)\,J^A_i(0)
    + \frac{1}{2E} \int du\,C^B_j(E,u)\,J^B_j(u) + \dots
\eea
with operators $J^A_i$, $J^B_j$ as determined in the previous section for
the appropriate quantum numbers. We denote coefficient functions in position
space with a tilde. In the second line, we have used translational invariance
and defined the momentum-space coefficients (without a tilde) as
\bea
   C^A_i(E) &=& \int ds\,e^{is\nb\cdot P}\,\tilde C^A_i(s) \,, \nl
   C^B_j(E,u) &=& \int dr\,ds\,e^{i(us+\bar u r)\nb\cdot P}\,
    \tilde C^B_j(s,r) \,.
\eea
Here $P=P_{\rm out}-P_{\rm in}$ is the total hard-collinear momentum of 
external states (strictly speaking this is a momentum operator), and 
$E=v\cdot P_-=(n\cdot v)(\bar n\cdot P)/2$. Reparameterization invariance 
ensures that the momentum-space coefficient functions depend on the 
combination $(n\cdot v)(\bar n\cdot P)=2E$, not $\nb\cdot P$. The variable 
$u\in [0,1]$ is the fraction of the large momentum component $\nb\cdot P$ 
carried by the fields in $\bar\X$ (the dressed outgoing hard-collinear quark 
field), and $\bar u\equiv 1-u$ is the corresponding momentum fraction carried 
by the fields in $\A_\perp$ (the dressed outgoing hard-collinear gluon 
field). The object $J^B_j(u)$ in (\ref{Jexpansion}) denotes the 
Fourier-transformed current operator
\be\label{eq:JBG}
   J_j^B(u) = \nb\cdot P \int {ds\over 2\pi}\,
   e^{-i u\nb\cdot P s}\, J_j^B(s,0) \,.
\ee
Note that both $J_i^A(0)$ and $J_j^B(u)$ also depend on the large energy 
scale $E$ as well as on the renormalization scale $\mu$. These dependences 
will be suppressed for simplicity.

The matching conditions for the momentum-space Wilson coefficient functions 
$C_i^A$ and $C_j^B$ at tree level follow from an analysis of current matrix 
elements in QCD and SCET$_{\rm I}$. The only subtlety is that a non-zero 
matching contribution is obtained from graphs where a hard-collinear gluon is 
emitted from the hard-collinear quark line. This might seem surprising at 
first sight, because the resulting propagator is close to the mass-shell. 
This contribution is present because two of the four components of the 
hard-collinear quark spinor are removed when QCD is matched onto 
SCET$_{\rm I}$. To see how it arises, consider the following diagram: 
\begin{equation}
\label{eq:decompose}
\begin{aligned}
\includegraphics[height=0.12\textwidth]{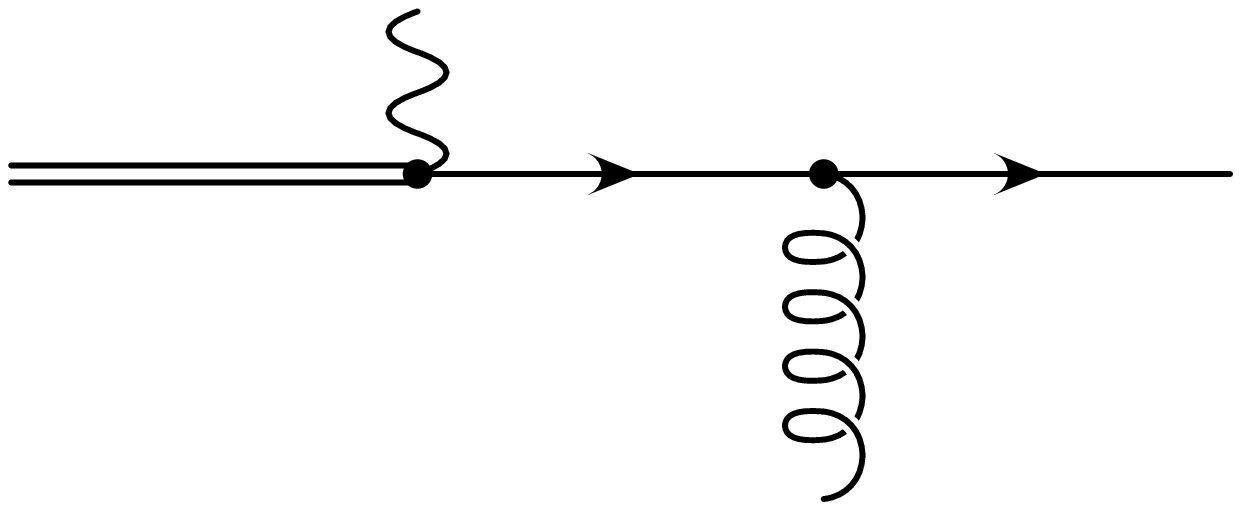}
 &&\raisebox{0.073\textwidth}{=}
 &&\includegraphics[height=0.12\textwidth]{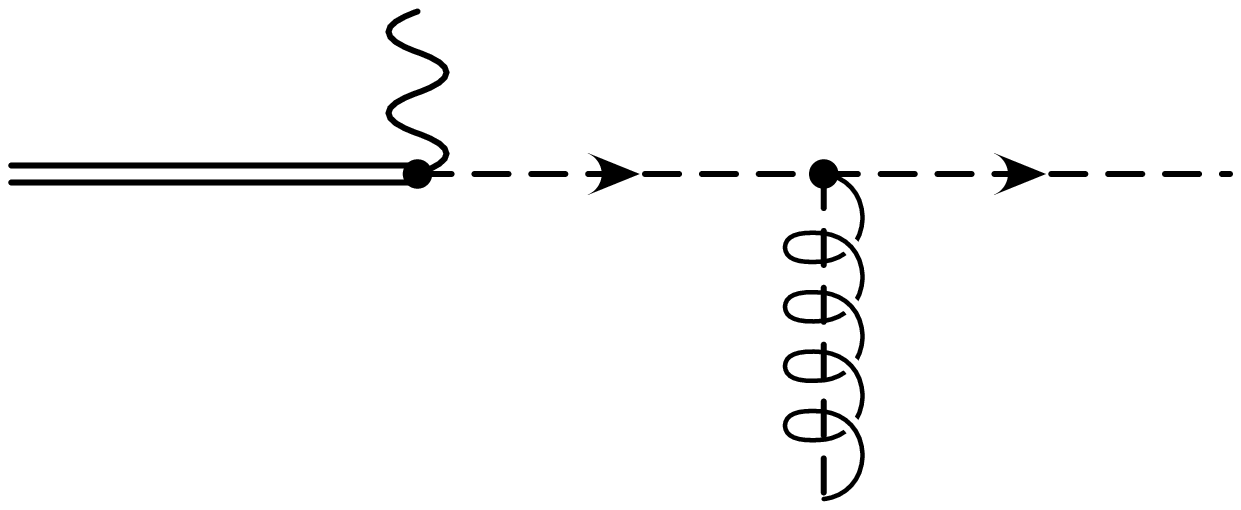}
 &&\raisebox{0.073\textwidth}{+}
 &&\includegraphics[height=0.12\textwidth]{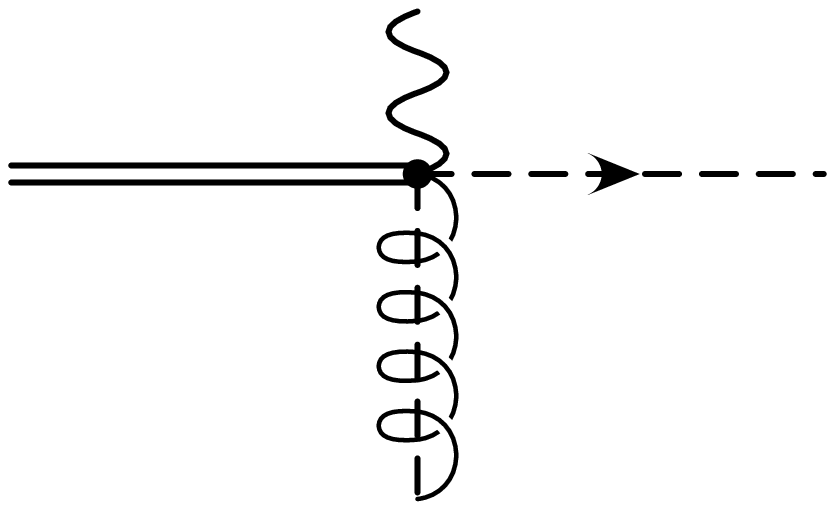} \\
 \gamma_\mu\,\frac{1}{\sla{p}}\,\Gamma\phantom{aaaaa}
 &&=&&\gamma_\mu\,\frac{\sla{n}}{2}\,\frac{1}{n\cdot p}\,\Gamma\phantom{aa}
 &&+&&\gamma_\mu\,\frac{\sla{\nb}}{2}\,\frac{1}{\nb\cdot p}\,\Gamma
\end{aligned}
\end{equation}
For vanishing transverse momentum ($p_\perp=0$), the intermediate quark
propagator takes the form shown in the second line. In the effective theory, 
the first term on the right hand side is represented by a graph where 
the gluon is emitted from a hard-collinear quark line. The second one, 
however, corresponds to a graph where the gluon is emitted from the $B$-type 
current, and contributes to the matching coefficient. Our tree-level results 
are given as follows.

\paragraph{\rm Scalar current:}

\be\label{eq:scalarmatching_tree}
   C^A_S = 1 \,, \qquad C^B_S = - 1 \,.
\ee

\paragraph{\rm Vector current:}

\bea\label{eq:vectormatching}
   C^A_{V1} &=& 1 \,, \qquad C^A_{V2,3} = 0 \,, \nl
   C^B_{V1} &=& 1 \,, \qquad C^B_{V2} = -2 \,, \qquad C^B_{V3} = -x \,,
    \qquad C^B_{V4}=0 \,.
\eea

\paragraph{\rm Tensor current:}

\bea\label{eq:tensormatching}
   C^A_{T1} &=&  1 \,, \!\quad\qquad C^A_{T2,3,4} = 0 \,, \nl
   C^B_{T1} &=& -1 \,, \qquad C^B_{T2} = -4 \,, \qquad C^B_{T3} = -2x \,,
    \qquad C^B_{T4,5,6} = 0 \,, \qquad C^B_{T7} = -4x \,.
\eea
Here $x=2E/m_b$. 
Our tree-level matching coefficients agree with previous results, for $C_i^A$ 
in \cite{Bauer:2000yr}, and for $C_j^B$ in 
\cite{Beneke:2002ph,Pirjol:2002km}.~\footnote{Note that we use square brackets 
around two or more indices to denote antisymmetrization 
($\gamma^{[\mu}\gamma^{\nu]}=\frac12(\gamma^\mu\gamma^\nu-\gamma^\nu\gamma^\mu)$, 
etc.), whereas in \cite{Pirjol:2002km} square brackets around two indices denote 
a commutator  
($\gamma^{[\mu}\gamma^{\nu]} = \gamma^\mu\gamma^\nu-\gamma^\nu\gamma^\mu$, etc.).}

\begin{figure}
\begin{center}
\begin{tabular}{cc}
\includegraphics[width=0.35\textwidth]{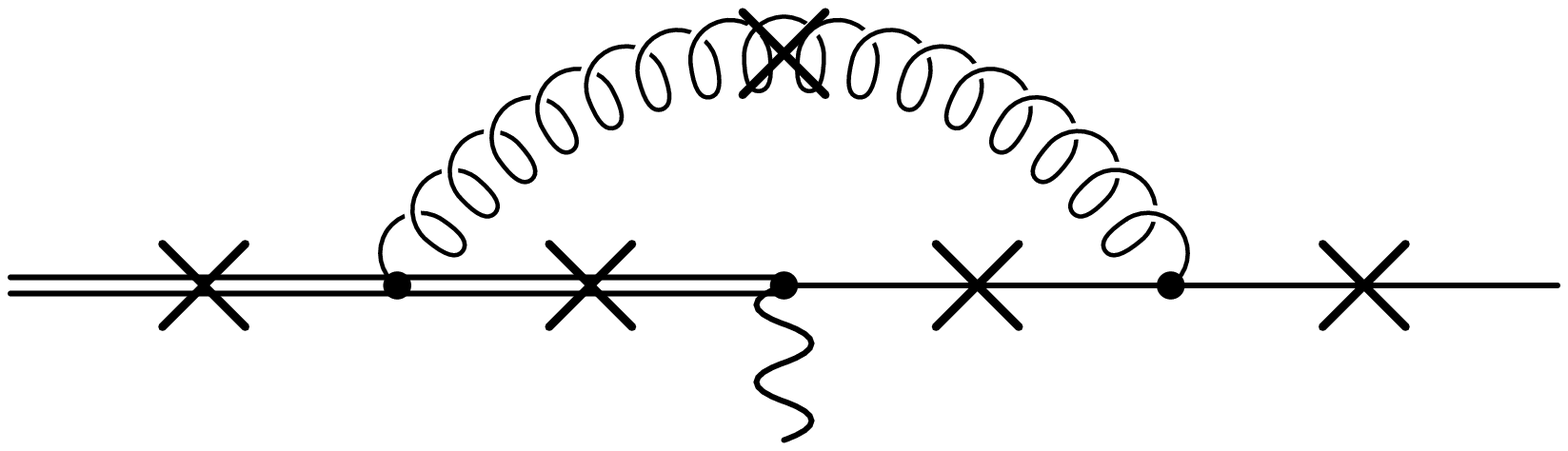} &
\includegraphics[width=0.35\textwidth]{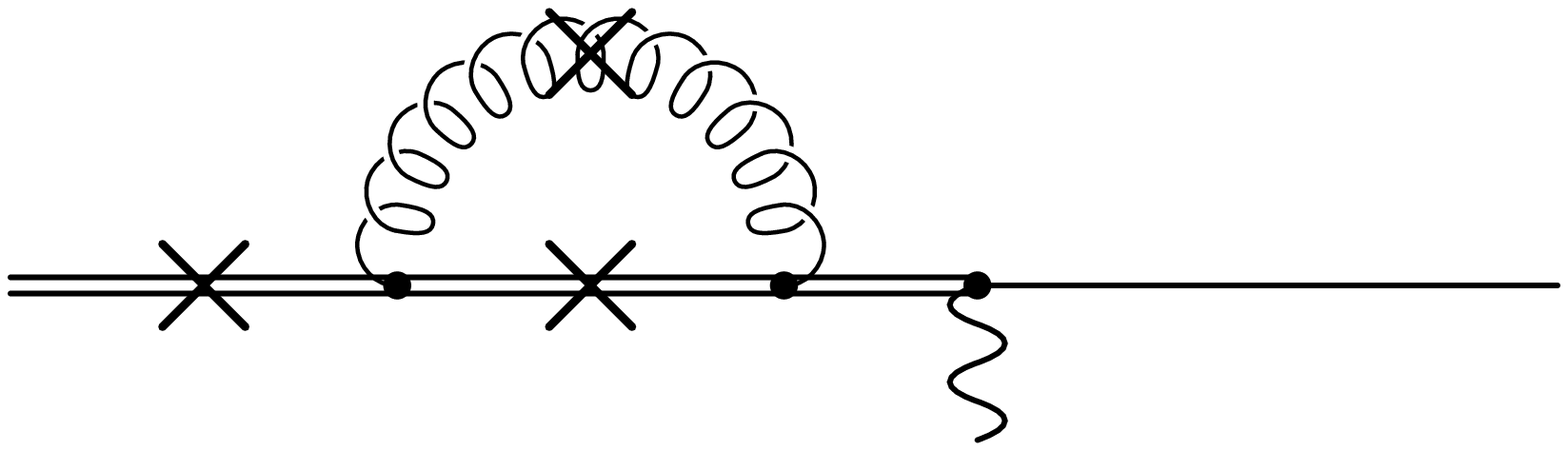}
\end{tabular}
\end{center}
\vspace{-0.5cm}
\caption{One-loop QCD diagrams contributing to the matching calculation for
the subleading scalar current. The external gluon can be attached at any of 
the places marked by a cross.}
\label{fig:full}
\end{figure}

For the resummation of the leading logarithms, the tree-level Wilson 
coefficients are sufficient. However, for the physical value of the $b$-quark 
mass the one-loop matching corrections may turn out to be comparable to the 
effect of leading-order running. To address this question, we evaluate as 
an example the one-loop matching corrections for the scalar current. We 
calculate the decay of a $b$-quark to an energetic light quark and an 
energetic gluon at one-loop order. Throughout this paper we use dimensional 
regularization with $d=4-2\epsilon$ dimensions and employ the 
$\overline{\rm MS}$ scheme to remove ultra-violet (UV) singularities. We 
use the background-field method and perform the calculation in an arbitrary 
covariant gauge. The eight one-loop diagrams contributing to the matrix 
element are shown in Figure~\ref{fig:full}. Each diagram involves at least 
one heavy-quark propagator; diagrams involving only massless propagators 
vanish in dimensional regularization once the external lines are put on the 
mass shell. Note that there is again a contribution in Figure~\ref{fig:full} 
corresponding to gluon emission from a nearly on-shell hard-collinear quark 
line, which is decomposed as in (\ref{eq:decompose}). To correctly identify 
the two parts of the QCD loop diagram, one first considers it for $p^2\ne 0$ 
and then expands around $p^2=0$. This expansion must be done {\em before\/} 
performing loop integrations. This ensures that all effective-theory loop 
diagrams vanish and only the hard part of the QCD diagram is left, which is 
exactly the part that has to be absorbed into the Wilson coefficients. We set 
all perpendicular components of the external momenta to zero and equate the 
effective-theory expression to the QCD result for the three-point function. 
Depending on the polarization of the background gluon field, we thus 
determine either the Wilson coefficient $C_S^A$ of the leading-order current 
operator (for an $A_-$ gluon), or the coefficient $C_S^B$ of the subleading 
current (for an $A_\perp$ gluon). After $\overline{\rm MS}$ subtractions, we 
obtain in the first case (again with $x=2E/m_b$)
\be\label{eq:CSAoneloop}
   C^A_S(E) = 1 + \frac{C_F\alpha_s}{4\pi} \left[ - 2\ln^2\frac{2E}{\mu}
   + 5\ln\frac{2E}{\mu} + 6\ln\frac{\mu_{\rm QCD}}{2E}
   + \frac{3-x}{1-x}\,\ln x - 2\,\mbox{Li}_2(1-x) - \frac{\pi^2}{12}
   \right] ,
\ee
where $\alpha_s\equiv\alpha_s(\mu)$. The fact that this expression, which 
was extracted from a three-point function, agrees with the result obtained 
from the heavy-to-light two-point function \cite{Bauer:2000yr} follows from 
invariance under collinear gauge transformations. Performing the calculation 
for the case of perpendicular gluon polarization, we obtain for the 
coefficient of the operator $J^B_S$
\bea\label{eq:CSBoneloop}
   C^B_S(E,u)
   &=& - C_S^A(E)
    + \frac{C_F\alpha_s}{4\pi}\,\Bigg\{ 4\ln\frac{2E}{\mu} - 4
    + \frac{2x}{1-x}\,\ln x \nl
   &&\mbox{}- \frac{x}{1-x\bar u} \left[ 1
    + \left( \frac{1}{1-x\bar u} + \frac{3-x}{1-x} \right) \ln(x\bar u)
    \right]
    + \frac{2(2-x)}{1-x}\,\frac{\ln\bar u}{u} \Bigg\} \nl
   &+& \left( C_F - \frac{C_A}{2} \right)
    \frac{\alpha_s}{4\pi}\,
    \Bigg\{ \left( 2\ln\frac{2E}{\mu} - 1 + \ln u \right)
    \frac{2\ln u}{\bar u}
    + \frac{2x}{1-x\bar u}\,\ln(x\bar u) - \frac{2\ln\bar u}{u} \nl
   &&\mbox{}+ \frac{2}{x u\bar u} \left[ (1-x u)\,\Big[
    {\rm Li}_2(1-x) - {\rm Li}_2(1-x u) \Big]
    - {\rm Li}_2(1-x\bar u) + \frac{\pi^2}{6} \right] \Bigg\} \,,
\eea
where $u$ corresponds to the fraction of the longitudinal hard-collinear momentum 
carried by the final-state quark. 
We note that at the endpoints, $C_S^B$ diverges 
only logarithmically: $C_S^B\sim\ln^2 u$ at $u=0$ and $C_S^B\sim\ln\bar u$ at 
$\bar u=0$. Also, despite appearances, there is no singularity at $x=1$ or 
$x=1/\bar u$. The scale dependence of $C_S^B$ agrees with a direct analysis 
of operator renormalization given in the following section. We postpone a 
detailed discussion on the relative importance of matching and RG running 
until Section~\ref{sec:solution}, when a solution to the RG equations is at 
hand.

As a final remark before ending this section, we note that the scalar current 
is not an independent operator but can be related to the vector current using 
the equation of motion for the quark fields,
$i\partial_\nu V^\nu=\overline{m}_b(\mu_{\rm QCD})\,S(\mu_{\rm QCD})$, where 
$S(\mu_{\rm QCD})$ denotes the scalar QCD current in (\ref{eq:currentDefs}) 
renormalized at scale $\mu_{\rm QCD}$, and $\overline{m}_b(\mu_{\rm QCD})$ is 
the running $b$-quark mass, both defined in the $\overline{\rm MS}$ scheme. 
Applying this identity to (\ref{Jexpansion}), it follows that to all orders 
in $\alpha_s$ and at leading order in $\Lambda_{\rm QCD}/m_b$
\bea \label{eq:coeffRelations}
   C^A_{V1} + \left( 1 - {E\over m_b} \right) C^A_{V2} + C^A_{V3}
   &=& \frac{\overline{m}_b(\mu_{\rm QCD})}{m_b}\,C^A_S \,, \nl
   C^B_{V1} + \left( 1 - {E\over m_b} \right) C^B_{V2} + C^B_{V3}
   &=& \frac{\overline{m}_b(\mu_{\rm QCD})}{m_b}\,C^B_S \,,
\eea
where at this order there is no difference between the meson mass $m_B$ and
the $b$-quark pole mass $m_b$. It is readily seen from 
(\ref{eq:scalarmatching_tree}) and (\ref{eq:vectormatching}) that these 
relations hold at tree level. At one-loop order
\be\label{eq:ZmZS}
   \frac{\overline{m}_b(\mu_{\rm QCD})}{m_b}
   = 1 + {C_F\alpha_s\over 4\pi} \left( 6\ln{m_b\over\mu_{\rm QCD}} - 4
   \right) .
\ee
Note that the dependence on the scale $\mu_{\rm QCD}$ cancels between the 
running mass $\overline{m}_b(\mu_{\rm QCD})$ and the scalar coefficients 
$C_S^A$ and $C_S^B$ on the right-hand side of the relations in 
(\ref{eq:coeffRelations}). The first of these relations can be verified to 
hold at one-loop order using the results of \cite{Bauer:2000yr}. In 
Section~{\ref{sec:applic}}, we will use these results to deduce the one-loop 
matching coefficients relevant to the vector form factor $F_0(q^2)$, using 
only scalar-current matching calculations.

\section{Anomalous dimensions}
\label{sec:anom}

The matching results derived in the previous section can be trusted at a
high renormalization point $\mu=\mu_h\sim 2E$, at which the Wilson 
coefficients are free of large logarithms and so can be reliably computed 
using fixed-order perturbation theory. In order to evolve the coefficients
down to lower values of $\mu$, one needs to solve the RG equation for the
SCET$_{\rm I}$ current operators. 

The currents $J_i^A$ and $J_j^B$ do not mix under renormalization. To see 
this, we first note that the operators $J_j^B$ cannot mix into $J_i^A$, as 
they are of higher order in power counting than the leading terms of $J_i^A$. 
In principle, the operators $J_i^A$ could mix into $J_j^B$ via time-ordered 
products of either the leading terms of $J_i^A$ with the subleading 
SCET$_{\rm I}$ Lagrangian, or the subleading terms of $J_i^A$ with the 
leading SCET$_{\rm I}$ Lagrangian. However, when $v_\perp^\mu=0$, in both 
cases the resulting time-ordered products have a structure different from 
the $J_j^B$ operators, containing additional perpendicular partial derivatives 
acting on hard-collinear fields. Finally, the various $J^A_i$ operators 
do not mix among themselves, since time-ordered products with the 
SCET$_{\rm I}$ Lagrangian cannot mix the different Dirac structures 
appearing in their definition. As a result, the effective-theory currents 
obey the integro-differential RG equations (summation over $k$ is understood
in the second equation)
\bea\label{eq:evolutionB}
   {d\over d\ln\mu}\,J_i^A(0) &=& - \gamma^A\,J_i^A(0) \,, \nl
   {d\over d\ln\mu}\,J_j^B(u) &=& - \int_0^1\!dv\,\gamma^B_{jk}(u,v)\,
    J_k^{B}(v) \,,
\eea
and the corresponding momentum-space Wilson coefficients satisfy the 
equations
\bea\label{eq:evolutionB_coeff}
   {d\over d\ln\mu}\,C_i^A(E) &=& \gamma^A\,C_i^A(E) \,, \nl
   {d\over d\ln\mu}\,C_j^B(E,v) &=& \int_0^1\!du\,\gamma^B_{kj}(u,v)\,
    C_k^{B}(E,u) \,.
\eea
The anomalous dimensions $\gamma^A$ and $\gamma_{jk}^B$ are calculated by 
isolating the UV divergences in SCET$_{\rm I}$ loop diagrams. The one-loop 
expression for $\gamma^A$ has been calculated previously 
\cite{Bauer:2000yr}, with the result
\be\label{eq:gamma_A}
   \gamma^A = - \Gamma_{\rm cusp}(\alpha_s)\,\ln{\mu\over 2E}
    + \tilde{\gamma}(\alpha_s)
   = {C_F\alpha_s\over\pi} \left( - \ln{\mu\over 2E} - {5\over 4} \right)
    + \order(\alpha_s^2) \,.
\ee
In \cite{Bosch:2003fc} it was argued that the first equality in 
(\ref{eq:gamma_A}) is valid to all orders in $\alpha_s$. The appearance of 
$\ln\mu$ in the anomalous dimension is explained by the theory of light-like 
Wilson loops with cusps. Only a single logarithm appears at any order in the 
strong coupling, with a coefficient governed by the universal cusp anomalous 
dimension $\Gamma_{\rm cusp}$ \cite{Korchemsky:wg}.

Unlike the situation for the $A$-type operators, the anomalous dimensions of 
the $B$-type currents depend on the specific Dirac structure, and there is
non-trivial operator mixing. The renormalization properties of all $B$-type
operators can be discussed by defining two operators
\be
   \J_1(s) = \bar\X(s\bar n)\,\A_{\perp\mu}(0)\,\Gamma_\perp^\mu\,h(0) \,,
    \qquad
   \J_2(s) = \bar\X(s\bar n)\,\A_{\perp\mu}(0)\,\gamma_\perp^\mu
    \gamma_{\perp\nu}\,\Gamma_\perp^\nu\,h(0) \,,
\ee
and their Fourier transforms as defined in (\ref{eq:JBG}). The Dirac
structure $\Gamma_\perp^\mu$ is determined by the specific current
under consideration. It follows from the Feynman rules of SCET$_{\rm
I}$ and the projection properties of the two-component spinor $\X$
that the two operators $\J_i$ close under renormalization. Defining
renormalization constants via $\J_i^{\rm ren}=Z_{ij}\,\J_j^{\rm
bare}$, we obtain
\be
   \bm{Z} = \left( \begin{array}{cc}
    Z_{11} & Z_{12} \\
    0 & ~~Z_{11} + 2(1-\epsilon) Z_{12}
   \end{array} \right)_{\rm \!\!poles} ,
\ee
where, in the $\overline{\rm MS}$ scheme, it is understood that only pole 
terms in the dimensional regulator $\epsilon=2-d/2$ are kept in 
$\bm{Z}-\bm{1}$. We define two anomalous-dimension functions
\be
   \gamma_1 = 2\alpha_s\,\frac{\partial}{\partial\alpha_s}\,Z^{(1)}_{11}
    \,, \qquad
   \gamma_{2} = 2\alpha_s\,\frac{\partial}{\partial\alpha_s}\,Z^{(1)}_{12}
    \,,
\ee
where $\bm{Z}^{(1)}$ is the coefficient of the $1/\epsilon$ pole in $\bm{Z}$. 
Then, at one-loop order, the corresponding anomalous dimension matrix 
determining the running of the currents $\J_i$ reads
\be
   \bm{\gamma}_{\rm 1-loop} = \left( \begin{array}{cc}
   \gamma_1 & \gamma_{2} \\
   0 & ~~\gamma_1 + 2\gamma_{2}
   \end{array} \right) .
\ee
At higher order, the $\order(\epsilon)$ terms arising from the Dirac
algebra have a non-trivial effect on the anomalous dimensions, as discussed 
in \cite{Becher:2000nm}.

In terms of these functions, the anomalous dimension of the scalar current 
is (from now on all results refer to the one-loop approximation)
\be\label{eq:smatrix}
   \gamma^B_S = \gamma_1 + 2\gamma_{2} \,.
\ee
For the vector current, we find the $4 \times 4$ anomalous-dimension matrix
\be\label{eq:vmatrix}
   \bm{\gamma}^B_V = \left( \begin{array}{ccc|c}
    \gamma_1 + 2\gamma_{2} & 0 & 0 & 0 \\
    0 & \gamma_1 + 2\gamma_{2} & 0 & 0 \\
    0 & 0 & \gamma_1+ 2\gamma_{2} & 0 \\
    \hline
    \gamma_{2} & 0 & -\gamma_{2} & \gamma_1
    \end{array} \right) .
\ee
The operators $J_{V1,2,3}^B$ are multiplicatively renormalized with anomalous 
dimension $\gamma_1+2\gamma_{2}$, whereas the operator $J_{V4}^B$ mixes with 
$J_{V1}^B$ and  $J_{V3}^B$. For the form factor analysis, it will be 
convenient to work with the linear combinations 
\bea\label{eq:vectorprime}
  J_{V1}^{B'\mu} &=& J_{V1}^{B\,\mu} - J_{V3}^{B\,\mu}
   = \bar\X(s\nb)\,\calAslash_\perp(r\nb)\,\gamma_\perp^\mu\,h(0) \,, \nl
  J_{V2}^{B'\mu} &=& J_{V2}^{B\,\mu}
   = \bar\X(s\nb)\,\calAslash_\perp(r\nb)\,v^\mu\,h(0) \,, \nl
  J_{V3}^{B'\mu} &=& J_{V3}^{B\,\mu} 
   = \bar\X(s\nb)\,\calAslash_\perp(r\nb)\,\frac{n^\mu}{n\cdot v}\,h(0) \,,
   \nl
  J_{V4}^{B'\mu} &=& 2J_{V4}^{B\,\mu} - J_{V1}^{B\,\mu} + J_{V3}^{B\,\mu}
   = \bar\X(s\nb)\,\gamma_\perp^\mu\,\calAslash_\perp(r\nb)\,h(0) \,,
\eea
which are multiplicatively renormalized. $J_{V1,2,3}^{B'}$ have anomalous 
dimension $\gamma_1+2\gamma_2$, while $J_{V4}^{B'}$ has anomalous dimension 
$\gamma_1$. The corresponding combinations of Wilson coefficients, which are
multiplicatively renormalized with the same anomalous dimensions as the 
respective currents $J_j^{B'}$, read
\vspace{-0.2cm}
\bea\label{Cprime1}
   C_{V1}^{B'} &=& C_{V1}^B + \frac12\,C_{V4}^B \,, \qquad
    C_{V2}^{B'} = C_{V2}^B \,, \qquad
    C_{V3}^{B'} = C_{V3}^B + C_{V1}^B \,, \nl 
   C_{V4}^{B'} &=& \frac12\,C_{V4}^B \,.
\eea
Similarly, for the tensor current we find that the operators $J_{T1,2,3,4}^B$
are multiplicatively renormalized with anomalous dimension 
$\gamma_1+2\gamma_2$, whereas the operators $J_{T5,6,7}^B$ mix with 
$J_{T1,2,3,4}^B$. As in the vector case, it will be convenient to change the 
operator basis, defining 
\bea\label{eq:tensorprime}
  J_{T1}^{B'\mu\nu} &=& J_{T1}^{B\,\mu\nu} + 2J_{T3}^{B\,\mu\nu}
   - 2J_{T4}^{B\,\mu\nu}
   = \bar\X(s\nb)\,\calAslash_\perp(r\nb)\,
   \gamma_\perp^{[\mu}\gamma_\perp^{\nu]}\,h(0) \,, \nl
  J_{T2}^{B'\mu\nu} &=& J_{T2}^{B\,\mu\nu} + J_{T4}^{B\,\mu\nu}
   = \bar\X(s\nb)\,\calAslash_\perp(r\nb)\,
   v^{[\mu}\gamma_\perp^{\nu]}\,h(0)\,, \nl
  J_{T3}^{B'\mu\nu} &=& J_{T3}^{B\,\mu\nu}
   = \bar\X(s\nb)\,\calAslash_\perp(r\nb)\,
   {n^{[\mu}\gamma_\perp^{\nu]}\over n\cdot v}\,h(0) \,, \nl
  J_{T4}^{B'\mu\nu} &=& J_{T4}^{B\,\mu\nu}
   = \bar\X(s\nb)\,\calAslash_\perp(r\nb)\,{n^{[\mu}v^{\nu]}\over n\cdot v}\,
   h(0) \,,\nl
  J_{T5}^{B'\mu\nu} &=& -2 J_{T5}^{B\,\mu\nu} + J_{T1}^{B\,\mu\nu}
   + 2 J_{T3}^{B\,\mu\nu} - 2 J_{T4}^{B\,\mu\nu} + 2 J_{T7}^{B\,\mu\nu}
   = \bar\X(s\nb)\,\A_{\perp\alpha}(r\nb)\,
   \gamma_\perp^{[\alpha}\gamma_\perp^\mu\gamma_\perp^{\nu]}\,h(0) \,, 
   \nl[0.2cm]
  J_{T6}^{B'\mu\nu} &=& -2 J_{T6}^{B\,\mu\nu} - J_{T2}^{B\,\mu\nu}
   - J_{T4}^{B\,\mu\nu}
   = \bar\X(s\nb)\,v^{[\mu}\gamma_\perp^{\nu]}\,\calAslash_\perp(r\nb)\,
   h(0) \,, \nl
  J_{T7}^{B'\mu\nu} &=& -2 J_{T7}^{B\,\mu\nu} - J_{T3}^{B\,\mu\nu}
   = \bar\X(s\nb)\,\frac{n^{[\mu}\gamma_\perp^{\nu]}}{n\cdot v}\,
   \calAslash_\perp(r\nb)\,h(0) \,. 
\eea
These operators are multiplicatively renormalized, $J_{T1,2,3,4}^{B'}$ with 
anomalous dimension $\gamma_1+2\gamma_2$, and $J_{T5,6,7}^{B'}$ with 
anomalous dimension $\gamma_1$. The corresponding combinations of Wilson 
coefficients are
\bea\label{Cprime2}
   C_{T1}^{B'} &=& C_{T1}^B + \frac12\,C_{T5}^B \,, \qquad
    C_{T2}^{B'} =  C_{T2}^B - \frac12\,C_{T6}^B \,, \qquad
    C_{T3}^{B'} =  C_{T3}^B - 2 C_{T1}^B - \frac12\,C_{T5}^B
    - \frac12\,C_{T7}^B \,, \nl
   &&\hspace{3.32cm}
    C_{T4}^{B'} = C_{T4}^B + 2 C_{T1}^B - C_{T2}^B \,, \nl
   C_{T5}^{B'} &=& -\frac12\,C_{T5}^B \,, \qquad
    C_{T6}^{B'} = -\frac12\,C_{T6}^B \,, \qquad
    C_{T7}^{B'} = -\frac12\,C_{T7}^B - \frac12\,C_{T5}^B \,.
\eea
Note that the tree-level matrix elements of the ``evanescent'' operator 
$J_{T5}^{B'}$ vanish in $d=4$ dimensions, and after additional finite 
renormalizations the same is true once loop corrections are included. This 
operator will not be relevant for our leading-order analysis, but it would 
enter a next-to-leading order calculation employing dimensional 
regularization. We will mention evanescent operators again in 
Section~\ref{sec:SCETII}, when we discuss the basis of SCET$_{\rm II}$
four-quark operators relevant to the hard-scattering contributions. 

\begin{figure}
\begin{center}
\includegraphics[width=0.7\textwidth]{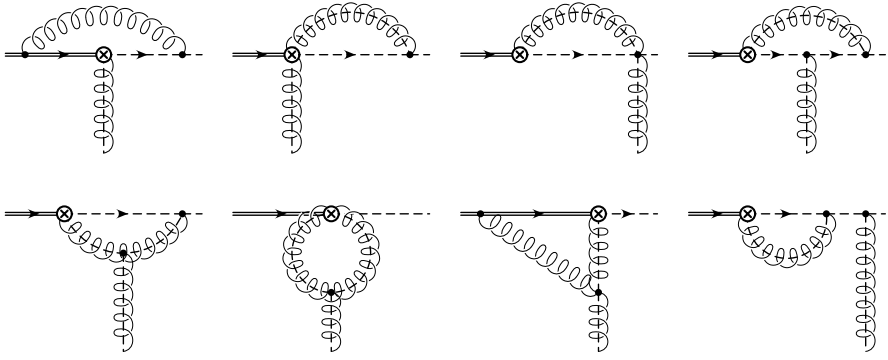}
\end{center}
\vspace{-0.5cm}
\caption{$\mbox{SCET}_{\rm I}$ graphs contributing to the anomalous 
dimensions of the subleading heavy-collinear currents $J_j^B$, represented by 
a crossed circle. Full lines denote soft fields, dashed lines hard-collinear 
fields.}
\label{fig:graphs}
\end{figure}

To calculate the one-loop anomalous dimensions, we evaluate the UV poles of 
the diagrams shown in Figure~\ref{fig:graphs} in dimensional regularization, 
treating the external gluon as a background field. The resulting expressions 
for the integral operators $\gamma_1$ and $\gamma_2$ can be written as
\bea \label{eq:gammaiDef}
   \gamma_1(u,v) &=& u\,V_1(u,v) + \delta(u-v)\,W(E,u) \,, \nl
   2\gamma_{2}(u,v) &=& u\,V_2(u,v) \,,
\eea
where
\bea\label{eq:oneloopVW}
   V_1(u,v) &=& \left( C_F - \frac{C_A}{2} \right) {\alpha_s\over\pi}
    \left( - {\theta(1-u-v)\over \bar{u}\bar{v}} \right) \nl
   &&\mbox{}- {C_A\over 2}{\alpha_s\over \pi} \left\{ {1\over uv}
    \left[ u\,{\theta(v-u)\over (v-u)} + v\,{\theta(u-v)\over (u-v)}
    \right]_+ - {\theta(v-u)\over v\bar{u}} - {\theta(u-v)\over u\bar{v}}
    \right\} , \nl
   V_2(u,v) &=& \left( C_F - \frac{C_A}{2} \right) {\alpha_s\over \pi}
    \left[ \left( 1 + {1\over\bar{u}} + {1\over\bar{v}} \right)
    \theta(1-u-v) + {\bar{u}\bar{v}\over uv}\,\theta(u+v-1) \right] \nl
   &&\mbox{}-{C_A\over 2}{\alpha_s\over \pi} \left[ \frac{\bar{u}}{u}
    \left( 1 + \frac{1}{\bar{v}} \right) \theta(u-v)
    + \frac{\bar{v}}{v} \left( 1 + \frac{1}{\bar{u}} \right) \theta(v-u)
    \right] , \nl
   W(E,u) &=& C_F{\alpha_s\over\pi} \left( -\ln{\mu\over 2E} - {5\over 4}
    + \ln u \right) - {C_A\over 2}{\alpha_s\over \pi} \left(
    \ln\frac{u}{\bar{u}} - 1 \right) .
\eea
For symmetric functions $g(u,v)$ the plus distribution is defined to act on 
test functions $f(v)$ as
\be\label{eq:plusDistrib}
   \int dv\,[g(u,v)]_+\,f(v) = \int dv\,g(u,v)\,\big[ f(v) - f(u) \big] \,.
\ee
It is remarkable that the functions $V_i(u,v)$ are symmetric in their 
arguments. This fact is not accidental, but can be traced back to a residual 
conformal symmetry in the effective theory when interactions with the soft 
heavy quark are ignored. Also, since soft gluons are unable to change the 
large component of hard-collinear momenta, the one-loop conformal-symmetry 
breaking term $\delta(u-v)W(E,u)$ is local with respect to $u$. Details of 
the conformal-symmetry arguments will be presented in 
Section~\ref{section:conformal}.

From (\ref{eq:evolutionB}), it follows that the evolution equation for the 
current eigenvectors $J_j^{B'}$ takes the form
\be\label{eq:JBevolve}
   {d\over d\ln{\mu}}\,J_j^{B'}(u)
   = - \int_0^1\!dv\,u\,V_\Gamma(u,v)\,J_j^{B'}(v) - W(E,u)\,J_j^{B'}(u) \,,
\ee
where $V_\Gamma$ is a linear combination of $V_1$ and $V_2$ determined by the 
anomalous dimension of the operator $J_j^{B'}$. For reasons that will become 
clear later, we use the notation $\Gamma=\,\parallel,\perp$ and denote 
$V_\parallel=V_1+V_2$ for operators with anomalous dimension 
$\gamma_1+2\gamma_2$, and $V_\perp=V_1$ for operators with anomalous 
dimension $\gamma_1$. The corresponding equation for the eigenvectors 
$C_j^{B'}$ of Wilson coefficients reads
\be\label{eq:CBevolve}
   {d\over d\ln\mu}\,C_j^{B'}(E,v)
   = \int_0^1\!du\,u\,V_\Gamma(u,v)\,C_j^{B'}(E,u)
   + W(E,v)\,C_j^{B'}(E,v) \,.
\ee
To gain more insight into the structure of the conformal-symmetry breaking 
term $W(E,u)$, we may consider the field redefinitions 
\cite{Bauer:2001yt,Becher:2003kh}
\be
   \xi_{hc}(x) = S_n(x_-)\,\xi_{hc}^{(0)}(x) \,, \qquad
   A_{hc}^\mu(x) = S_n(x_-)\,A_{hc}^{(0)\mu}(x)\,S_n^\dagger(x_-) \,, \qquad
   h(x) = S_v(x)\,h^{(0)}(x) \,,
\ee
which decouple soft gluons from the leading-order hard-collinear and 
heavy-quark Lagrangians. Here $S_n$ and $S_v$ are soft Wilson lines in the 
$n$ and $v$ directions, respectively. From (\ref{BBs}), it follows that in 
terms of the new fields the $B$-type operators take the form
\be
   J^B_j(s,r) = \bar\X^{(0)}(s\nb)\,\A_{\perp\mu}^{(0)}(r\nb)\,
   \Gamma_{\perp}^\mu \big[ S_n^\dagger\,S_v \big](0)\,h^{(0)}(0) \,.
\ee
The combination $[S_n^\dagger\,S_v](0)$ represents a closed loop with a cusp 
at $x=0$ formed by two Wilson lines in the $v$ and $n$ directions. The 
anomalous dimension of this object is given by the universal cusp anomalous 
dimension times a logarithm of the soft scale \cite{Korchemsky:wg}. After 
adding a contribution from the hard-collinear sector necessary to
eliminate the dependence on the infra-red regulator $(-p_{hc}^2)$
(c.f. \cite{Becher:2003kh,Bauer:2000ew},), the result is
\be
   \Gamma_{\rm cusp} \left[ \ln{ 2v\cdot p_{hc}\,\mu\over(-p_{hc}^2)}
   + \ln{(-p_{hc}^2)\over\mu^2} \right]
   = \Gamma_{\rm cusp}\,\ln{2v\cdot p_{hc}\over\mu} \,,
\ee
with $p_{hc}$ a hard-collinear momentum. From these considerations, we 
conclude that to all orders in perturbation theory 
\be\label{eq:decompW}
   W(E,u) = - \Gamma_{\rm cusp}(\alpha_s)\,\ln{\mu\over 2E}
   + w(u,\alpha_s),
\ee
with the one-loop expression for $w$ determined from (\ref{eq:oneloopVW}).

\section{Renormalization-group evolution in SCET$_{\bf I}$}
\label{sec:solution}

The anomalous dimensions obtained in the previous section allow us to solve 
the RG evolution equations (\ref{eq:JBevolve}) and (\ref{eq:CBevolve}) at 
leading order in RG-improved perturbation theory. At this order the leading
double and single logarithmic contributions are resummed to all orders in 
perturbation theory. Technically, this means that one must compute the 
matching coefficients at tree level, the anomalous dimension kernels 
$V_i(u,v)$ at one-loop order, and the cusp anomalous dimension entering the 
function $W(E,u)$ at two-loop order. 
In practice, one wants to impose matching conditions for the coefficient functions 
$C_j^{B'}$ at a high scale $\mu_h\sim 2E$ and then evolve the result down to an 
intermediate scale 
$\mu_i\sim\sqrt{2E\Lambda_{\rm QCD}}$, at which SCET$_{\rm I}$ is matched 
onto the low-energy effective theory SCET$_{\rm II}$. While the 
integro-differential evolution equations can be solved numerically, we find 
it instructive to also discuss a formal analytic solution to these 
equations.

\subsection{Eigenfunctions and eigenvalues of the evolution operator}
\label{section:appendix}

It will be convenient to rewrite (\ref{eq:JBevolve}) and (\ref{eq:CBevolve}) 
in the somewhat obscure form
\bea\label{eq:evolution_JC}
   {d\over d\ln{\mu}}\,{J_j^{B'}(u)\over u\bar{u}}
   &=& - \int_0^1\!dv\,v\bar v^2\,{V_\Gamma(u,v)\over\bar{u}\bar{v}}\,
    {J_j^{B'}(v)\over v\bar{v}} - W(E,u)\,{J_j^{B'}(u)\over u\bar{u}} \,, \nl
   {d\over d\ln{\mu}}\,{C_j^{B'}(E,u)\over \bar{u}}
   &=& \int_0^1\!dv\,v\bar v^2\,{V_\Gamma(v,u)\over\bar{u}\bar{v}}\,
    {C_j^{B'}(E,v)\over\bar{v}} + W(E,u)\,{C_j^{B'}(E,u)\over\bar{u}} \,.
\eea
Because of the symmetry property $V_i(u,v)=V_i(v,u)$, it follows that the 
operator eigenfunctions $[J_n^{B'}(u)/u\bar{u}]$ have the same form as the 
coefficient eigenfunctions $[C_n^{B'}(u)/\bar{u}]$, but come with 
eigenvalues of the opposite sign. We consider first the hypothetical case 
where $W(E,u)=0$ and focus on the non-diagonal terms in the evolution 
kernels, $V_i(u,v)$ with $i=1,2$. We will show that the general solution to 
the eigenvalue equation
\be\label{eq:eigenvalue_eqn}
   \int_0^1\!du\,u\bar u^2\,{V_i(u,v)\over\bar{u}\bar{v}}\,\psi_n(u)
   = \lambda_{i,n}\,\psi_n(v)
\ee
is given by
\be\label{eq:eigenfunctions}
   \psi_n(u) = \sqrt{2(n+2)(n+3)\over n+1}\,P_n^{(2,1)}(2u-1) \,,
\ee
where $P_n^{(2,1)}$ are Jacobi (hyper-geometric) polynomials, and the 
eigenfunctions $\psi_n$ are normalized according to
\be\label{normint}
   \langle\psi_n|\psi_m\rangle
   \equiv \int_0^1\!du\,u\bar u^2\,\psi_n(u)\,\psi_m(u) = \delta_{nm} \,.
\ee
The eigenvalues for the kernels $V_1$ and $V_2$ may be expressed in the 
closed form
\bea\label{eq:eigenvalues}
   \lambda_{1,n} &=& \left( C_F - \frac{C_A}{2} \right) {\alpha_s\over\pi}\,
    {(-1)^{n+1}\over n+2} - {C_A\over 2}\,{\alpha_s\over\pi}
    \left( 1-2H_{n+1} - {1\over n+2} \right) , \nl
   \lambda_{2,n} &=& \left( C_F - \frac{C_A}{2} \right) {\alpha_s\over\pi}\,
    {(-1)^n(n^2+4n+5)\over(n+1)(n+2)(n+3)}
    - {C_A\over 2}\,{\alpha_s\over\pi}\,{2\over(n+1)(n+3)} \,,
\eea
where $H_n=\sum_{m=1}^n 1/m$ are the harmonic numbers. Using the solution to 
the eigenvalue equation for $V_i$, we then present a formal algebraic 
solution to the evolution equation for the SCET$_{\rm I}$ coefficient 
functions.

Let us now prove the statements just made. 
We proceed along lines similar to the diagonalization of the evolution kernel for 
the pion LCDA, as analyzed  in \cite{Lepage:1980fj} (see in particular Appendix D). 
From the eigenvalue equation 
(\ref{eq:eigenvalue_eqn}) and the symmetry of $V_i(u,v)$ it follows that 
eigenfunctions belonging to different eigenvalues must be orthogonal in the 
measure $u\bar u^2\,du$ on the interval $0\le u\le 1$, as shown in 
(\ref{normint}). To show that the eigenfunctions take the form 
(\ref{eq:eigenfunctions}), we consider the integrals
\be
   {\cal I}_{i,n}(v) = \int_0^1\!du\,u\bar u^2\,
   {V_i(u,v)\over\bar{u}\bar{v}}\,(2u-1)^n \,.
\ee
We will show below that ${\cal I}_{i,n}(v)$ is a polynomial in $(2v-1)$ of 
degree $n$. Hence, in operator notation, with 
$\langle u|n\rangle\equiv (2u-1)^n$ and 
$\langle u|V_i|n\rangle\equiv{\cal I}_{i,n}(u)$, we have 
\be
   V_i|n\rangle = \sum_{m=0}^n (V_i)_{mn}|m\rangle
\ee
with $(V_i)_{mn}=0$ for $m>n$. In particular, $|0\rangle$ is an 
eigenfunction with eigenvalue $(V_i)_{00}$. By induction, for each $n\ge 1$ 
there is a corresponding eigenvalue given by the diagonal matrix element 
$(V_i)_{nn}$ with eigenfunction proportional to the linear combination of 
$|0\rangle,\dots,|n\rangle$ that is orthogonal to each of 
$|0\rangle,\dots,|n-1\rangle$. The result (\ref{eq:eigenfunctions}) follows 
since the Jacobi polynomials $P_n^{(2,1)}(2u-1)$ form the unique extension 
of the constant function to a basis of orthogonal polynomials with measure 
$u\bar{u}^2\,du$ on the unit interval. As a byproduct of this analysis, the 
coefficient of $(2v-1)^n$ in the expansion of ${\cal I}_{i,n}(v)$ is 
identified with the $n$-th eigenvalue $\lambda_{i,n}$.

To evaluate ${\cal I}_{i,n}$, it is convenient to introduce new variables 
$x=2u-1$ and $y=2v-1$. The resulting integrals yield
\bea\label{eq:I_In}
   {\cal I}_{1,n} &=& \left( C_F - \frac{C_A}{2} \right) {\alpha_s\over\pi}\,
    (-1)^{n+1}\sum_{m=0}^n {m+1\over (n+1)(n+2)}\,y^m \nl
   &&\mbox{}- {C_A\over 2}\,{\alpha_s\over\pi} \left[ (1 - 2 H_{n+1})\,y^n
    + \sum_{m=0}^{n-1}
    {\left[ 1 + (-1)^{n-m} \right]\,(m+1)\over(n-m)(n+1)}\,y^m
    - \!\sum_{m=0}^n {m+1\over(n+1)(n+2)}\,y^m \right] , \nl
   {\cal I}_{2,n} &=& \left( C_F - \frac{C_A}{2} \right) {\alpha_s\over\pi}\,
    {(-1)^n \over 2(n+1)(n+2)(n+3)} \sum_{m=0}^n \left[
    (-1)^{n-m} + 9 + 4(n+m) + 2nm \right] y^m \nl
   &&\mbox{}- {C_A\over 2}\,{\alpha_s\over\pi}\,{1\over 2(n+1)(n+2)(n+3)}
    \sum_{m=0}^n \left[ (5+2n)(-1)^{n-m} + 3 + 2m \right] y^m \,.
\eea
Inspection of these results shows that the coefficient of $y^m$ in
${\cal I}_{i,n}$ indeed vanishes for $m>n$, while the coefficient of $y^n$ 
gives the eigenvalues (\ref{eq:eigenvalues}).

\subsection{Solution to the evolution equation}

Due to the presence of the conformal-symmetry breaking term $W(E,u)$, the
eigenfunctions of the full anomalous-dimension operators
$\gamma_1+2\gamma_2$ and $\gamma_1$ cannot be written in 
closed form. We will now show how a formal algebraic solution can be 
obtained. We begin by isolating the strong $\mu$ dependence of the 
coefficient functions due to $\Gamma_{\rm cusp}$, writing the solution to the 
evolution equation (\ref{eq:evolution_JC}) as
\be\label{eq:decomp_C}
   C_j^{B'}(E,u,\mu) = \left({2E\over \mu_h}\right)^{a(\mu_h,\mu)}
   e^{S(\mu_h,\mu)} \int_0^1 dv\,U_\Gamma(u,v,\mu_h,\mu)\,
   C_j^{B'}(E,v,\mu_h) \,,
\ee
where
\be\label{eq:defS}
   S(\mu_1,\mu_2) = -\int_{\alpha_s(\mu_1)}^{\alpha_s(\mu_2)}
   {d\alpha\over\beta(\alpha)}\,\Gamma_{\rm cusp}(\alpha)
   \int_{\alpha_s(\mu_1)}^\alpha {d\alpha^\prime\over\beta(\alpha^\prime)} \,,
\ee
with $\beta(\alpha_s)\equiv d\alpha_s/d\ln\mu$, is a universal Sudakov factor, 
and 
\be\label{eq:def_a}
   a(\mu_1,\mu_2) = \int_{\alpha_s(\mu_1)}^{\alpha_s(\mu_2)}
   {d\alpha\over \beta(\alpha)}\,\Gamma_{\rm cusp}(\alpha) \,. 
\ee
Note that $S(\mu_1,\mu_2)$ is negative, since 
$\Gamma_{\rm cusp}(\alpha_s)\ge 0$ \cite{Belitsky:2003ys}. At leading order 
in RG-improved perturbation theory one finds \cite{Bosch:2003fc}
\bea\label{VHres}
   S(\mu_h,\mu) 
   &=& \frac{\Gamma_0}{4\beta_0^2} \left[ \frac{4\pi}{\alpha_s(\mu_h)}
    \left( 1 - \frac{1}{r_1} - \ln r_1 \right)
    + \frac{\beta_1}{2\beta_0}\,\ln^2 r_1
    - \left( \frac{\Gamma_1}{\Gamma_0} - \frac{\beta_1}{\beta_0} \right) 
    (r_1-1-\ln r_1) \right] , \nl
  a(\mu_h,\mu) &=& - \frac{\Gamma_0}{2\beta_0}\,\ln r_1 \,,
\eea
where $r_1=\alpha_s(\mu)/\alpha_s(\mu_h)\ge 1$. The relevant RG coefficients 
arising in the perturbative expansion of the cusp anomalous dimension and 
the $\beta$ function,
\be\label{eq:expansion_coeffs}
   \beta(\alpha_s) = -2\alpha_s\sum_{n=0}^\infty \beta_n
    \left( {\alpha_s\over 4\pi} \right)^{n+1} , \qquad
   \Gamma_{\rm cusp}(\alpha_s) = \sum_{n=0}^\infty \Gamma_n
    \left( {\alpha_s\over 4\pi} \right)^{n+1} ,
\ee
are $\Gamma_0=4C_F$, 
$\Gamma_1=4C_F\,[({67\over 9}-{\pi^2\over 3})\,C_A-{20\over 9}\,T_F n_f]$, 
and $\beta_0={11\over 3}\,C_A-{4\over 3}\,T_F n_f$, 
$\beta_1={34\over 3}\,C_A^2-{20\over 3}\,C_A T_F n_f-4C_F T_F n_f$. We take 
$n_f=4$ as the number of light quark flavors, even at low renormalization 
scales. 

The remaining evolution is governed by a function $U_\Gamma(u,v,\mu_h,\mu)$ 
with initial condition $U_\Gamma(u,v,\mu_h,\mu_h)=\delta(u-v)$ at the 
high-energy matching scale $\mu=\mu_h$. From (\ref{eq:evolution_JC}) we find 
the corresponding RG equation
\be\label{eq:evolution_fu}
   {d\over d\ln\mu}\,\frac{U_\Gamma(u,v,\mu_h,\mu)}{\bar u}
   = \int_0^1\!dy\,y\bar y^2\,\frac{V_\Gamma(y,u)}{\bar y\bar u}\,
   \frac{U_\Gamma(y,v,\mu_h,\mu)}{\bar y}
   + w(u)\,\frac{U_\Gamma(u,v,\mu_h,\mu)}{\bar u} \,,
\ee
where $w(u)$ is defined via (\ref{eq:decompW}). As before, the subscript 
$\Gamma=\,\parallel,\perp$ serves as a reminder that we must distinguish 
two cases of evolution functions corresponding to the different eigenvalues 
of the anomalous-dimension matrices. A formal solution of this equation may 
be constructed by expanding $U_\Gamma(u,v,\mu_h,\mu)$ on the basis of 
eigenfunctions $\psi_n(u)$,
\be\label{Udecomp}
   \frac{U_\Gamma(u,v,\mu_h,\mu)}{\bar u}
   = \sum_n \psi_n(u)\,u_n(v,\mu_h,\mu) \,, 
\ee
where
\be\label{un}
   u_n(v,\mu_h,\mu) = \int_0^1\!du\,u\bar u^2\,\psi_n(u)\,
   \frac{U_\Gamma(u,v,\mu_h,\mu)}{\bar u}
\ee
follows from (\ref{normint}). Inserting this expansion into the evolution 
equation, and projecting out the coefficients $u_n$, we obtain
\be\label{unevol}
   {d\over d\ln\mu}\,u_m(v,\mu_h,\mu) = \lambda_{\Gamma,m}\,u_m(v,\mu_h,\mu)
   + \sum_n w_{mn}\,u_n(v,\mu_h,\mu) \,, 
\ee
where $\lambda_{\Gamma,n}=\lambda_{1,n}+\lambda_{2,n}$ or 
$\lambda_{\Gamma,n}=\lambda_{1,n}$ depending on the anomalous-dimension 
eigenvalue, and the quantities $w_{mn}$ are given by the overlap integrals
\be
   w_{mn} = \int_0^1\!dv\,v\bar v^2\,\psi_m(v)\,w(v)\,\psi_n(v) \,.
\ee
Collecting the elements $w_{mn}$ into an infinite-dimensional matrix 
$\bm{w}$, and the eigenvalues $\lambda_{\Gamma,n}$ into the diagonal matrix 
$\bm{\lambda}_\Gamma$, we can write the formal solution to (\ref{unevol}) in 
the form
\be\label{eq:finalC}
   u_m(v,\mu_h,\mu) = \sum_n v\bar v\,\psi_n(v) 
   \left[ \mbox{P}\exp\left(
   \int_{\alpha_s(\mu_h)}^{\alpha_s(\mu)} \frac{d\alpha}{\beta(\alpha)}\,
   \big[ \bm{\lambda}_{\Gamma}(\alpha) + \bm{w}(\alpha) \big] \right) 
   \right]_{mn} ,
\ee
where the initial condition at the matching scale $\mu=\mu_h$ follows from
(\ref{un}).\footnote{The expansion of $C^{B^\prime}_j(u)/\bar u$ on the 
truncated basis of eigenfunctions $\{\psi_n(u),\,0\le n\le N\}$ converges in 
the limit $N\to\infty$, with the norm (\ref{normint}), provided that 
$\int_0^1\!du\,u\bar{u}^2\,|C^{B^\prime}(u)/\bar{u}|^2<\infty$. Since it is 
only the convolution over the product of $C_j^{B^\prime}$ with the jet 
functions and meson LCDAs which appears in the final expression for the form 
factors, cf.~(\ref{eq:factorization}) and (\ref{eq:resumT}), this condition 
is more restrictive than necessary on the endpoint behavior of the 
coefficient functions, and this ``norm'' sense of convergence is stronger 
than we require.}
The symbol ``P'' denotes coupling-constant ordering, with 
$\bm{\lambda}_\Gamma(\alpha)+\bm{w}(\alpha)$ appearing to the left of 
$\bm{\lambda}_\Gamma(\alpha')+\bm{w}(\alpha')$ when $\alpha>\alpha'$. 
Combining this result with (\ref{Udecomp}) provides a complete solution for 
the evolution function.

For a leading-order solution, we may expand the anomalous dimensions and 
$\beta$ function to one-loop order, defining as usual
\be
   \lambda_{i,n}(\alpha_s) = {\alpha_s\over 4\pi}\,\lambda_{i,n}^{(0)}
    + \dots \,, \qquad
   \bm{w}(\alpha_s) = {\alpha_s\over 4\pi}\,\bm{w}^{(0)} + \dots \,.
\ee
Then the ordered exponential in (\ref{eq:finalC}) may be written as
\be
   \mbox{P}\exp\left(
   \int_{\alpha_s(\mu_h)}^{\alpha_s(\mu)} \frac{d\alpha}{\beta(\alpha)}\,
   \big[ \bm{\lambda}_{\Gamma}(\alpha) + \bm{w}(\alpha) \big] \right) 
   = \bm{U} \exp\bigg[ 
   - \frac{(\bm{\lambda}_{\Gamma}^{(0)}+\bm{w}^{(0)})_{\rm diag}}{2\beta_0}\,
   \ln r_1 \bigg] \bm{U}^{-1} \,,
\ee
where $\bm{U}$ is the unitary matrix that diagonalizes 
$\bm{\lambda}_\Gamma^{(0)}+\bm{w}^{(0)}$, i.e.\ 
$\bm{\lambda}_\Gamma^{(0)}+\bm{w}^{(0)}%
=\bm{U}\,(\bm{\lambda}_\Gamma^{(0)}+\bm{w^{(0)}})_{\rm diag}\,\bm{U}^{-1}$. 
Since $\bm{\lambda}_\Gamma^{(0)}+\bm{w}^{(0)}$ is a real symmetric matrix 
it can be diagonalized with real eigenvalues, which we collect in the 
diagonal matrix $(\bm{\lambda}_\Gamma^{(0)}+\bm{w}^{(0)})_{\rm diag}$. 
Finally, we can simplify the answer further by using that at tree level the 
initial conditions for the Wilson coefficients at the matching scale 
$\mu=\mu_h$ collected in 
(\ref{eq:scalarmatching_tree})--(\ref{eq:tensormatching}) are independent of 
$u$: $C_j^{B^\prime}(E,u,\mu_h)\equiv C_j^{B^\prime}(E,\mu_h)$. We then 
obtain the final result (valid at leading order in RG-improved perturbation 
theory)
\be \label{eq:CB_LO}
   C_j^{B'}(E,u,\mu) = \left({2E\over \mu_h}\right)^{a(\mu_h,\mu)}\,
    e^{S(\mu_h,\mu)}\,U_\Gamma(u,\mu_h,\mu)\,
   C_j^{B'}(E,\mu_h) \,,
\ee
where
\bea\label{ULO}
   U_\Gamma(u,\mu_h,\mu)
   &\equiv& \int_0^1\!dv\,U_\Gamma(u,v,\mu_h,\mu) \\
   &&\hspace{-2.1cm}
    = \sum_{m,n} \bar u\,\psi_m(u)\,\sqrt{\frac{2}{(n+1)(n+2)(n+3)}}
    \left( \bm{U} \exp\left[ 
    - \frac{(\bm{\lambda}_{\Gamma}^{(0)}+\bm{w}^{(0)})_{\rm diag}}
           {2\beta_0}\,
    \ln r_1 \right] \bm{U}^{-1} \right)_{mn} . \nonumber
\eea
In practice we will truncate the basis of eigenfunctions, so that 
$\bm{\lambda}_{\Gamma}^{(0)}+\bm{w}^{(0)}$ becomes a 
finite-dimensional matrix, which can be diagonalized without much 
difficulty. 

\begin{figure}
\begin{center}
\begin{tabular}{cc}
\psfrag{x}{$u$}\psfrag{y}[b]{$U_\Gamma$}
\includegraphics[width=0.48\textwidth]{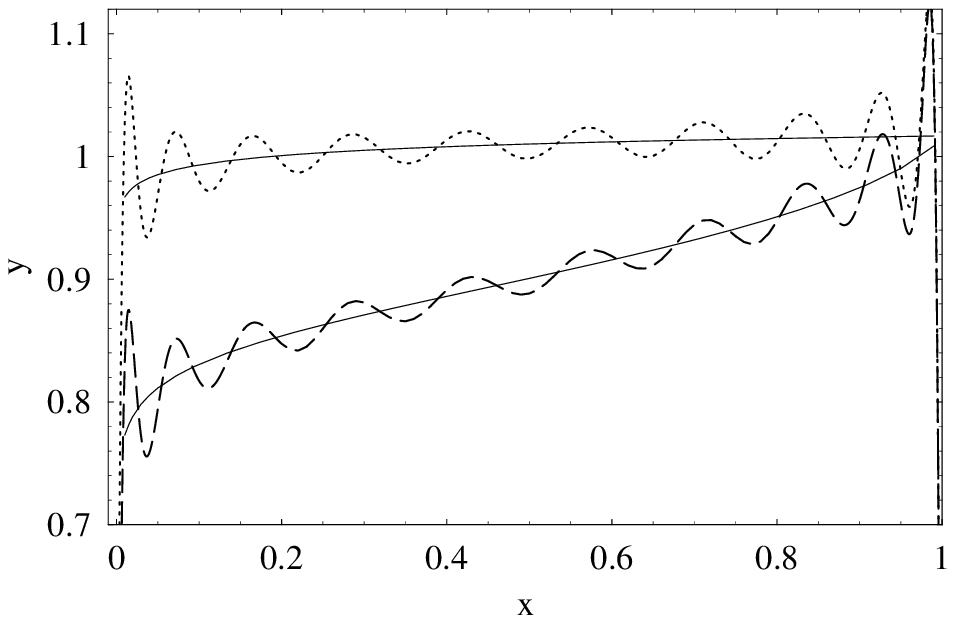} &
\psfrag{x}{$u$}\psfrag{y}[b]{$U_\Gamma$}
\includegraphics[width=0.48\textwidth]{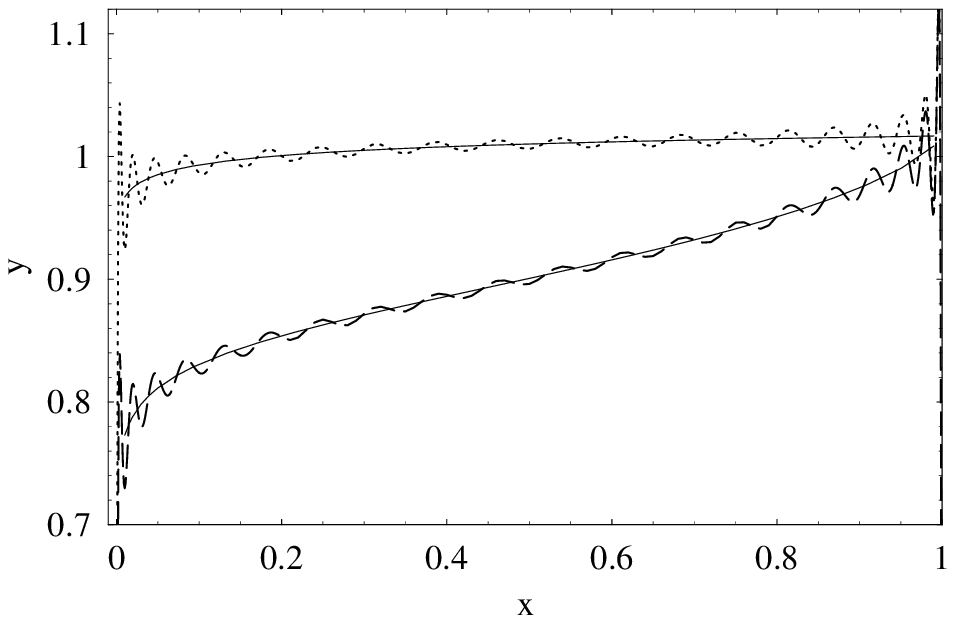}
\end{tabular}
\end{center}
\vspace{-0.5cm}
\caption{Functions $U_\Gamma(u,\mu_h,\mu)$ in (\ref{ULO}) for 
$\mu_h=4.8$\,GeV and $\mu=1.55$\,GeV. The curves correspond to RG evolution 
with $\gamma_1+2\gamma_2$ (dotted) and $\gamma_1$ (dashed). The left and 
right plots show results derived by using 20 and 40 basis functions $\psi_n$, 
respectively. The solid lines were obtained by numerical integration of the 
evolution equation.}
\label{fig:tree}
\end{figure}

Figure~\ref{fig:tree} illustrates the effects of RG evolution. We show 
numerical results for the two evolution functions $U_\Gamma(u,\mu_h,\mu)$ 
corresponding to the cases with anomalous dimensions $\gamma_1+2\gamma_2$ 
(dotted curves) and $\gamma_1$ (dashed curves). The plots are obtained with 
$\mu_h=2E=m_b=4.8$\,GeV and $\mu=\mu_i=\sqrt{2E\Lambda_h}=1.55$\,GeV, where
$\Lambda_h\approx 0.5$\,GeV serves as a typical hadronic scale. In addition 
to the solution obtained with 20 and 40 basis functions, the figure also shows 
results derived from a numerical integration of the evolution equation (solid 
lines). For the numerical solution, the evolution to a lower scale is 
performed in discrete steps of $\Delta\ln\mu=0.02$. To calculate the change 
in $U_\Gamma(u,\mu_h,\mu_n)$ in the evolution step from $\mu_n$ to 
$\mu_{n+1}$, the convolution integral with the evolution kernel is evaluated 
for one hundred different $u$ values, and the function 
$U_\Gamma(u,\mu_h,\mu_{n+1})$ is obtained from a fit to these values. The 
results from the two different methods agree nicely: the curves obtained with 
a finite number of basis polynomials oscillate about the numerical results, 
the number of turning points being equal to the order of the highest basis 
polynomial. The amplitude of the oscillations decreases once more basis 
polynomials are included. To obtain the Wilson coefficients at the low scale, 
these results must still be multiplied with the universal Sudakov factor 
$(2E/\mu_h)^{a(\mu_h,\mu)}\,e^{S(\mu_h,\mu)}\approx 0.89$ (for our choice of 
parameters). We observe that the additional, $u$-dependent resummation effects 
described by the functions $U_\Gamma(u,\mu_h,\mu)$ are very small for the 
coefficients with anomalous dimension $\gamma_1+2\gamma_2$, whereas they are 
more sizeable for those with anomalous dimension $\gamma_1$, reaching about 
$-20\%$ for the smallest values of $u$.

To summarize our results, we compile the Wilson coefficients for the $B$-type 
current operators obtained at leading order in RG-improved perturbation 
theory. We introduce the short-hand notation $C_\Gamma(E,u,\mu)%
=(2E/\mu_h)^{a(\mu_h,\mu)}\,e^{S(\mu_h,\mu)}\,U_\Gamma(u,\mu_h,\mu)$, where
$\Gamma=\,\parallel$ or $\Gamma=\,\perp$ depending on whether the anomalous 
dimension is $\gamma_1+2\gamma_2$ or $\gamma_1$, respectively. We will see 
later that these two cases are in one-to-one correspondence with the nature 
of the light final-state meson in $B\to P,V$ transitions. The case 
$\Gamma=\,\parallel$ applies for transitions into a pseudoscalar or 
longitudinally polarized vector meson ($P$ or $V_\parallel$), while the case 
$\Gamma=\,\perp$ applies for transitions into a transversely polarized vector 
meson ($V_\perp$). Our results are given as follows.

\paragraph{\rm Scalar current:}

\be
   C_S^B(\mu) = - C_\parallel(E,u,\mu) \,.
\ee

\paragraph{\rm Vector current:}

\bea
   C_{V1}^B(\mu) &=& C_\parallel(E,u,\mu) \,, \qquad\hspace{1.0cm}
    C_{V2}^B(\mu) = -2 C_\parallel(E,u,\mu) \,, \nl
   C_{V3}^B(\mu) &=& - \frac{2E}{m_b}\,C_\parallel(E,u,\mu) \,, \qquad
    C_{V4}^B(\mu) = 0 \,.
\eea

\paragraph{\rm Tensor current:}

\bea
   C_{T1}^B(\mu) &=& - C_\parallel(E,u,\mu) \,, \qquad
    C_{T2}^B(\mu) = -4 C_\parallel(E,u,\mu) \,, \qquad
    C_{T4,5,6}^B(\mu) = 0 \,, \nl
   C_{T3}^B(\mu) &=& - \frac{4E}{m_b}\,C_\perp(E,u,\mu) \,, \qquad
    C_{T7}^B(\mu) = - \frac{8E}{m_b}\,C_\perp(E,u,\mu) \,.
\eea
The corresponding Wilson coefficients in the primed basis follow from 
(\ref{Cprime1}) and (\ref{Cprime2}). At leading order, the results for the 
primed coefficients $C_{V1,2,4}^{B'}$ and $C_{T1,2,5,6}^{B'}$ are given by 
the same expressions as in the original basis, i.e.\ $C_j^{B'}=C_j^B$, while 
$C_{V3}^{B'}(\mu)=(1-2E/m_b)\,C_\parallel(E,u,\mu)$, and 
$C_{T3}^{B'}(\mu)=C_{T4}^{B'}(\mu)=2 C_\parallel(E,u,\mu)$, 
$C_{T7}^{B'}(\mu)=(4E/m_b)\,C_\perp(E,u,\mu)$.  

\begin{figure}
\begin{center}
\begin{tabular}{cc}\psfrag{x}{$u$}\psfrag{y}[B]{$C_S^B$}
\includegraphics[width=0.55\textwidth]{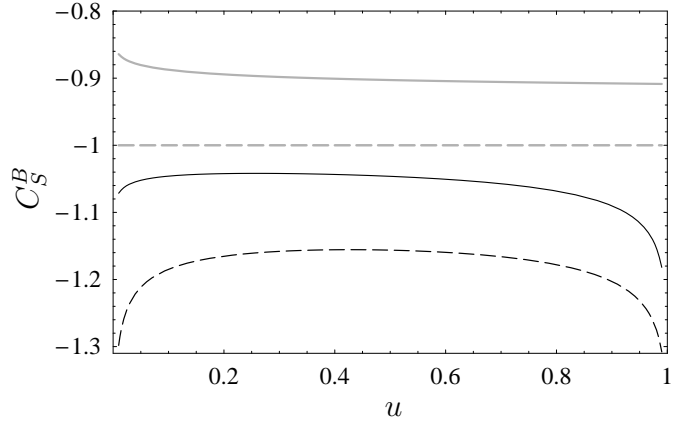}
\end{tabular}
\end{center}
\vspace{-0.5cm}
\caption{Results for the Wilson coefficient $C_S^B(E,u,\mu_i)$ at $E=m_b/2$. 
The dashed lines represent the tree-level (gray) and one-loop (black) 
coefficients at the high scale $\mu_h=4.8$\,GeV. The solid lines are obtained 
after evolving both coefficients with the leading-order anomalous dimension 
down to the intermediate scale $\mu_i=1.55$\,GeV.}
\label{fig:loop}
\end{figure}

Figure~\ref{fig:loop} illustrates our results for the scalar current. While
one should use tree-level matching combined with one-loop running for a 
consistent treatment at leading order in RG-improved perturbation theory, the
figure also shows results obtained by including the one-loop matching 
corrections presented in (\ref{eq:CSBoneloop}). We solve the evolution 
equation (\ref{eq:CBevolve}) by direct numerical integration, expanding the 
cusp anomalous dimension entering $W(E,u)$ to two-loop order, but using 
one-loop expressions for $V_i(u,v)$ and the remaining terms in $W(E,u)$. We 
again take $\mu_h=2E=m_b$ and $\mu_i=\sqrt{2E\Lambda_h}$ (with 
$\Lambda_h=0.5$\,GeV). The dashed curves give the one-loop matching results 
at the high scale $\mu_h$, while the solid lines show the result of RG 
evolution to the intermediate scale $\mu_i$. Comparing the black and gray 
curves, we observe that the one-loop matching corrections are of the same 
order of magnitude as the effects of RG evolution. This fact provides 
motivation for an extension of our anomalous-dimension calculation to the 
two-loop order, which would be necessary for a systematic treatment at 
next-to-leading order in RG-improved perturbation theory.

For completeness, we briefly discuss also the solution to the RG equation 
for the coefficients of the $A$-type currents, given in the first line in 
(\ref{eq:evolutionB_coeff}). The solution is  
\be\label{eq:decomp_CA}
   C^A_i(E,\mu) = \left( {2E\over\mu_h} \right)^{a(\mu_h,\mu)}
   \exp\bigg[ S(\mu_h,\mu) + \int_{\alpha_s(\mu_h)}^{\alpha_s(\mu)}
   {d\alpha\over\beta(\alpha)}\,\tilde\gamma(\alpha) \bigg]\,C^A_i(E,\mu_h)
   \,,  
\ee
with $S(\mu_h,\mu)$ and $a(\mu_h,\mu)$ as defined in (\ref{eq:defS}) and 
(\ref{eq:def_a}). At leading order in RG-improved perturbation theory we 
may use the expansions (\ref{VHres}) together with 
\be
   \int_{\alpha_s(\mu_h)}^{\alpha_s(\mu)}{d\alpha\over\beta(\alpha)}\,
   \tilde\gamma(\alpha) = - {\tilde{\gamma}_0\over 2\beta_0}\ln{r_1} \,,  
\ee
where $\tilde\gamma(\alpha_s)$ is expanded similarly to 
(\ref{eq:expansion_coeffs}), and $\tilde{\gamma}_0=-5C_F$. As an illustration 
of the size of one-loop matching corrections, we may consider again the 
scalar case, where at tree level $C_S^A=1$. At leading order in RG-improved 
perturbation theory, with tree-level matching, we find 
$C_S^A(E,\mu_i)\approx 1.097$ at $\mu_i=\sqrt{2E\Lambda_h}$ and $E=m_b/2$.
From (\ref{eq:CSAoneloop}), the coefficient at the high scale $\mu_h=2E=m_b$ 
including one-loop corrections is $C_S^A(E,\mu_h)\approx 0.934$. Leading-order 
RG evolution to the intermediate scale then yields 
$C_S^A(E,\mu_i)\approx 1.025$. 

\subsection{Constraints from conformal symmetry} 
\label{section:conformal}

The conformal invariance of QCD at the classical level can be used to 
simplify the solutions to the evolution equations for the hard-scattering 
kernels and light-meson LCDAs. As a consequence of this approximate symmetry, 
the evolution equations become diagonal at leading order once they are 
written in a basis of eigenfunctions of definite conformal spin. In the case 
of the pion LCDA, these basis functions are the Gegenbauer polynomials 
$C_n^{3/2}$~\cite{Lepage:1980fj}. For the heavy-light current the situation 
is more complicated: the conformal symmetry is not only violated by quantum 
effects, but broken explicitly by the presence of the heavy quark. 
Interactions with the soft sector of the effective theory destroy the 
conformal invariance of the hard-collinear sector. However, since the soft 
interactions do not change the large momentum fractions of the hard-collinear 
particles, the breaking of the symmetry can only occur in the local part of 
the anomalous dimensions, i.e., in the term 
$\delta(u-v)\,W(E,u)$ in (\ref{eq:gammaiDef}). We focus here on the 
hard-collinear sector of the effective theory, showing in particular that the 
eigenfunctions of the non-local terms $V_i(u,v)$ in (\ref{eq:gammaiDef}) must 
take the form (\ref{eq:eigenfunctions}). For the remainder of this section we 
will refer to hard-collinear modes simply as ``collinear''.  

Before discussing the properties of heavy-light currents in more detail, we 
briefly recall some aspects of conformal symmetry~\cite{Braun:2003rp}. The 
full conformal algebra consists of the generators ${\bm P}_\mu$, 
${\bm J}_{\mu\nu}$ of translations and Lorentz transformations, augmented by 
the generators ${\bm D}$ of scale transformations, $x^\mu\to\lambda x^\mu$, 
and ${\bm K}_\mu$ of special conformal transformations, 
\begin{equation} 
   x^\mu \to {x^\mu + a^\mu x^2\over 1 + 2a\cdot x + a^2 x^2} \,.
\end{equation} 
The action of the generators on a field of scaling dimension $l$ and 
arbitrary spin is
\begin{align} 
   i[{\bm P}_\mu, \Phi(x)]
   &= \partial_\mu \Phi(x) \,, 
   & i[{\bm J}_{\mu\nu}, \Phi(x)]
   &= \left( x_\mu\partial_\nu - x_\nu\partial_\mu - \Sigma_{\mu\nu} \right)
    \Phi(x) \,, \\ 
   i[{\bm D}, \Phi(x)] 
   &= \left( x\cdot\partial + l \right) \Phi(x) \,,
   & i[{\bm K}_\mu, \Phi(x)]
   &= \left( 2x_\mu x\cdot \partial - x^2 \partial_\mu + 2lx_\mu
    - 2x^\nu \Sigma_{\mu\nu} \right) \Phi(x) \,. \nonumber
\end{align} 
The spin operator $\Sigma_{\mu\nu}$ on scalar, fermion, and vector fields is 
given by
\be
   \Sigma_{\mu\nu}\,\phi = 0 \,, \qquad
   \Sigma_{\mu\nu}\,\psi = {i\over 2}\sigma_{\mu\nu}\,\psi \,, \qquad
   \Sigma_{\mu\nu}\,A_\alpha = g_{\nu\alpha}\,A_\mu - g_{\mu\alpha}\,A_\nu 
   \,.
\ee
The collinear part of the SCET operators is given by products of fields 
smeared along the light ray $x^\mu=r\nb^\mu$, i.e.\ 
$\Phi(x)=\Phi(r\nb)\equiv\Phi(r)$. The ${\rm SL}(2,R)$ subgroup that maps 
this light-ray onto itself is called the collinear conformal group. It is 
obtained from the four generators $\nb\cdot{\bm P}$, $n\cdot{\bm K}$, 
$n^\mu\nb^\nu {\bm J}_{\mu\nu}$, and ${\bm D}$. To classify operators under 
this group, it is convenient to work with the linear combinations
\begin{align}
   {\bm L}_{+1}
   &= -i\,\nb\cdot {\bm P} \,,
   &  {\bm L}_{-1}
   &=\frac{i}{4}\,n\cdot {\bm K} \,, \nonumber\\
   {\bm L}_0 &={i\over 2}({\bm D} + {1\over 2}n^\mu\nb^\nu {\bm J}_{\mu\nu})
    \,,\hspace{-2.0cm}
   & {\bm E} &= {i\over 2}({\bm D} - {1\over 2}n^\mu \nb^\nu 
    {\bm J}_{\mu\nu}) \,.
\end{align}
The generator ${\bm E}$ counts the twist $t$ and commutes with the remaining 
three generators, which fulfill the angular-momentum commutation relations
\be
   [{\bm L}_{+1},{\bm L}_{-1}] = 2 {\bm L}_0 \,, \qquad
   [{\bm L}_0,{\bm L}_{\pm 1}] = \pm {\bm L}_{\pm 1} \,.
\ee
The action of these operators on the fields $\Phi(r)$ is
\bea\label{eq:confspinrep}
   {[{\bm L}_{+1},\Phi(r)]} &=& - \partial_r\,\Phi(r) \,,  \nl
   {[{\bm L}_{-1},\Phi(r)]} &=& (r^2\,\partial_r + 2jr)\,\Phi(r) \,, \nl
   {[{\bm L}_{0},\Phi(r)]} &=& (r\,\partial_r + j)\,\Phi(r) \,,
\eea
where the quantum number $j$ is referred to as conformal spin. The twist $t$ 
and the conformal spin $j$ of a field $\Phi$ are related to the scaling 
dimension $l$ and the spin projection $s$ through the relations $t=l-s$ and 
$j=\frac12(l+s)$, where $s$ is defined by
\begin{equation}
   \hat{\Sigma}\,\Phi\equiv {1\over 2}\nb^\mu n^\nu\Sigma_{\mu\nu}\,\Phi
   = s\,\Phi \,.
\end{equation}

To express the SCET fields in terms of (primary) conformal fields of 
definite spin and twist, we introduce the field-strength tensor
\be 
   g {\cal G}_{\mu\nu} = W^\dagger gG_{\mu\nu} W
   = i[{\cal D}_\mu, {\cal D}_\nu] \,, 
\ee
where ${\cal D}_\mu=\partial_\mu- i \A_\mu$.  In particular, 
\be
   \nb\cdot\partial\,\A_\perp^\mu
   = \nb^\alpha g_\perp^{\beta\mu} g{\cal G}_{\alpha\beta}
   \equiv g{\cal G}_{\nb\perp}^\mu \,, \qquad
   \nb\cdot\partial\,n\cdot\A = \nb^\alpha n^\beta g{\cal G}_{\alpha\beta} 
   \equiv g{\cal G}_{\nb n} \,. 
\ee 
Then for the collinear SCET fields at the origin, we find for the twist and 
conformal spin eigenvalues 
\begin{align}\label{eq:confspins}
   [{\bm E},\X(0)] &= \X(0) \,,
   & [{\bm L}_0,\,\X(0)] &= \X(0) \,, \nonumber\\
    [{\bm E}, {\cal G}_{\nb\perp}^\mu(0)] &= {\cal G}_{\nb\perp}^\mu(0) \,,
   & [{\bm L}_0, {\cal G}_{\nb\perp}^\mu(0)]
   &= \frac{3}{2}\,{\cal G}_{\nb\perp}^\mu(0) \,, \nonumber\\
    [{\bm E}, {\cal G}_{\nb n}(0)] &= 2\, {\cal G}_{\nb n}(0) \,,
   & [{\bm L}_0, {\cal G}_{\nb n}(0)] &= {\cal G}_{\nb n}(0)  \,. 
\end{align}

In the following, we want to decompose a given operator into components with 
definite conformal spin. The construction of the corresponding basis is done 
in two steps. First, one identifies the operator of minimal conformal spin 
${\cal O}_0$, i.e., the operator at the bottom of an irreducible conformal 
tower, defined by $[{\bm L}_{-1},{\cal O}_0]=0$. The complete basis of 
conformal operators is then given by repeated application of the raising 
operator ${\bm L}_{+1}$ to the highest-weight operator 
${\cal O}_0$:\,\footnote{It is standard terminology to refer to the operator 
of {\em minimal\/} conformal spin as the highest-weight operator.} 
\vspace{-0.2cm}
\begin{align}
   {\cal O}_{k} &= \underbrace{[{\bm L}_{+1},[{\bm L}_{+1},
    [\dots,[{\bm L}_{+1}},{\cal O}_0]\dots]]] \,. \\[-0.3cm]
   &\hspace{1.7cm} k~{\rm times} \nonumber
\end{align}
A trivial example of this procedure is the expansion of $\X(r)$ 
into operators with definite conformal spin. In this case the highest-weight 
operator is ${\cal O}_0=\X(0)$, and the raising operator ${\bm L}_{+1}$ acts 
as a derivative with respect to $r$. The expansion of $\X(r)$
in conformal spin thus coincides with the Taylor expansion about $r=0$. 
In order to analyze the SCET currents $J^B$, 
we now consider the decomposition of the product of two fields, 
which we assume have individual conformal spins $j_1$ 
and $j_2$, into operators of definite conformal spin. We first rewrite each 
of the two component fields $\Phi_{j1}(r_1)$ and $\Phi_{j2}(r_2)$ in the 
conformal spin basis. In other words, we expand the product in a Taylor 
series about $r_1=r_2=0$. At the $n$-th order, this leaves us with 
operators of the form
\begin{equation}
   \partial_{r_1}^{n_1} \partial_{r_2}^{n_2}\,\Phi_{j1}(r_1)\,\Phi_{j2}(r_2)
   \big|_{r_1=r_2=0} \,,
\end{equation}
where $n_1+n_2=n$. In general, these operators do not have definite 
conformal spin. The minimal conformal spin of an operator built from two 
fields of conformal spin $j_1$ and $j_2$ with $n$ derivatives on the fields 
is $j=j_1+j_2+n$, and the highest-weight operator for this case 
is~\cite{Makeenko:bh}
\be\label{eq:Mak}
   {\cal O}_0^{(n,j_1,j_2)}(r) = \partial_r^n \bigg\{ 
   \Phi_{j_1}(r)\, P_n^{(2j_1-1, 2j_2-1)}\left( 
   {\partial_r -\ov{\partial_r}\over \partial_r+\ov{\partial_r}} \right)
   \Phi_{j_2}(r) \bigg\} \,.
\ee
The Jacobi polynomials $P_n^{(\alpha, \beta)}$ appear as the Clebsch-Gordan 
coefficients of the collinear conformal group. 

As an illustration, the operators relevant for leading-twist light meson 
LCDAs have the form $\bar\X(r_1)\,\Gamma\,(\sla{\nb}/2)\,\X(r_2)$. 
From the general result (\ref{eq:Mak}), and the conformal spin eigenvalues (\ref{eq:confspins}), 
it follows that the highest-weight operators are
\begin{equation}\label{eq:JacobiP11}
   {\cal O}_0^{(n,1,1)}(r) = (i\nb\cdot\partial)^n \bigg\{ 
   \bar\X(r\nb)\,\Gamma\,\frac{\sla{\nb}}{2}\,P_n^{(1, 1)}\!\!
   \left( {i\nb\cdot\partial -i\nb\cdot\ov{\partial}\over
          i\nb\cdot\partial+i\nb\cdot\ov{\partial}} \right) \X(r\nb)
   \bigg\} \,.
\end{equation} 
For example, with $\Gamma=\gamma_5$, the pion-to-vacuum matrix 
element yields a moment of the pion LCDA,
\be
   \langle 0|{\cal O}_0^{(n,1,1)}(0)|\pi\rangle
   \propto \int_0^1\!du\,P_n^{(1,1)}(u-\bar u)\,\phi_\pi(u,\mu) \,,
\ee
which projects out the component of $\phi_\pi(u,\mu)$ proportional to 
$u\bar u\,P_n^{(1,1)}(2u-1)$, where $P_n^{(1,1)}\propto C_n^{3/2}$. 
Similarly, we may expand the collinear fields $\A_\perp^\mu(r_1)\,\X(r_2)$ 
appearing in the current operators $J_j^B$ into operators of definite 
conformal spin. For this case we find that the highest-weight operators are 
\begin{equation}\label{eq:JacobiP21}
   {\cal O}_0^{(n,2,1)}(r) = (i\nb\cdot\partial)^n \bigg\{ 
   [ i\nb\cdot\partial\,\A_{\perp\mu}](r\nb)\,
   P_n^{(2, 1)}\!\!\left( {i\nb\cdot\partial-i\nb\cdot\ov{\partial}
                           \over i\nb\cdot\partial+i\nb\cdot\ov{\partial}}
   \right) \X(r\nb) \bigg\} \,,
\end{equation}
and the corresponding eigenfunctions for $J_j^B$ are proportional to 
$u\bar u\,P_n^{(2,1)}(2u-1)$, in agreement with (\ref{eq:eigenfunctions}).  
From the discussion following (\ref{eq:evolution_JC}), the
eigenfunctions for the Wilson coefficients $C_j^B$ must then be proportional to 
$\bar u\,P_n^{(2,1)}(2u-1)$.

In the conformal-symmetry limit $\bm{L}^2$, $\bm{L}_0$, and $\bm E$ are 
conserved charges, and operators with different conformal spin or twist 
therefore do not mix. For the collinear fields appearing in the heavy-light 
currents the situation is especially simple, since there is only one 
operator for a given set of quantum numbers. Operators with different 
collinear field content or with $\partial_\perp^\mu$ or $n\cdot\partial$ 
derivatives on one of the two fields have higher twist. Also, by 
construction, each one of the operators ${\cal O}_k^{(n,2,1)}$ has unique 
quantum numbers, namely
\bea
   {[{\bm L}^2,{\cal O}_k^{(n,2,1)}]}
   &=& \left( n + \frac{5}{2} \right) \left( n + \frac{3}{2} \right)
    {\cal O}_k^{(n,2,1)} \,, \nl
   {[{\bm L_0},{\cal O}_k^{(n,2,1)}]}
   &=& \left( n + k + \frac{5}{2} \right) {\cal O}_k^{(n,2,1)} \,.
\eea
To the extent that conformal symmetry is preserved, the operators 
${\cal O}_k^{(n,2,1)}$ are thus multiplicatively renormalized. This explains 
why the eigenfunctions of the non-diagonal part of the leading-order 
evolution kernel take the form of Jacobi polynomials $P_n^{(2,1)}$, as was 
shown by explicit computation in Section~\ref{section:appendix}.

\section{Renormalization-group evolution in SCET$_{\bf II}$}
\label{sec:SCETII}

After RG evolution down to the scale $\mu_i\sim\sqrt{2E\Lambda_{\rm QCD}}$, 
the intermediate effective theory SCET$_{\rm I}$ is matched onto 
SCET$_{\rm II}$. In this section, we present a complete operator basis for 
the hard-scattering contributions in SCET$_{\rm II}$ and compute the 
corresponding matching coefficients (called jet functions). To 
investigate the size of loop corrections, we present the complete 
one-loop matching corrections to these coefficients and show that they can be 
expressed in terms of only two universal functions. We then consider 
resummation in SCET$_{\rm II}$ and present the general solution to the RG 
equation. In the following Section~\ref{sec:applic}, we will apply these 
results to obtain RG-improved expressions for the hard-scattering 
contributions to $B\to P,V$ form factors.

\subsection{Low-energy representation of the \boldmath$B$-type 
operators\unboldmath}

Below the intermediate scale $\mu_i\sim \sqrt{2E\Lambda_{\rm QCD}}$, the 
effective-theory description is in terms of soft, collinear, and 
soft-collinear messenger modes of SCET$_{\rm II}$ 
\cite{Hill:2002vw,Becher:2003qh}. The Lagrangian in the soft sector is the 
same as in (\ref{eq:scetLagrangian}), while the collinear Lagrangian becomes
\be
   {\cal L}_c = \bar\xi_c\,{\sla{\nb}\over 2}
   \left( in\cdot D_{c} + i\Dslash_{c\perp}\,{1\over i\nb\cdot D_{c}}\,
   i\Dslash_{c\perp} \right) \xi_c \,.
\ee
Soft-collinear messenger modes may be decoupled via field redefinitions in 
the leading-order soft and collinear Lagrangians and in the four-quark 
operators representing the hard-scattering terms \cite{Becher:2003kh}, and we 
do not display them here. In terms of the decoupled fields, it is convenient 
to introduce the gauge-invariant combinations \cite{Hill:2002vw,Bauer:2001ct} 
\bea
   \X_c &=& [W_c^\dagger\,\xi_c]\sim \lambda \,, \qquad
   \Q_s = [S^\dagger q_s]\sim \lambda^{3/2} \,, \qquad
   \H = [S^\dagger h] \sim \lambda^{3/2} \,, \nl
   \A_c^\mu &=& W_c^\dagger\,(iD_c^\mu\,W_c)\sim (\lambda^2,0,\lambda) \,,
    \qquad
    \A_s^\mu = S^\dagger\,(iD_s^\mu\,S)\sim (0,\lambda,\lambda) \,,
\eea
where $W_c$ and $S$ are collinear and soft Wilson lines in the $\nb$ and $n$ 
directions, respectively. To account for non-localities generated by 
integrating out hard-collinear modes, collinear and soft fields are smeared 
along light-like directions, i.e.\ $\phi_c(s\nb)$ and $\phi_s(tn)$ 
\cite{Hill:2002vw}.

Form factors arising in semileptonic and radiative $B$ decays derive from
current operators in the effective weak Hamiltonian mediating the decay of a 
heavy $b$ quark into a left-handed light quark. The relevant operators are
\be
   V_L^\mu = \bar q\,\gamma^\mu (1-\gamma_5)\,b \,, \qquad
   T_L^{\mu\nu} = (-i)\,\bar q\,\sigma^{\mu\nu}(1+\gamma_5)\,b
    = \bar q\,\gamma^{[\mu}\gamma^{\nu]}\,(1+\gamma_5)\,b \,.
\ee
We find it instructive when discussing radiative corrections to consider also 
the scalar current, 
\be\label{eq:scalarLeft}
   S_L = \bar q\,(1+\gamma_5)\,b \,, 
\ee
which can be related to the vector current by QCD equations of motion. In the 
NDR scheme with anti-commuting $\gamma_5$, the preceding analysis of operator 
mixing and renormalization is unchanged by the insertion of $(1+\gamma_5)$ 
next to the light quark. From now on, the replacement 
$\bar q\to\bar q\,(1+\gamma_5)$ in the SCET$_{\rm I}$ currents $J_j^{B'}$ 
will be understood implicitly. The $B$-type current operators of 
${\rm SCET}_{\rm I}$ then match onto four-quark operators in 
${\rm SCET}_{\rm II}$ of the form 
\cite{Hill:2002vw,Bauer:2002aj,Beneke:2003pa,Lange:2003pk}
\be\label{eq:4quarkop}
   O_i(s,t) = \bar{\X}_c(s\nb)\,(1+\gamma_5)\,\Gamma_{c,i}\,
   {\sla{\nb}\over 2}\,\X_c(0)\,
   \,\bar{\Q}_s(tn)\,(1\pm\gamma_5)\,{\sla{n}\over 2}\,\Gamma_{s,i}\,\H(0) \,.
\ee
The chirality of the field $\bar{\Q}_s(tn)$ is the same as that of $\X_c(0)$, which in 
turn is determined by the Dirac structure $\Gamma_{c,i}$. We restrict 
attention to the case where the matrix elements of these operators are 
evaluated with an initial-state pseudoscalar $B$-meson. We then construct a 
basis of four-quark operators in which the soft component fields arise in 
the combination $\bar{\Q}_s\,(1\pm\gamma_5)\,(\sla{n}/2)\,\H$. (In a more 
general case, operators containing $\gamma_\perp^\mu$ next to $\bar{\Q}_s$, 
and structures with more than one perpendicular Lorentz index, would also 
have to be considered.) For the scalar current $S_L$ there is only one 
possible structure,  
\be\label{eq:OS}
   O_S = \bar\X_c\,(1+\gamma_5)\,{\sla{\nb}\over 2}\,\X_c\,\,
   \bar\Q_s\,(1+\gamma_5)\,{\sla{n}\over 2}\,\H \,,
\ee
while for the vector current we find the three operators
\bea\label{eq:OV}
   O_{V1}^\mu &=& {n^\mu\over n\cdot v}\,
    \bar{\X}_c\,(1+\gamma_5)\,{\sla{\nb}\over 2}\,\X_c\,\,
    \bar{\Q}_s\,(1+\gamma_5)\,{\sla{n}\over 2}\,\H \,, \nl
   O_{V2}^\mu &=& \hspace{0.44cm} v^\mu\,
    \bar{\X}_c\,(1+\gamma_5)\,{\sla{\nb}\over 2}\,\X_c\,\,
    \bar{\Q}_s\,(1+\gamma_5)\,{\sla{n}\over 2}\,\H \,, \nl
   O_{V3}^\mu &=& \hspace{0.41cm}
    \bar{\X}_c\,(1+\gamma_5)\,\gamma_\perp^\mu\,{\sla{\nb}\over 2}\,\X_c\,\,
    \bar{\Q}_s\,(1-\gamma_5)\,{\sla{n}\over 2}\,\H \,.
\eea
For the tensor current we only keep terms that are non-zero when contracted 
with $q_\nu$, where $q$ with $q_{\perp\nu}=0$ is the momentum transfer. This
yields
\bea\label{eq:OT}
   O_{T1}^{\mu\nu} &=& \hspace{0.33cm} {n^{[\mu} v^{\nu]}\over n\cdot v}\,
    \bar{\X}_c\,(1+\gamma_5)\,{\sla{\nb}\over 2}\,\X_c\,\,
    \bar{\Q}_s\,(1+\gamma_5)\,{\sla{n}\over 2}\,\H \,, \nl
   O_{T2}^{\mu\nu} &=& {n^{[\mu}\over n\cdot v}\,
    \bar{\X}_c\,(1+\gamma_5)\,\gamma_\perp^{\nu]}\,{\sla{\nb}\over 2}\,\X_c
    \,\,\bar{\Q}_s\,(1-\gamma_5)\,{\sla{n}\over 2}\,\H \,, \nl
   O_{T3}^{\mu\nu} &=& %\hspace{0.38cm} 
    v^{[\mu}\,
    \bar{\X}_c\,(1+\gamma_5)\,\gamma_\perp^{\nu]}\,{\sla{\nb}\over 2}\,\X_c
    \,\,\bar{\Q}_s\,(1-\gamma_5)\,{\sla{n}\over 2}\,\H \,.
\eea
Color octet-octet operators have vanishing matrix elements between
physical meson states, and also do not mix into the color singlet-singlet 
sector \cite{Becher:2003kh}. We may therefore restrict attention 
throughout to color singlet-singlet operators.

In constructing the most general basis, we notice that adjacent 
$\gamma_\perp$ matrices can always be avoided. Using
\be
   \{ \gamma_\perp^\mu, \gamma_\perp^\nu \} = 2g_\perp^{\mu\nu} \,,
\ee
any such structure can be reduced to combinations of $g_\perp^{\mu\nu}$ and 
totally anti-symmetric combinations,
\be
   \gamma_\perp^{\mu_1\dots \mu_n}
   \equiv \gamma_\perp^{[\mu_1}\!\dots\gamma_\perp^{\mu_n]} \,.
\ee
In four dimensions $\gamma_\perp^{\mu_1\dots \mu_n}$ vanishes for $n\ge 3$, 
while for $n=2$ (we use $\epsilon^{0123}=-1$)
\be
   \gamma_\perp^{[\mu}\gamma_\perp^{\nu]}
   = {i\over 2}{\epsilon}^{\mu\nu\rho\sigma} \nb_\rho n_\sigma 
   \left( {\sla{\nb}\sla{n}\over 4} - {\sla{n}\sla{\nb}\over 4} \right)
   \gamma_5 
   \equiv i\epsilon_{\perp}^{\mu\nu}
   \left( {\sla{\nb}\sla{n}\over 4} - {\sla{n}\sla{\nb}\over 4} \right)
   \gamma_5 \,.
\ee
In dimensional regularization, so-called ``evanescent'' operators containing 
structures $\gamma_\perp^{\mu_1\dots \mu_n}$ for $n\ge 3$ appear once 
radiative corrections are taken into account. A regularization scheme 
including the effects of these operators must be employed for 
next-to-leading order calculations. This is the two-dimensional 
analogue, in the space of transverse directions, of the procedure employed in 
four dimensions \cite{Dugan:1990df,Buras:1989xd,Herrlich:1994kh}.

\subsection{Matching calculations}
\label{sec:SCETIImatching}

We introduce the momentum-space coefficient functions
\be\label{eq:momentum_D}
   D_i(E,\omega,u,\mu)\equiv \int ds\,dt\,e^{-i\omega n\cdot v t}\,
   e^{ius\nb\cdot P}\,\tilde D_i(s,t,\mu) \,,
\ee
where $\tilde D_i(s,t,\mu)$ is the coefficient multiplying the operator 
$O_i(s,t)$ in (\ref{eq:4quarkop}), and the dependence of $D_i$ on 
$\nb\cdot P$ and $n\cdot v$ is only through the product 
$(\nb\cdot P)(n\cdot v)=2E$. The final step of matching 
${\rm SCET}_{\rm I}$ onto ${\rm SCET}_{\rm II}$ is described by so-called jet 
functions ${\cal J}_{ij}$ defined via the relation
\be
\label{eq:jet_function}
   D_i(E,\omega,u,\mu_i) = {1\over 2E\omega} \int_0^1\!dy\,
   {\cal J}_{ij}\bigg(u,y,\ln{2E\omega\over\mu_i^2},\mu_i \bigg)\,
   C^{B}_j(E,y,\mu_i) \,,
\ee
where $C^B_j$ are the coefficient functions of the ${\rm SCET}_{\rm I}$ 
currents $J_j^B$, and $\mu_i\sim\sqrt{2E\Lambda_{\rm QCD}}$ denotes the 
intermediate matching scale. Since all interactions between collinear fields 
and the $b$ quark have been integrated out already in SCET$_{\rm I}$, the jet 
functions are independent of the heavy-quark velocity $v$. Dimensional 
analysis and rescaling invariance then imply that they depend on $\omega$ 
only through the dimensionless ratio 
$2E\omega/\mu_i^2$. 
Furthermore, since $\mu_i$ only 
appears logarithmically, the same must be true for this ratio of scales, to 
any finite order in perturbation theory \cite{Bosch:2003fc}. This explains 
why above we have written the third argument of the jet functions as 
$L=\ln(2E\omega/\mu_i^2)$. 

For the calculation of the jet functions, it is convenient to work in the 
primed 
basis of current eigenvectors $J_j^{B'}$ defined in (\ref{eq:vectorprime}) 
and (\ref{eq:tensorprime}). From the structure of these currents, as well as
the structure of the resulting four-quark operators in SCET$_{\rm II}$ 
displayed in (\ref{eq:OS})--(\ref{eq:OT}), we notice that 
\vspace{-0.2cm}
\bea
   O_{V1}^\mu &=& {n^\mu\over n\cdot v}\,O_S \,, \qquad\,
    O_{V2}^\mu = v^\mu\,O_S \,, \qquad\,\,\;
    O_{T1}^{\mu\nu} = {n^{[\mu}v^{\nu]}\over n\cdot v}\,O_S \,, \nl
   J_{V3}^{B'\mu} &=& {n^\mu\over n\cdot v}\,J_S^B \,, \qquad
    J_{V2}^{B'\mu} = v^\mu J_S^B \,, \qquad
    J_{T4}^{B'\mu\nu} = {n^{[\mu}v^{\nu]}\over n\cdot v}\,J_S^B \,,
\eea
so that the SCET$_{\rm II}$ operators $O_{V1,2}$, $O_{T1}$ and their 
SCET$_{\rm I}$ counterparts $J^{B'}_{V3,2}$, $J^{B'}_{T4}$ are related to the 
scalar operators $O_S$ and $J^B_S$ simply by overall factors involving 
$n^\mu$ and $v^\mu$. It follows that, e.g., the jet function arising in the
matching of $J^{B'}_{V3}$ onto $O_{V1}$ is precisely the same as that arising
in the matching of $J^{B}_S$ onto $O_S$. Similarly, we find that 
\bea
   O_{T2}^{\mu\nu} &=& {1\over n\cdot v}\,{n^{[\mu}}\,O_{V3}^{\nu]} \,,
    \qquad~~\; O_{T3}^{\mu\nu} = v^{[\mu}\,O_{V3}^{\nu]} \,, \nl
   J_{T7}^{B'\mu\nu} &=& {1\over n\cdot v}\,{n^{[\mu}} J_{V4}^{B'\nu]} \,,
    \qquad J_{T6}^{B'\mu\nu} = v^{[\mu} J_{V4}^{B'\nu]} \,, 
\eea
so that the matching of $J_{T7}^{B'}$ ($J_{T6}^{B'}$) onto $O_{T2}$ 
($O_{T3}$) is precisely the same as that of $J_{V4}^{B'}$ onto $O_{V3}$. The 
operators $J_{V1}^{B'}$ and $J_{T2,3}^{B'}$ match onto SCET$_{\rm II}$ 
four-quark operators with vanishing projection onto the $B$-meson, while 
$J_{T1,5}^{B'}$ contain two perpendicular Lorentz indices. These operators are 
therefore not relevant to the $B\to P,V$ form factors, and we have thus not 
listed the corresponding SCET$_{\rm II}$ operators in (\ref{eq:OV}) and 
(\ref{eq:OT}). Collecting the elements ${\cal J}_{ij}$ into matrices 
$\bm{{\cal J}}$, these observations imply that the jet functions can be 
expressed in terms of only two functions ${\cal J}_{\parallel}$ and 
${\cal J}_{\perp}$. The result is  
\be\label{Jmatrs}
   \bm{{\cal J}}_{S} = {\cal J}_\parallel \,, \qquad 
   \bm{{\cal J}}_{V}^\prime 
   = \left( \begin{array}{rrr|r}
     0 & 0 & {\cal J}_\parallel & 0 \\
     0 & {\cal J}_\parallel & 0 & 0 \\
     \hline
     0 & 0 & 0 & J_\perp
    \end{array} \right) , \qquad 
   \bm{{\cal J}}_{T}^\prime
   = \left( \begin{array}{rrrr|rrr}
    0 & \,\,0 & \,\,0 & {\cal J}_\parallel & \,\,0 & \,\,0 & \,\,0 \\
    \hline
     0 & 0 & 0 & 0 & 0 & 0 & {\cal J}_\perp \\
     0 & 0 & 0 & 0 & 0 & {\cal J}_\perp & 0
   \end{array} \right) , 
\ee
where the primes remind us that these results refer to the basis of the
SCET$_{\rm I}$ currents $J_j^{B'}$. At tree level, we find
\be\label{eq:jetTree}
   {\cal J}_{\parallel}(u,v,L,\mu_i)_{\rm tree}
   = {\cal J}_{\perp}(u,v,L,\mu_i)_{\rm tree}
   = - {4\pi C_F\alpha_s(\mu_i)\over N}\,{1\over 2E\bar u}\,\delta(u-v) \,.
\ee 

\begin{figure}[t]
\begin{center}
\includegraphics[width=0.9\textwidth]{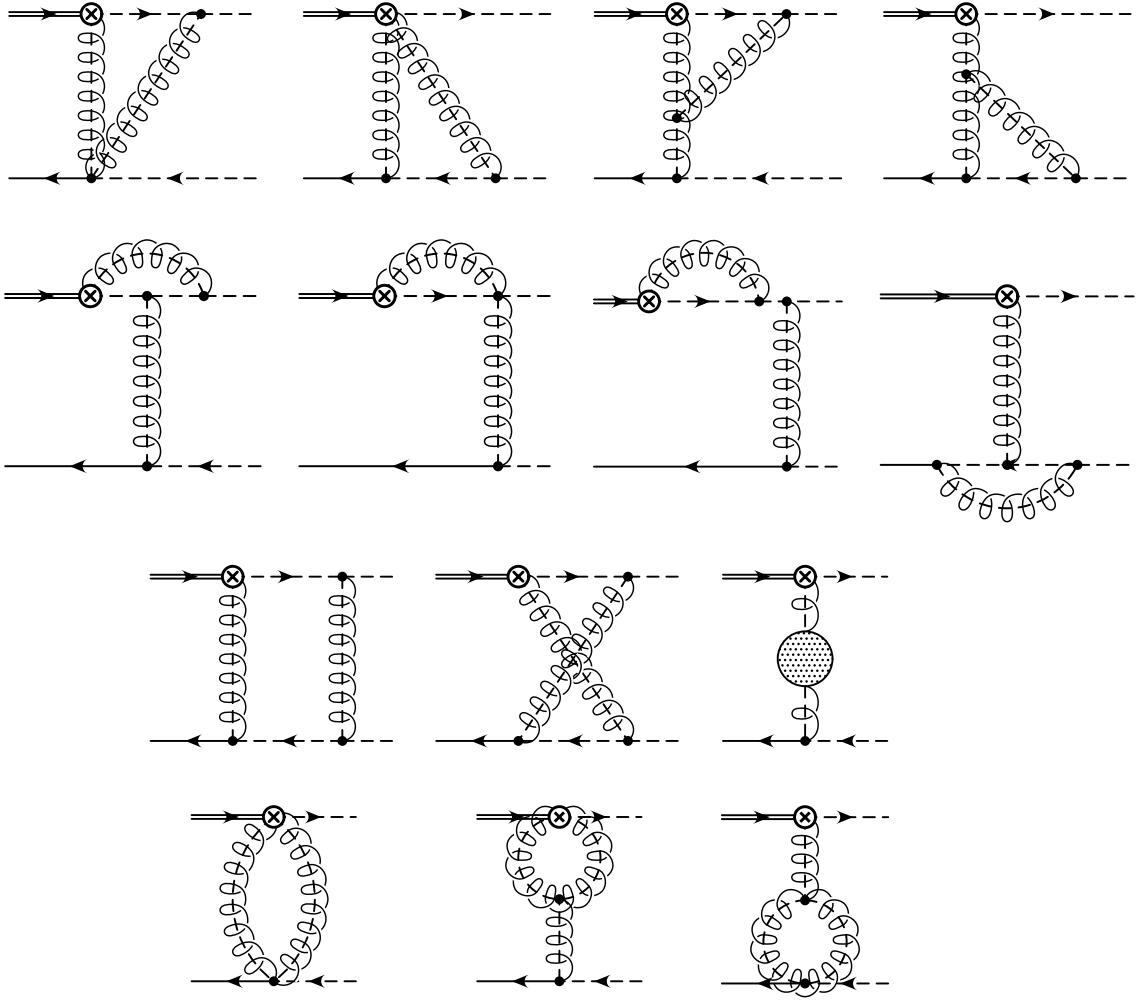}
\end{center}
\vspace{-0.5cm}
\caption{$\mbox{SCET}_{\rm I}$ graphs contributing to the matching of the 
currents $J_j^{B'}$ (crossed circle) onto the $\mbox{SCET}_{\rm II}$ 
four-quark operators $O_i$. Full lines denote soft fields, dashed lines 
hard-collinear fields. Diagrams with soft gluons or scaleless loops are not 
shown.}
\label{fig:jet}
\end{figure}

In order to study the effect of next-to-leading order matching corrections, 
we now calculate the one-loop matching corrections to the jet functions 
${\cal J}_\parallel$ and ${\cal J}_\perp$. The relevant diagrams are depicted 
in Figure~\ref{fig:jet}. Contributions involving soft gluons cancel in the 
matching once the corresponding SCET$_{\rm II}$ diagrams are evaluated, so 
that only hard-collinear SCET$_{\rm I}$ diagrams need be considered. 
The $1/\epsilon$ poles are canceled by the renormalization constants of the 
SCET$_{\rm I}$ and SCET$_{\rm II}$ operators. Details of this calculation are 
presented in \cite{toappear}. The results take the form
\bea\label{j12def}
   {\cal J}_\parallel(u,v,L,\mu_i)
   &=& {4\pi C_F\alpha_s(\mu_i)\over N}\,{1\over 2E\bar u} \left[
    - \delta(u-v) + \frac{\alpha_s(\mu_i)}{4\pi}\,
    j_\parallel(u,v,L) + \order(\alpha_s^2) \right] , \nl
   {\cal J}_\perp(u,v,L,\mu_i)
   &=& {4\pi C_F\alpha_s(\mu_i)\over N}\,{1\over 2E\bar u} \left[
   - \delta(u-v) + \frac{\alpha_s(\mu_i)}{4\pi}\,j_{\perp}(u,v,L)
   + \order(\alpha_s^2) \right] ,
\eea
where $L=\ln(2E\omega/\mu_i^2)$, and the functions $j_\parallel$ and $j_\perp$ are 
given by 
\bea\label{1loopjet}
   j_\parallel(u,v,L) &=& j_1(u,v,L) + j_2(u,v,L)+j_3(u,v) \,,\nl
   j_\perp(u,v,L) &=& j_1(u,v,L) - j_3(u,v) \,,
\eea
with
\bea
j_1(u,v,L) &=& \delta(u-v)\,\Bigg\{ T_F\,n_f
    \left[ {20\over 9} - {4\over 3}\,(L+\ln\bar u) \right] \nl 
   &&\qquad\mbox{}+ \left( C_F - {C_A\over 2} \right)
    \left[ - (L+\ln u)^2 - 3(L+\ln\bar u) + 8 - {\pi^2\over 2} \right] \nl
   &&\qquad\mbox{}- {C_A\over 2} \left[ (L+\ln\bar u)^2
    - {13\over 3}\,(L+\ln\bar u) + {80\over 9} - {\pi^2\over 6} \right]
    \Bigg\} \nl
   &&\mbox{}+ \left( C_F - {C_A\over 2} \right) \Bigg\{
    - 2 \left[ {\theta(v-u)\over v-u}\,(L+\ln(v-u)) 
    + {\theta(u-v)\over u-v}\,(L+\ln(u-v)) \right]_+ \nl
   &&\qquad\mbox{}+ \theta(1-u-v) \left[ {2v\over\bar u\bar v}
    \left( L + \ln{v(1-u-v)\over\bar u} \right)
    + {2(1-u-v)\over u\bar u\bar v} \right] \nl
   &&\qquad\mbox{}+ \theta(v-u)\left[ {2\over\bar u}\,(L+\ln(v-u))
    + {2\over v-u}\,{\bar v\over\bar u}\,\ln{\bar u\over\bar v}
    - {2\over u\bar u} \right] \nl
   &&\qquad\mbox{}+ \theta(u-v) \left[ {2\over u\bar v}\,(L+\ln(u-v))
    + {2\over u-v}\,{v\bar u\over u\bar v}\,\ln{u\over v}
    - {2\over u\bar v} \right] \Bigg\} \nl 
   &&\mbox{}+ C_F \left\{ {v(1+u)\over u\bar v} 
    - \theta(v-u) \left[ {2\over\bar u}
    \left( L + \ln{(v-u)\bar v\over\bar u} \right)
    + {(v-u)(1+u)\over u\bar u\bar v} \right] \right\} , \nl
   j_2(u,v,L)
   &=& 2 \left( C_F - {C_A\over 2} \right) \Bigg\{
    {1\over u} - \theta(u-v)\,{\bar u\over u\bar v}
    \left( L + \ln{(u-v)v\over u} \right) \nl
   &&\qquad\mbox{}- \theta(v-u)\,{\bar v\over u\bar u}
    \left( L + \ln{(v-u)\bar v\over\bar u} \right) \nl
   &&\qquad\mbox{}+ \theta(1-u-v)
    \left[ {(1-u-v)^2-uv\over u\bar u\bar v}
    \left( L + \ln{v(1-u-v)\over\bar u} \right) - {1\over u} \right]
    \Bigg\}\nl
   &&\mbox{}+ 2 C_F \left[ {v\bar u\over u}\,(L+\ln v\bar v) - v 
    - \theta(v-u)\,{(v-u)\over u\bar u}
    \left( L + \ln{(v-u)\bar v\over\bar u} \right) \right] ,\nl
   j_3(u,v)
   &=& C_F \left[ -\frac{\bar{u} v}{u \bar{v}}\,\theta(u-v) - \theta(v-u) \right] .
\eea
The plus distribution is defined as in (\ref{eq:plusDistrib}). At 
one-loop order there are also matching coefficients onto evanescent 
operators. We adopt a renormalization scheme in which the matrix 
elements of these operators between physical states vanish
\cite{Dugan:1990df}, and so do not list their matching coefficients
here. However, the choice of evanescent operators is not unique, and
the jet functions as well as the matrix elements of the physical
SCET$_{\rm II}$ four-quark operators depend on this choice. We have 
chosen the evanescent operators such that the matrix elements of the 
physical operators are given by the $\overline{\rm MS}$ LCDAs of the 
$B$ meson and light meson \cite{toappear}.\footnote{We are indebted to 
M.~Beneke and D.~s.~Yang for pointing out that this was not true with 
the choice of evanescent operators adopted in the preprint version of 
this paper.}

While the numerical impact of the one-loop matching corrections to the
jet functions will be studied later, it is interesting at this point
to use the explicit expressions in (\ref{1loopjet}) to learn something
about the ``natural scale'' to be used in the perturbative expansions
for the jet functions.  In the BLM scale-setting prescription, the
scale in the coupling constant of the leading-order terms is chosen so
as to absorb all terms proportional to
$\beta_0=\frac{11}{3}\,C_A-\frac43\,T_F n_f$ in the next-to-leading
order corrections~\cite{Brodsky:1982gc}.  In the present case, this
means that (for fixed $u$ and $\omega$ values) we should replace
$\alpha_s(\mu_i)\to\alpha_s(\mu_{\rm BLM})$, where
\be
   \mu_{\rm BLM}^2 = e^{-\frac53}\cdot 2\bar u E\omega
\ee
for both jet functions, which agrees with previous results reported in
\cite{Neubert:2002ix}. When applied to a specific process, such as the 
hard-scattering contributions to heavy-to-light form factors to be considered 
in Section~\ref{sec:applic}, the BLM scale setting should be done after the
convolution integrals over the LCDAs have been performed, which is equivalent
to using mean values $\bar u\to e^{\langle\ln\bar u\rangle}$ and 
$\omega\to e^{\langle\ln\omega\rangle}$, weighted by the corresponding 
tree-level amplitudes, to define an effective scale for each decay process. 
The resulting BLM scales, while parametrically of order $\mu_i$, are 
numerically rather low, typically $\mu_{\rm BLM}\approx 0.7$\,GeV or less. 
The smallness of these scales appears to suggest a poor convergence of the 
perturbative expansions of the jet functions. However, it turns out 
that the BLM prescription cannot be trusted in the present 
case.  In the limit where we neglect the mild $y$-dependence of the coefficients 
$C_j^{B'}(E,y,\mu_i)$, 
it follows from (\ref{eq:jet_function}) and (\ref{j12def}) that 
the convolution integrals relevant to the 
hard-scattering contributions governed by the longitudinal jet function 
${\cal J}_\parallel$ yield
\bea\label{eq:integrated_jet}
   &&\int_0^\infty\!{d\omega\over\omega}\,\phi_B(\omega,\mu_i)
    \int_0^1\!\frac{du}{\bar u}\,\phi_M(u,\mu_i)\,\alpha_s(\mu_i)
    \left[ 1 - \frac{\alpha_s(\mu_i)}{4\pi} \int_0^1\!dy\,
    j_\parallel(u,y,\ln(2E\omega/\mu_i^2)) \right] \quad \nl
   &&\propto \alpha_s(\mu_i) \left\{ 1 + {\alpha_s(\mu_i)\over 4\pi}
    \left[ {4\over 3} \left\langle \ln^2{2E\omega\over\mu_i^2} \right\rangle
    + \left( - {19\over 3} + {\pi^2\over 9} \right)
    \left\langle \ln{2E\omega\over \mu_i^2} \right\rangle + 3.99 \right]
    \right\} ,
\eea
where $\langle\dots\rangle$ denotes an average over the $B$-meson LCDA
$\phi_B(\omega,\mu_i)$ with measure $d\omega/\omega$, and for
simplicity we have assumed the asymptotic form $\phi_M(u,\mu_i)=6u\bar
u$ for the LCDA of the light meson. The corresponding result for the
jet function ${\cal J}_\perp$ is obtained by replacing
$j_{\parallel}\to j_\perp$, in which case the coefficient of the
single logarithm changes to $(-6+\frac{\pi^2}{9})$, and the constant
term changes to 0.73. If scale $\mu_i$ is chosen such that the
logarithms appearing in this expression are not too large, then (at
least through one-loop order) the perturbative corrections are of
moderate size. For comparison, keeping only the BLM terms of order
$\beta_0\alpha_s^2$, we would obtain
\be\label{eq:integrated_jetBLM}
   \alpha_s(\mu_i) \left\{ 1 + {\alpha_s(\mu_i)\over 4\pi}
   \left[ -{25\over 3} \left\langle \ln{2E\omega\over \mu_i^2} \right\rangle
   + 26.39 \right] \right\}
\ee
instead of the expression in the second line of (\ref{eq:integrated_jet}). 
We conclude that the BLM prescription overestimates the size of the 
$\order(\alpha_s^2)$ corrections by large amounts. The perturbative 
expansions of the jet functions at a scale 
$\mu_i=\sqrt{2E\Lambda_h}\approx 1.5$\,GeV appear to be reasonably well 
behaved.

\subsection{Sudakov resummation in \boldmath SCET$_{\rm II}$\unboldmath}
\label{sec:resumscetii}

Having performed the matching onto SCET$_{\rm II}$, we now wish to evolve 
the momentum-space coefficient functions $D_i$ in (\ref{eq:momentum_D})
from the intermediate scale $\mu_i\sim\sqrt{2E\Lambda_{\rm QCD}}$ down to a 
hadronic scale $\mu$, which is independent of the large energy $E$. This can 
be achieved by solving the integro-differential RG equation
\cite{Becher:2003kh}
\be\label{eq:SCET_II_RG}
   {d\over d\ln\mu}\,D_i(E,\omega',u',\mu) = \int_0^\infty\!d\omega
   \int_0^1\!du\,\gamma_\Gamma(\omega,\omega',u,u',\mu)\,
   D_i(E,\omega,u,\mu) \,.
\ee
The anomalous dimension $\gamma_\Gamma$ depends on the spin-parity 
quantum numbers of the light final-state meson $M$. As before, we must
distinguish two cases: $\Gamma=\,\parallel$ for $M=P,V_\parallel$, and 
$\Gamma=\,\perp$ for $M=V_\perp$. The anomalous dimension may be decomposed 
as
\be
   \gamma_\Gamma(\omega,\omega', u,u',\mu)
   = \delta(\omega-\omega')\,K_\Gamma(u,u')
   + \delta(u-u')\,H(\omega,\omega',\mu) \,,
\ee
where
\bea
   v\bar v\,K_\Gamma(u,v)
   &=& {C_F\alpha_s\over\pi}\,\Bigg\{
    - \left[ u\bar v\,{\theta(v-u)\over v-u}
    + v\bar u\,{\theta(u-v)\over u-v} \right]_+ + {1\over 2}\,u\bar u\,
    \delta(u-v) \nl
   &&\qquad\mbox{}- c_\Gamma \Big[ u\bar v\,\theta(v-u)
    + v\bar u\,\theta(u-v) \Big] \Bigg\} + \order(\alpha_s^2) \,,
\eea
with $c_\parallel=1$, $c_\perp=0$ is the Brodsky-Lepage kernel 
\cite{Lepage:1980fj}, and
\bea\label{eq:gamma_Bmeson}
   H(\omega,\omega',\mu)
   &=& \left[ \Gamma_{\rm cusp}(\alpha_s)\,\ln{\mu\over\omega}
    + \gamma(\alpha_s) \right] \delta(\omega-\omega')
    + \omega\,\Gamma(\omega,\omega',\alpha_s) \\
   &=& {C_F\alpha_s\over\pi}\,\Bigg\{
    \left( \ln{\mu\over\omega} - {5\over 4} \right)
    \delta(\omega-\omega')
    - \omega \left[ {\theta(\omega-\omega')\over\omega(\omega-\omega')}
    + {\theta(\omega'-\omega)\over\omega'(\omega'-\omega)} \right]_+
    \Bigg\} + \order(\alpha_s^2) \nonumber
\eea
is the corresponding kernel governing the evolution of the $B$-meson LCDA
\cite{Bosch:2003fc,Lange:2003ff}. 

As a consequence of the conformal symmetry of QCD at the classical level, the
kernel $v\bar v\,K_\Gamma(u,v)$ is symmetric in $u$ and $v$, and $K_\Gamma$ 
is diagonalized by Gegenbauer polynomials (cf.\ (\ref{eq:JacobiP11}) in 
Section~\ref{section:conformal}). We write
\be
   \int_0^1\!du\,K_\Gamma(u,v)\,\varphi_n(u)
   = \kappa_{\Gamma,n}\,\varphi_n(v) \,,
\ee
where at one-loop order
\be
   \kappa_{\Gamma,n} = {C_F\alpha_s\over\pi} \left[
   2(H_{n+1}-1) + {1\over 2} - {c_\Gamma\over(n+1)(n+2)} \right] .
\ee
We choose the normalization of $\varphi_n$ such that they are orthonormal in 
the measure $u\bar u\,du$ on the unit interval,
\bea
   \varphi_n(u) = 2\sqrt{2n+3\over(n+1)(n+2)}\,C_n^{(3/2)}(2u-1)
   = \sqrt{(2n+3)(n+2)\over n+1}\,P_n^{(1,1)}(2u-1) \,.
\eea
After expanding the coefficients $D_i(E,\omega,u,\mu)$ on the basis of 
eigenfunctions $\varphi_n(u)$, the RG equation for each Gegenbauer moment is 
solved as in \cite{Bosch:2003fc}. The result reads
\bea\label{eq:decomp_D}
   D_i(E,\omega,u,\mu)
   &=& \frac{1}{2E\omega} \left( {\omega\over\mu} \right)^{-a(\mu_i,\mu)}
    e^{S(\mu,\mu_i)}\,\sum_{n=0}^\infty \varphi_n(u) \\
   &\times& \int_0^1\!dy\,C^{B'}_j(E,y,\mu_i)
    \int_0^1\!dx\,x\bar x\,\varphi_n(x)\,
    {\cal J}'_{ij}(x,y,\nabla_\eta,\mu_i)\,
    \bigg( {2E\omega\over \mu_i^2} \bigg)^\eta\,
    e^{V_{\Gamma,n}(\mu_i,\mu,\eta)} \bigg|_{\eta=0} , \nonumber
\eea
where $S(\mu,\mu_i)$ and $a(\mu_i,\mu)$ are given in (\ref{eq:defS}) and 
(\ref{eq:def_a}), the jet functions ${\cal J}^\prime_{ij}$ in the primed 
basis are obtained from (\ref{eq:jet_function}) by replacing 
$2E\omega/\mu_i^2\to\nabla_\eta$, and
\be
   V_{\Gamma,n}(\mu_i,\mu,\eta)
   = \int_{\alpha_s(\mu_i)}^{\alpha_s(\mu)} {d\alpha\over\beta(\alpha)}\,
   \bigg[ {\cal F}(\eta - a(\mu_i,\mu(\alpha)),\alpha)
   + \kappa_{\Gamma,n}(\alpha) + \gamma(\alpha) \bigg] \,, 
\ee
with 
\bea
   {\cal F}(\eta,\alpha_s)
   &\equiv& \int_0^\infty\!d\omega\,\omega'\,\Gamma(\omega,\omega',\alpha_s)
    \left( {\omega\over\omega'} \right)^\eta \nl
   &=& \Gamma_{\rm cusp}^{\rm 1-loop}(\alpha_s)\,\Big[ \psi(1+\eta)
    + \psi(1-\eta) + 2\gamma_E \Big] 
    + \order(\alpha_s^2) \,. 
\eea
For a complete leading-order solution, we require the two-loop expression for 
$\Gamma_{\rm cusp}$ appearing in $S(\mu,\mu_i)$, one-loop expressions for the 
remaining anomalous dimensions in $a(\mu_i,\mu)$ and 
$V_{\Gamma,n}(\mu_i,\mu,\eta)$, and tree-level matching conditions for the jet 
functions. At this order, the relevant expansions are 
\bea\label{VLres}
   S(\mu,\mu_i) 
   &=& \frac{\Gamma_0}{4\beta_0^2} \left[ \frac{4\pi}{\alpha_s(\mu)}
    \left( 1 - {r_2} + \ln r_2 \right)
    + \frac{\beta_1}{2\beta_0}\,\ln^2 r_2
    - \left( \frac{\Gamma_1}{\Gamma_0} - \frac{\beta_1}{\beta_0} \right) 
    \left({1\over r_2}-1+\ln r_2\right) \right] , \nl
   a(\mu_i,\mu) 
   &=& - \frac{\Gamma_0}{2\beta_0}\,\ln r_2 \,, \nl
   V_{\Gamma,n}(\mu_i,\mu,0)
   &=& \ln{\Gamma\Big( 1 - {\Gamma_0\over 2\beta_0}\,\ln{r_2} \Big)\over
           \Gamma\Big( 1 + {\Gamma_0\over 2\beta_0}\,\ln{r_2} \Big)}
    - \gamma_E\,{\Gamma_0\over\beta_0}\,\ln{r_2}
    - {1\over 2\beta_0} \left( \kappa_{\Gamma,n}^{(0)} + \gamma_0
    \right) \ln{r_2} \,,  
\eea
where $r_2=\alpha_s(\mu)/\alpha_s(\mu_i)\ge 1$, and as usual we have expanded 
\be
   \kappa_{\Gamma,m} = \kappa_{\Gamma,m}^{(0)}\,{\alpha_s\over 4\pi}
   + \dots \,,
   \qquad \gamma = \gamma_0\,{\alpha_s\over 4\pi} + \dots \,.
\ee

\section{Application to heavy-to-light form factors}
\label{sec:applic}

In this section we apply our results to obtain resummed expressions for the 
hard-scattering contributions to heavy-to-light form factors at large recoil 
energy. We begin by recalling the definitions of the ten form factors 
describing $B$ decays into pseudoscalar and vector mesons. Equating these 
definitions to the corresponding effective-theory expressions shows that at 
leading order in $1/E$ only eight of the form factors are independent. As 
shown in (\ref{eq:soft_plus_hard}), in this approximation the form factors 
are given by the sum of a soft-overlap contribution, expressed in terms of 
universal non-perturbative matrix elements $\zeta_M(E)$ (where $M=P$, 
$V_\parallel$, $V_\perp$ depending on the spin-parity quantum numbers of the 
light final-state meson), and a hard-scattering contribution, given by a 
convolution integral over meson LCDAs. We focus on the second term and 
present general, resummed expressions for the corresponding hard-scattering 
kernels. The results are particularly simple when hard matching corrections 
are ignored. In this case, for a given final-state meson $M$, the 
hard-scattering contributions to all $B\to M$ form factors can be 
parameterized in terms of a universal quantity $H_M$. The Sudakov suppression 
factors are mild in all cases. In order to investigate the relative 
importance of running and matching corrections, we apply our previous 
one-loop matching results for the scalar current to obtain the one-loop 
corrected hard-scattering contribution to the vector-current form factor
$F_0$, and to the form factor $F_+$ at maximum recoil ($q^2=0$).

\subsection{Form-factor definitions and factorization formulae}

We begin by recalling the definitions of the form factors parameterizing $B$ 
decays into pseudoscalar ($P$) and vector ($V$) mesons, following the 
conventions of \cite{Beneke:2000wa} (we use the sign convention 
$\epsilon^{0123}=-1$):
\bea\label{eq:formfactordefs}
   \langle P(p^\prime)| \bar{q}\,\gamma^\mu b |\bar{B}(p)\rangle
   &=& F_+(q^2) \left( p^\mu + p^{\prime\mu}
    - {m_B^2 - m_P^2 \over q^2}\,q^\mu \right)
    + F_0(q^2)\,{m_B^2 - m_P^2\over q^2}\,q^\mu \,, \nl
   \langle P(p^\prime)| \bar{q}\,\sigma^{\mu\nu}q_\nu\, b |\bar{B}(p)\rangle
   &=& {iF_T(q^2)\over m_B + m_P} \left[
    q^2\,(p^\mu + p^{\prime\mu}) - (m_B^2-m_P^2)\,q^\mu \right] , \nl
   \langle V(p^\prime,\eta)| \bar{q}\,\gamma^\mu b |\bar{B}(p)\rangle
   &=& {2iV(q^2)\over m_B + m_V}\,{\epsilon}^{\mu\nu\rho\sigma}
    \eta^*_\nu\,p^\prime_\rho\,p_\sigma \,, \nl
   \langle V(p^\prime,\eta)| \bar{q}\,\gamma^\mu\gamma_5\,b
   |\bar{B}(p)\rangle
   &=& 2m_V A_0(q^2)\,{\eta^*\cdot q\over q^2}\,q^\mu
    + (m_B+m_V)\,A_1(q^2) \left( \eta^{*\mu}
    - {\eta^*\cdot q\over q^2}\,q^\mu \right) \nl
   &&\mbox{}- A_2(q^2)\,{\eta^*\cdot q\over m_B + m_V}
    \left( p^\mu + p^{\prime\mu} - {m_B^2-m_V^2\over q^2}\,q^\mu \right) ,
    \nl
   \langle V(p^\prime,\eta)| \bar{q}\,\sigma^{\mu\nu} q_\nu\,b 
   |\bar{B}(p)\rangle
   &=& -2 T_1(q^2)\,{\epsilon}^{\mu\nu\rho\sigma}
    \eta^*_\nu\,p^\prime_\rho\,p_\sigma \,, \nl
   \langle V(p^\prime,\eta)| \bar{q}\,\sigma^{\mu\nu}\gamma_5\,q_\nu\,b
   |\bar{B}(p)\rangle
   &=& -i\,T_2(q^2)\left[ (m_B^2-m_V^2)\,\eta^{*\mu} - \eta^*\cdot q\,
    (p^\mu + p^{\prime\mu}) \right] \nl
   &&\mbox{}-i\,T_3(q^2)\,\eta^*\cdot q \left[ q^\mu
    - {q^2\over m_B^2-m_V^2}\,(p^\mu + p^{\prime\mu}) \right] .
\eea
Here $q=p-p^\prime$ is the momentum transfer, and $\eta$ denotes the 
polarization vector of the vector meson, which satisfies $\eta\cdot p'=0$ 
and $\eta\cdot\eta^*=-1$. The polarization vector for longitudinally 
polarized vector mesons is given by
\be
  \eta_\parallel^\mu = {p^{\prime\,\mu}\over m_V}
  - {m_V \over\nb\cdot p^\prime}\,\nb^\mu
  \simeq {E\over m_V}\,\frac{n^\mu}{n\cdot v} \,.
\ee

For the evaluation of the hadronic matrix elements entering the 
hard-scattering contributions to the form factors, we need the definitions of 
the leading-twist LCDAs for the light mesons in terms of SCET$_{\rm II}$ 
fields. They are \cite{Beneke:1999br}
\bea
   \langle P(p^\prime)| \bar{\X}_c(s\nb)\,\Gamma\,{\sla{\nb}\over 2}\,\X_c(0)
   |0\rangle
   &=& {if_P\over 4}\,\nb\cdot p^\prime\,\,
    {\rm tr} \left( \gamma_5\,\Gamma\,{\sla{\nb}\sla{n}\over 4} \right)
    \int_0^1\!du\,e^{ius\nb\cdot p^\prime}\,\phi_P(u,\mu) \,, \nl
   \langle V_{\parallel}(p^\prime)| \bar{\X}_c(s\nb)\,\Gamma\,
   {\sla{\nb}\over 2}\,\X_c(0) |0\rangle
   &=& - {if_{V_\parallel}\over 4}\,\nb\cdot p^\prime\,\,
    {\rm tr} \left( \Gamma\,{\sla{\nb}\sla{n}\over 4} \right)
    \int_0^1\!du\,e^{ius\nb\cdot p^\prime}\,\phi_{V_\parallel}(u,\mu) \,, \nl 
   \langle V_\perp(p^\prime, \eta)| \bar{\X}_c(s\nb)\,\Gamma\,
   {\sla{\nb}\over 2}\,\X_c(0) |0\rangle
   &=& {if_{V_\perp}(\mu) \over 4}\,\nb\cdot p^\prime\,\,
    {\rm tr} \left( \sla{\eta^*_\perp}\,\Gamma\,{\sla{\nb}\sla{n}\over 4}
    \right)
    \int_0^1\!du\,e^{ius\nb\cdot p^\prime}\,\phi_{V_\perp}(u,\mu) \,,
    \qquad
\eea
where the $\phi_M$ are normalized according to 
$\int_0^1\!du\,\phi_M(u,\mu)=1$. Note that the transverse vector-meson decay 
constants are scale dependent. Below, we will sometimes write $f_M(\mu)$ as a 
generic notation for the light-meson decay constants, keeping in mind that a
scale dependence is present only in the case of $f_{V_\perp}$. The LCDA for 
the $B$-meson is defined as \cite{Grozin:1996pq,Lange:2003ff}
\be\label{phiBdef}
   \langle 0| \bar{\Q}_s(tn)\,{\sla{n}\over 2}\,\Gamma\,\H(0)
   |\bar{B}_v\rangle
   = - {i\sqrt{m_B}\,F(\mu)\over 2}\,\,
   {\rm tr} \left( {\sla{n}\over 2}\,\Gamma\,{1+\sla{v}\over 2}\,\gamma_5
   \right)
   \int_0^\infty\!d\omega\,e^{-i\omega tn\cdot v}\,\phi_B(\omega,\mu) \,,
\ee
where $\bar B_v$ denotes the $B$-meson state defined in HQET, and $F$ is 
related to the asymptotic value of the $B$-meson decay constant in the 
heavy-quark limit (see below). 

For completeness, we list also the soft-overlap contributions to form 
factors, parameterized in terms of universal matrix elements $\zeta_M$~\cite{Charles:1998dr,Beneke:2000wa}. 
At leading order in $1/E$, we define\footnote{Our 
convention for $\zeta_{V_\parallel}$ is the same as in \cite{Charles:1998dr}, 
and is such that all $\zeta_M$ functions have the same power counting. 
This differs from the corresponding function
$\xi_\parallel$ in \cite{Beneke:2000wa}, given by $\zeta_{V_\parallel}=(E/m_V)\,\xi_\parallel$.}
\bea\label{eq:zetas}
   \langle P(p^\prime)| \bar{\X}_{hc}\,\Gamma\,h |\bar{B}_v\rangle
   &=& 2E\,\zeta_P(E)\,\,{\rm tr} \left( {\sla{\nb}\sla{n}\over 4}\,
    \Gamma\,{1+\sla{v}\over 2} \right) , \nl
   \langle V_{\parallel}(p^\prime)| \bar{\X}_{hc}\,\Gamma\,h 
   |\bar{B}_v\rangle 
   &=& -2E\,\zeta_{V_\parallel}(E)\,\,
    {\rm tr}\left( {\sla{\nb}\sla{n}\over 4}\,\Gamma\,{1+\sla{v}\over 2}\,
    \gamma_5 \right) , \nl
   \langle V_{\perp}(p^\prime)| \bar{\X}_{hc}\,\Gamma\,h |\bar{B}_v\rangle 
   &=& 2E\,\zeta_{V_\perp}(E)\,\,{\rm tr}\left( \sla{\eta}^*_\perp\,
    {\sla{\nb}\sla{n}\over 4}\,\Gamma\,{1+\sla{v}\over 2}\,\gamma_5 \right) .
\eea
The definitions are given in terms of SCET$_{\rm I}$ currents. As shown in 
\cite{Lange:2003pk}, the functions $\zeta_M(E)$ can be decomposed further 
into matrix elements of SCET$_{\rm II}$ operators. These operators all 
satisfy the same symmetry relations, and the linear combinations contributing 
to the form factor behave under renormalization in precisely the same way as 
the SCET$_{\rm I}$ operators used in the definitions (\ref{eq:zetas}). 

To simplify the factorization formulae, it proves convenient to write the 
coefficient functions in the form 
\be\label{eq:DversusT}
   D_i(E,\omega,u,\mu)\equiv {1\over\omega}\,{4K_F(\mu)\over m_B f_B}\,
   T_i(E,\omega,u,\mu) \,,
\ee
which serves as a definition of the hard-scattering kernels $T_i$. Here 
$K_F$ relates the HQET parameter $F$ in (\ref{phiBdef}) to the physical 
$B$-meson decay constant, $f_B\sqrt{m_B}=K_F(\mu)\,F(\mu)$, up to 
corrections of order $\Lambda_{\rm QCD}/m_b$. At next-to-leading order
\be\label{eq:KF}
   K_F(\mu) = 1 + {C_F\alpha_s(\mu)\over 4\pi}
   \left( 3\ln{m_b\over \mu} - 2 \right) .
\ee

With all of the definitions in place, it is now a straightforward matter to 
equate QCD matrix elements to the corresponding SCET expressions. For the 
scalar and pseudoscalar currents, for example, we obtain
\bea\label{eq:scalarme}
   {1\over m_B}\,\langle P(p^\prime)| \bar{q}\,b |\bar{B}(p)\rangle  
   &=& {2E\over m_B}\,C^A_S(E,\mu)\,\zeta_P(E,\mu) \nl 
   &+& {2E\over m_B} \int_0^\infty\!{d\omega\over\omega}\,\phi_B(\omega,\mu)
    \int_0^1\!du\,f_P\,\phi_P(u,\mu)\,T_S(E,\omega,u,\mu) \nl
   &\equiv& {2E\over m_B} \left[ C^A_S\,\zeta_P 
    + \phi_B\otimes f_P\,\phi_P\otimes T_S \right] ,\nl
   {1\over m_B}\,\langle V_{\parallel}(p^\prime)| \bar{q}\,\gamma_5\,b
   |\bar{B}(p)\rangle
   &=& - {2E\over m_B} \left[ C^A_S\,\zeta_{V_\parallel} 
    + \phi_B\otimes f_{V_\parallel}\,\phi_{V_\parallel}\otimes T_S \right] , 
\eea
where in the second line we have introduced a symbolic notation for the 
convolution integrals. In general, the scale $\mu$ in the soft-overlap terms 
could be taken different from that in the hard-scattering terms. Similarly, 
from the matrix elements of vector and tensor currents we find the relations
\bea\label{eq:factorization}
   F_+ &=& \left( C^A_{V1} + {E\over m_B}\,C^A_{V2} + C^A_{V3} \right)
    \zeta_P + \phi_B\otimes f_P\,\phi_P\otimes 
    \left( T_{V1} + {E\over m_B}\,T_{V2} \right) , \nl
   \frac{m_B}{2E}\,F_0 &=& \left[ C^A_{V1}
    + \!\left( 1 - {E\over m_B} \right)\! C^A_{V2} +C^A_{V3} \right] \zeta_P
    + \phi_B\otimes f_P\,\phi_P\otimes\! \left[ T_{V1}
    + \!\left( 1 - {E\over m_B} \right)\! T_{V2} \right] \!,\nl
   \frac{m_B}{m_B+m_P}\,F_T
   &=& \left( C_{T1}^A - \frac12\,C_{T2}^A + \frac12\,C_{T4}^A \right)
    \zeta_P 
    + \phi_B\otimes f_P\,\phi_P\otimes \frac12\,T_{T1} \,, \nl
   A_0 &=& \!\left[ C^A_{V1} + \!\left(\! 1 - {E\over m_B} \right)\! C^A_{V2} 
    + C^A_{V3} \right]\!\zeta_{V_\parallel}
    + \phi_B\otimes f_{V_\parallel}\,\phi_{V_\parallel}\otimes
    \!\left[ T_{V1} + \!\left(\! 1 - {E\over m_B} \right)\! T_{V2} \right]
    \!, \nl
   \frac{m_B+m_V}{2E}\,A_1 &=& C^A_{V1}\,\zeta_{V_\perp} 
    + \phi_B\otimes f_{V_\perp}\,\phi_{V_\perp}\otimes T_{V3} \,, \nl
   \frac{m_B}{m_B+m_V}\,V &=& C^A_{V1}\,\zeta_{V_\perp} 
    + \phi_B\otimes f_{V_\perp}\,\phi_{V_\perp}\otimes T_{V3} \,, \nl
   \frac{E m_B\,(V-A_2)}{m_V(m_B\!+\!m_V)} 
   &=& \left( C^A_{V1} + {E\over m_B}\,C^A_{V2} + C^A_{V3} \right) 
    \zeta_{V_\parallel} 
    + \phi_B\otimes f_{V_\parallel}\,\phi_{V_\parallel}\otimes 
    \left( T_{V1} + {E\over m_B}\,T_{V2} \right) , \nl
   T_1 &=& \left[ C^A_{T1} - {1\over 2} \left( 1 - {E\over m_B} \right) 
    C^A_{T2} - {1\over 2}\,C^A_{T3} \right] \zeta_{V_\perp} \nl
   &&\mbox{}- \frac12\,\phi_B\otimes f_{V_\perp}\,\phi_{V_\perp}\otimes 
    \left[ T_{T2} + \left( 1 - {E \over m_B} \right) T_{T3} \right] , \nl
   \frac{m_B}{2E}\,T_2 &=& \left[ C^A_{T1}
    - {1\over 2} \left( 1 - {E\over m_B} \right) C^A_{T2}
    - {1\over 2}\,C^A_{T3} \right] \zeta_{V_\perp} \nl
   &&\mbox{}- \frac12\,\phi_B\otimes f_{V_\perp}\,\phi_{V_\perp}\otimes 
   \left[ T_{T2} + \left( 1 - {E\over m_B} \right) T_{T3} \right] , \nl
   \frac{m_B T_2 - 2E\,T_3}{2m_V}\!
   &=& \!\left( C^A_{T1}- {1\over 2}\,C^A_{T2} + {1\over 2}\,C^A_{T4}
    \right) \zeta_{V_\parallel}
    + \phi_B\otimes f_{V_\parallel}\,\phi_{V_\parallel}\otimes
    \frac12\,T_{T1} \,.
\eea
We have dropped terms of relative order $(m_M/E)^2$, where $m_M$ is the light 
meson mass. Recall that we have defined $E\equiv v\cdot P_-$, which differs 
from the true energy by an amount of order $m_M^2/E$.  Also, at leading order
in $\Lambda_{\rm QCD}/m_b$ there is no difference between $m_B$ and $m_b$. 
The results for $F_+$, $F_0$ and $F_T$ in (\ref{eq:factorization}) 
agree at tree level with the corresponding expressions in 
\cite{Pirjol:2002km}.\footnote{See also \cite{Pirjol:2003ef}. In equation (22) 
of this paper $-B_4^{(t)}$ should be replaced by $+B_4^{(t)}$ \cite{danprivate}.} 
Beyond tree level, and to any order in perturbation theory,  
we find that there is still only a single jet function for decays into a given light 
final-state meson, e.g. ${\cal J}_\parallel$ for a pseudoscalar meson. 
This implies that the two functions $J_a$ and $J_b$ introduced in 
\cite{Pirjol:2002km} are not independent. Our result has important implications for 
the universality of hard-scattering corrections, which we discuss in more detail in 
the following subsection. The general results (\ref{eq:factorization}) establish 
that the form-factor relations 
\be
   A_1(q^2) = {2E m_B\over (m_B+m_V)^2}\,V(q^2) \,, \qquad
   T_2(q^2) = {2E\over m_B}\,T_1(q^2) \,,
\ee
are rigorous predictions of QCD at leading order in $1/E$, and to all orders 
in $\alpha_s$. These relations were conjectured in \cite{Burdman:2000ku}, and 
they were shown to hold at one-loop order in \cite{Beneke:2000wa}. 

The scalar and pseudoscalar currents may be related to the vector and 
axial-vector currents using the equation of motion for the quark fields, as 
in (\ref{eq:coeffRelations}). When applied to the matrix elements appearing 
in (\ref{eq:formfactordefs}) and (\ref{eq:scalarme}), it follows that 
(setting as previously the light quark masses to zero)
\bea\label{eq:eom_ff}
   \frac{\overline{m}_b(\mu_{\rm QCD})}{m_B^2}\,\langle P(p^\prime)|
   [\bar{q}\,b](\mu_{\rm QCD}) |\bar{B}(p)\rangle  
   &=& \left( 1 - {m_P^2\over m_B^2} \right) F_0(q^2) 
    = F_0(q^2) + \dots \,, \nl
   \frac{\overline{m}_b(\mu_{\rm QCD})}{m_B^2}\,\langle V_{\parallel}(p')|
   [\bar{q}\,\gamma_5\,b](\mu_{\rm QCD}) |\bar{B}(p)\rangle
   &=& - {2m_V\,\eta^*_\parallel\cdot q\over m_B^2}\,A_0(q^2) 
    = - {2E\over m_B}\,A_0(q^2) + \dots \,, \qquad 
\eea
where the dots represent terms of order $m_M^2/m_B^2$. Comparison with 
(\ref{eq:factorization}) shows that to all orders in perturbation theory
\bea\label{eq:scetii_rel}
   C^A_{V1} + \left( 1 - {E\over m_b} \right) C^A_{V2} + C^A_{V3}
   &=& \frac{\overline{m}_b(\mu_{\rm QCD})}{m_b}\,C^A_S \,, \nl
   T_{V1} + \left( 1 - {E\over m_b} \right) T_{V2}
   &=& \frac{\overline{m}_b(\mu_{\rm QCD})}{m_b}\,T_S \,.  
\eea
These are the SCET$_{\rm II}$ realizations of the SCET$_{\rm I}$ coefficient 
relations found in (\ref{eq:coeffRelations}). 

\subsection{Sudakov resummation for the hard-scattering contributions}

We now apply the results of Sections \ref{sec:solution} and \ref{sec:SCETII} 
to obtain completely general, resummed expressions for the hard-scattering 
kernels $T_i$ appearing in the form-factors (\ref{eq:factorization}). Starting from the
defining relation (\ref{eq:DversusT}) for $T_i$, and combining it with the
result (\ref{eq:decomp_D}) for the coefficients $D_i$, we obtain
\bea\label{eq:resumT}
   T_i(E,\omega,u,\mu)
   &=& {m_B f_B\over 8E}\,{e^{V_F(\mu_h,\mu)}\over K_F(\mu_h)} 
    \left( {2E\over\mu_h} \right)^{a(\mu_h,\mu_i)}
    \left( {\omega\over\mu} \right)^{-a(\mu_i,\mu)}
    e^{S(\mu_h,\mu_i)+S(\mu,\mu_i)} \nl
   &\times& \sum_{n=0}^\infty \varphi_n(u) \int_0^1\!dy
    \int_0^1\!dv\,U_\Gamma(y,v,\mu_h,\mu_i)\,C_j^{B^\prime}(E,v,\mu_h) \nl
   &\times& \int_0^1\!dx\,x\bar x\,\varphi_n(x)\,
    {\cal J}'_{ij}(x,y,\nabla_\eta,\mu_i)
    \left( {2E\omega\over\mu_i^2} \right)^\eta
    e^{V_{\Gamma,n}(\mu_i,\mu,\eta)} \bigg|_{\eta=0} \,.
\eea
We have used (\ref{eq:decomp_C}) to express the coefficients 
$C_j^{B'}(E,y,\mu_i)$ at the intermediate scale, which enter in 
(\ref{eq:decomp_D}), in terms of their matching conditions at the high scale 
$\mu_h$. Also, we have rewritten $K_F(\mu)=e^{-V_F(\mu_h,\mu)}\,K_F(\mu_h)$,
where
\be\label{eq:VF_LO}
   V_F(\mu_h,\mu) = - \int_{\alpha_s(\mu_h)}^{\alpha_s(\mu)}
    {d\alpha\over\beta(\alpha)}\,\gamma_F(\alpha) 
   = {\gamma_{F,0}\over 2\beta_0}\,\ln{r_1r_2} + \dots
\ee
is the solution to the evolution equation
\be\label{eq:K_Frunning}
   {d\over d\ln{\mu}}\,K_F = \gamma_F\,K_F \,, \qquad
   \gamma_F = \sum_{n=0}^\infty \gamma_{F,n}
   \left( {\alpha_s\over 4\pi} \right)^{n+1} ,
\ee
with $\gamma_{F,0}=-3C_F$. As previously, 
$r_1=\alpha_s(\mu_i)/\alpha_s(\mu_h)$ and $r_2=\alpha_s(\mu)/\alpha_s(\mu_i)$.
In (\ref{eq:resumT}) all large logarithms are resummed into the various 
evolution functions. The matching coefficients $K_F(\mu_h)$, 
$C_j^{B'}(E,v,\mu_h)$, and ${\cal J}'_{ij}(x,y,\nabla_\eta,\mu_i)$ can be 
calculated reliably in fixed-order perturbation theory. The expressions for
the kernels $T_i$ are formally (to the order we are working) independent of
the two matching scales $\mu_h$ and $\mu_i$. A dependence on the low scale
$\mu$ remains, which will cancel against the scale dependence of the hadronic
matrix elements of the SCET$_{\rm II}$ operators.

The general result (\ref{eq:resumT}) simplifies considerably when hard 
matching corrections at the high scale $\mu_h\sim m_b$ are ignored. In this 
case $K_F(\mu_h)=1$, and using the fact that the tree-level matching 
conditions for $C_j^{B'}(E,v,\mu_h)$ are independent of $v$, it follows that
\be
   \int_0^1\!dv\,U_\Gamma(y,v,\mu_h,\mu_i)\,C_j^{B'}(E,v,\mu_h)
   = U_\Gamma(y,\mu_h,\mu_i)\,C_j^{B'}(E,\mu_h) \,,
\ee
where $U_\Gamma(y,\mu_h,\mu_i)$ are the functions shown in 
Figure~\ref{fig:tree}. Using the tree-level matching conditions compiled in 
(\ref{eq:scalarmatching_tree})--(\ref{eq:tensormatching}) along with the
relations (\ref{Cprime1}) and (\ref{Cprime2}), it is straightforward to 
evaluate the products ${\cal J}'_{ij}\,C_j^{B'}(E,\mu_h)$ for the various
kernels $T_i$, with the matrices ${\cal J}'_{ij}$ as given in (\ref{Jmatrs}). 
We find
\bea\label{simpleTs}
   &&T_S:~-{\cal J}_\parallel \,, \nl
   &&T_{V1}:~\left( 1 - \frac{2E}{m_b} \right) {\cal J}_\parallel \,, \qquad
    T_{V2}:~-2 {\cal J}_\parallel \,, \qquad
    T_{V3}:~0 \,, \nl[-0.2cm]
   &&T_{T1}:~2 {\cal J}_\parallel \,, \qquad
    T_{T2}:~\frac{4E}{m_b}\,{\cal J}_\perp \,, \qquad
    T_{T3}:~0 \,.
\eea
It follows that the convolution integrals for the hard-scattering 
contributions to the form factors in (\ref{eq:factorization}) may be written 
in terms of three universal functions
\bea\label{eq:DeltaFM}
   H_M &\equiv&
    - {E f_B\over 2m_B}\,\left( {2E\over\mu_h} \right)^{a(\mu_h,\mu_i)} 
    e^{S(\mu_h,\mu_i) + S(\mu,\mu_i) + V_F(\mu_h,\mu)}\,
    \sum_{n=0}^\infty \int_0^1\!du\,\varphi_n(u)\,f_M(\mu)\,\phi_M(u,\mu)
    \nl[-0.2cm]
   &\times& \int_0^\infty\!{d\omega\over\omega}\,\phi_B(\omega,\mu) 
    \left( {\omega\over\mu} \right)^{-a(\mu_i,\mu)}\!
    \int_0^1\!dy\,U_\Gamma(y,\mu_h,\mu_i) \nl 
   &\times& \int_0^1\!dx\,x\bar x\,\varphi_n(x)\,
    {\cal J}_\Gamma(x,y,\nabla_\eta,\mu_i)
    \left( {2E\omega\over\mu_i^2} \right)^\eta
    e^{V_{\Gamma,n}(\mu_i,\mu,\eta)} \bigg|_{\eta=0} \,,
\eea
where $\Gamma=\,\parallel$ for $M=P$ or $V_\parallel$, and $\Gamma=\,\perp$ 
for $M=V_\perp$. We have extracted a factor $-(m_B/2E)^2$ from the expression 
in (\ref{eq:resumT}) so as to remove any non-logarithmic $E$ dependence from
the quantities $H_M$ and ensure that they are positive. (Recall that the 
tree-level jet functions are proportional to $-1/E$.) Note that in 
(\ref{eq:DeltaFM}) the $\mu$ dependence of the kernels $T_i$ cancels against 
that of the hadronic quantities $\phi_B$, $\phi_M$, and $f_M$. We repeat that 
this expression neglects hard matching corrections at the scale 
$\mu_h\sim m_b$, but it allows for arbitrary corrections to the jet 
functions. If only the tree-level matching conditions for the jet functions 
given in (\ref{eq:jetTree}) are retained, the result simplifies further to
\bea\label{HMsimple}
   H_M &=& \frac{\pi C_F\alpha_s(\mu_i)}{N}\,\frac{f_B}{m_B}
    \left( {2E\over\mu_h} \right)^{a(\mu_h,\mu_i)} 
    e^{S(\mu_h,\mu_i) + S(\mu,\mu_i)}\,
    \sum_{n=0}^\infty \int_0^1\!du\,\varphi_n(u)\,f_M(\mu)\,\phi_M(u,\mu)
    \nl[-0.2cm]
   &\times& e^{V_F(\mu_h,\mu) + V_{\Gamma,n}(\mu_i,\mu,0)}
    \int_0^\infty\!{d\omega\over\omega}\,\phi_B(\omega,\mu) 
    \left( {\omega\over\mu} \right)^{-a(\mu_i,\mu)}\!
    \int_0^1\!dy\,y\,\varphi_n(y)\,U_\Gamma(y,\mu_h,\mu_i) \,.
\eea

From (\ref{eq:factorization}) and (\ref{simpleTs}), it is easy to read off 
the hard-scattering contributions $\Delta F_i$ to the various form factors.
For the three $B\to P$ form factors, we obtain
\bea
   \Delta F_+ &=& \left( \frac{m_B}{2E} \right)^2 
    \left( \frac{4E}{m_B} - 1 \right) H_P \,, \qquad
    \Delta F_0 = \frac{m_B}{2E}\,H_P \,, \nl
   \Delta F_T &=& - \left( \frac{m_B}{2E} \right)^2 
    \left( 1 + \frac{m_P}{m_B} \right) H_P \,.
\eea
The corresponding results for the seven $B\to V$ form factors read
\bea
   \Delta A_0 &=& \left( \frac{m_B}{2E} \right)^2 H_{V_\parallel} \,, \qquad
    \Delta A_1 = \Delta V = 0 \,, \nl
   \Delta A_2 &=& - \frac{2m_V}{m_B-m_V} \left( \frac{m_B}{2E} \right)^3
    \left( \frac{4E}{m_B} - 1 \right) H_{V_\parallel} \,,
\eea
and
\bea
   \Delta T_1 &=& \frac{m_B}{2E}\,H_{V_\perp} \,, \qquad
    \Delta T_2 = H_{V_\perp} \,, \nl
   \Delta T_3 &=& \frac{m_B}{2E} \left( H_{V_\perp}
    + \frac{m_V m_B}{2E^2}\,H_{V_\parallel} \right) .
\eea

\subsection{Numerical results}

We are finally in a position to study the phenomenological implications of 
our results. For the leading-twist LCDAs of the light mesons, we take for
simplicity the asymptotic forms
\be
   \phi_P(u) = \phi_{V_\parallel}(u) = \phi_{V_\perp}(u) = 6u(1-u) \,.
\ee
Then only the $n=0$ term contributes to the sum (\ref{eq:DeltaFM}), where 
$\varphi_0(u)=\sqrt{6}$. To proceed further we require
the $B$-meson LCDA, which is poorly known at present. As an illustrative 
model, we adopt the form \cite{Grozin:1996pq}
\be\label{eq:Bmodel}
   \phi_B(\omega,\mu_0) = {\omega\over\lambda_B^2}\,
   e^{-\omega/\lambda_B} \,, \qquad
   \lambda_B = {2\over 3}\,(m_B-m_b)\approx 0.32\,\mbox{GeV} \,,
\ee
where $\mu_0$ is a low hadronic scale, at which the model assumes the stated
functional form. The relevant convolution integral resulting from 
(\ref{HMsimple}) at $\mu=\mu_0$ is then 
\be\label{phiBmoment}
   \int_0^\infty {d\omega\over\omega}\,\phi_B(\omega,\mu_0)\,
   \left( {\omega\over\mu_0} \right)^{-a}
   = {1\over\lambda_B}\,\Gamma(1-a)
   \left( {\lambda_B\over\mu_0} \right)^{-a} .
\ee
Combining all pieces, we find at leading order in RG-improved perturbation 
theory
\bea
   H_M &=& \frac{3\pi C_F\alpha_s(\mu_i)}{N}\,
    \frac{f_B f_M(\mu_i)}{m_B\lambda_B}
    \left( {2E\over\mu_h} \right)^{a(\mu_h,\mu_i)}
    \left( {\lambda_B\over\mu_0} \right)^{-a(\mu_i,\mu_0)}
    e^{S(\mu_h,\mu_i) + S(\mu_0,\mu_i)} \nl
   &\times& e^{V_F(\mu_h,\mu_0) + V_0(\mu_i,\mu_0)}
    \int_0^1\!dy\,2y\,U_\Gamma(y,\mu_h,\mu_i) \,.
\eea
The quantity
\be
   V_0(\mu_i,\mu_0)
   = \ln\Gamma\Big( 1 - {\Gamma_0\over 2\beta_0}\,\ln{r_2} \Big)
   - \left( \gamma_E\,{\Gamma_0\over\beta_0} + 
   \frac{\gamma_0}{2\beta_0} \right) \ln{r_2} \,,  
\ee
where now $r_2=\alpha_s(\mu_0)/\alpha_s(\mu_i)$, results from the product of 
the $\Gamma$ function in (\ref{phiBmoment}) times $e^{V_{\Gamma,0}}$ from 
(\ref{VLres}), after the term involving $\kappa_{\Gamma,0}^{(0)}$ has been 
used to cancel the scale dependence of $f_M(\mu)$, replacing it with 
$f_M(\mu_i)$. We expand the RG factors $a(\mu_h,\mu_i)$ and $S(\mu_h,\mu_i)$ 
according to (\ref{VHres}), $a(\mu_i,\mu_0)$ and $S(\mu_0,\mu_i)$ according 
to (\ref{VLres}), and $V_F(\mu_h,\mu)$ as shown in (\ref{eq:VF_LO}). 

\begin{figure}
\begin{center}
\begin{tabular}{cc}
\psfrag{x}[]{$\mu_0$~[GeV]}
\psfrag{y}[b]{$H_{P,V_\parallel}/H_{P,V_\parallel}^{\rm tree}$}
\includegraphics[width=0.48\textwidth]{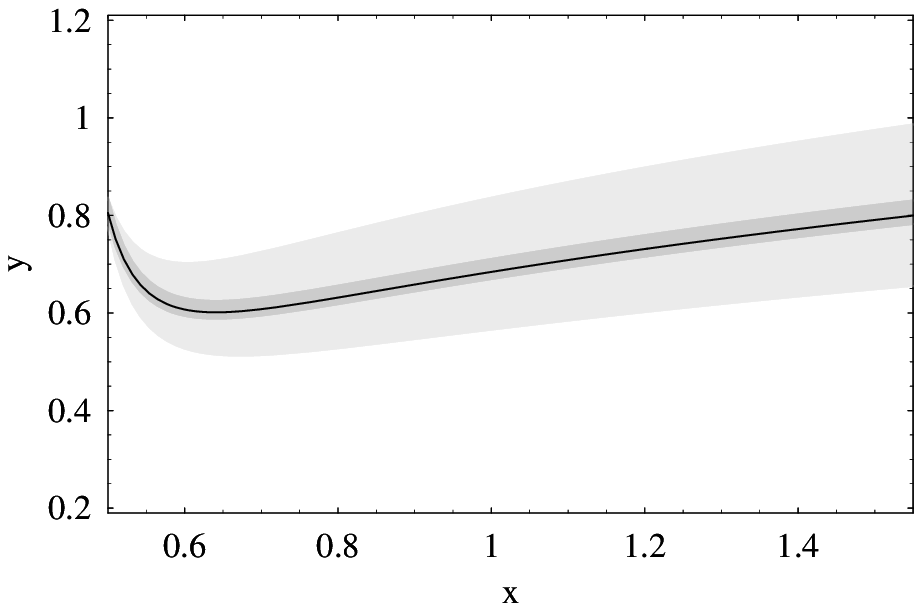} & 
\psfrag{x}[]{$\mu_0$~[GeV]}
\psfrag{y}[b]{$H_{V_\perp}/H_{V_\perp}^{\rm tree}$}
\includegraphics[width=0.48\textwidth]{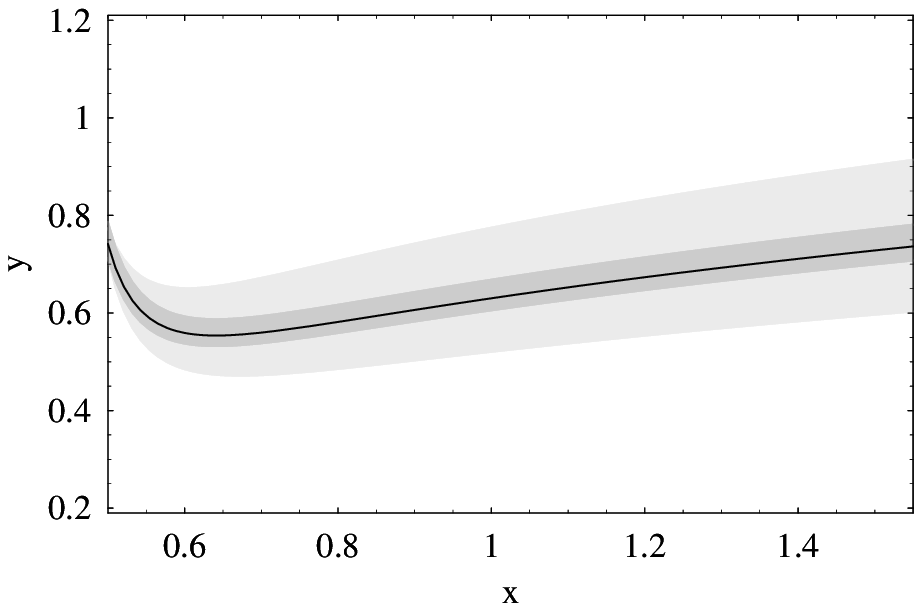}  
\end{tabular}
\end{center}
\vspace{-0.5cm}
\caption{\label{fig:simplified_RG}
Leading-order resummed results for the hard-scattering contributions to 
heavy-to-light form factors evaluated at $2E=m_B$, in units of the tree-level 
expressions for $H_M^{\rm tree}$ in (\ref{DFtree}).
The bands reflect the sensitivity to the choice of the two matching scales
$\mu_h$ (dark band) and $\mu_i$ (light band). They are the result of varying
$0.5m_b^2<\mu_h^2<2m_b^2$ and $0.5m_b\Lambda_h<\mu_i^2<2m_b\Lambda_h$, 
respectively.}
\end{figure}

It is convenient to present the results in terms of the tree-level 
expressions, which we define at a fixed reference scale 
$\mu_i=\sqrt{m_b\Lambda_h}$ with $\Lambda_h=0.5$\,GeV,  
\be\label{DFtree}
   H_M^{\rm tree}
   \equiv {3\pi C_F\alpha_s(\sqrt{m_b\Lambda_h})\over N}\,
   {f_B f_M(\sqrt{m_b\Lambda_h})\over m_B\lambda_B} \,.
\ee
As an example we may consider the pion (with $f_\pi=131$\,MeV) and the $\rho$
meson (with $f_{\rho_\parallel}=198$\,MeV and 
$f_{\rho_\perp}(\sqrt{m_b\Lambda_h})=152$\,MeV \cite{Ball:1998kk}) as 
representative pseudoscalar and vector mesons, respectively. Then with 
$f_B=180$\,MeV and $\lambda_B=0.32$\,GeV, we obtain 
$H_P^{\rm tree}\approx 0.021$, $H_{V_\parallel}^{\rm tree}\approx 0.032$, 
and $H_{V_\perp}^{\rm tree}\approx 0.025$. In Figure~\ref{fig:simplified_RG} 
we plot the dependence of the quantities $H_M$ on the model parameter $\mu_0$ 
and investigate their sensitivity to the matching scales $\mu_h$ and $\mu_i$. 
We observe that the results are rather stable for a wide range of $\mu_0$ 
values. If $\phi_B$ can be modeled accurately by (\ref{eq:Bmodel}) for some 
value of $\mu_0$ within this range, then the results are relatively 
insensitive to the precise value of $\mu_0$. At $\mu_0=1$\,GeV, the 
sensitivity to the matching scales is approximately $\pm 20\%$ for the 
considered range of $\mu_i$, and $\pm 5\%$ for the considered range of 
$\mu_h$.

\begin{figure}
\begin{center}
\begin{tabular}{cc}
\psfrag{x}[]{$E$~[GeV]}
\psfrag{y}[b]{$\Delta F_0/\Delta F_0^{\rm tree}$}
\includegraphics[width=0.55\textwidth]{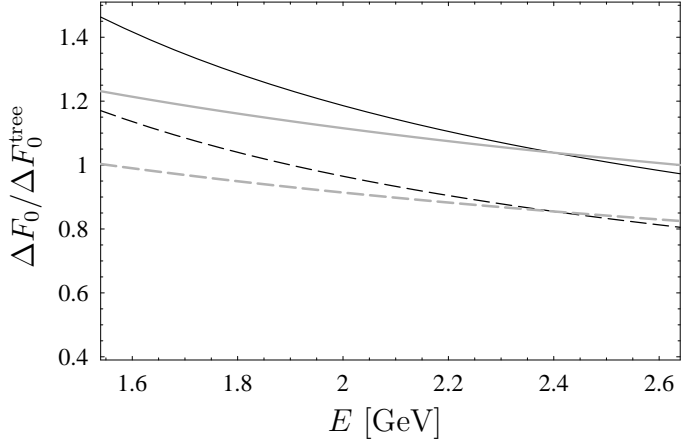}
\end{tabular}
\end{center}
\vspace{-0.5cm}
\caption{\label{fig:resummedF} 
Hard-scattering contribution to the form factor $F_0$ including one-loop 
matching correction, in units of the tree-level expression 
$\Delta F_0^{\rm tree}=\frac{m_B}{2E}\,H_P^{\rm tree}$. The solid and dashed 
curves correspond to a low-energy scale $\mu_0$ such that 
$\alpha_s(\mu_0)=0.5$ and 1.0, respectively. The black lines are obtained 
with matching scales $\mu_h=2E$ and $\mu_i=\sqrt{2E\Lambda_h}$ with 
$\Lambda_h=0.5$\,GeV, the gray lines with $\mu_h=m_b$ and 
$\mu_i=\sqrt{m_b\Lambda_h}$.}
\end{figure}

\begin{figure}
\begin{center}
\begin{tabular}{cc}
\psfrag{a}{$\alpha_s$}
\psfrag{x}[]{$\mu_0$~[GeV]}
\psfrag{y}[b]{$\Delta F_+/\Delta F_{+}^{\rm tree}|_{E=E_{\rm max}}$}
\includegraphics[width=0.55\textwidth]{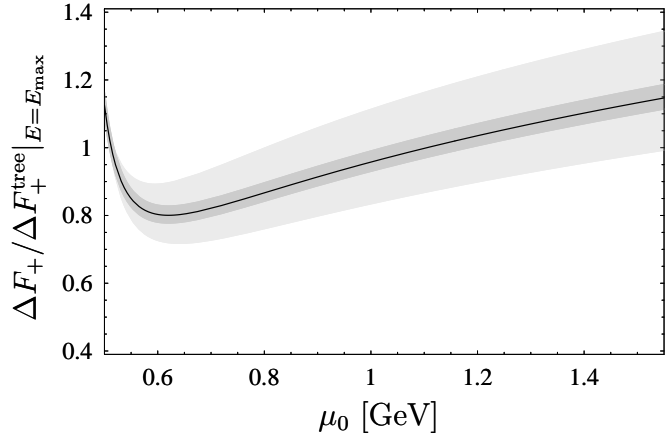}
\end{tabular}
\end{center}
\vspace{-0.5cm}
\caption{\label{fig:resummedFmu} 
Hard-scattering contribution to the form factor $F_+(0)=F_0(0)$, including 
one-loop matching corrections, as a function of the model parameter $\mu_0$. 
The bands reflect the sensitivity to the choice of the two matching scales 
$\mu_h$ (dark band) and $\mu_i$ (light band). They are the result of varying 
$0.5m_b^2<\mu_h^2<2m_b^2$ and $0.5m_b\Lambda_h<\mu_i^2<2m_b\Lambda_h$, 
respectively.}
\end{figure}

In order to investigate the size of one-loop matching corrections, we return 
to the example of the scalar current. Inserting the one-loop jet functions 
calculated in Section~\ref{sec:SCETIImatching} into the general relation 
(\ref{eq:resumT}), it is straightforward to project out the relevant moment 
of the light-meson LCDA. The second relation in (\ref{eq:scetii_rel}) then 
yields the hard-scattering contribution to the form factor $F_0$ including 
leading-order resummation effects and next-to-leading order matching 
corrections. Figure~\ref{fig:resummedF} shows the energy dependence of 
$\Delta F_0$ for two different values for the model parameter $\mu_0$. In 
Figure~\ref{fig:resummedFmu} we restrict attention to maximum recoil energy, 
$E=m_B/2$. In this case $F_+(0)=F_0(0)$, so that our analysis applies to both 
vector form factors. We again observe a mild dependence on the model 
parameter $\mu_0$, and a slightly reduced (with respect to the leading-order 
results in Figure~\ref{fig:simplified_RG}) sensitivity to the matching scales 
$\mu_h$ and $\mu_i$.

As a final note, we emphasize that the numerical results presented above 
depend on the model used for the $B$-meson LCDA, primarily through its first 
inverse moment, $\lambda_B$. However, the statement that leading-order 
resummation effects are encoded by universal functions $H_M$, which are the 
same for all form factors with the same light meson in the final state, is 
independent of any model assumptions. The one-loop matching analysis for the 
sample case of the form factor $F_0$ provides some insight into the size of 
next-to-leading order corrections and the convergence of the perturbative 
expansion of the jet functions. The one-loop matching corrections at the hard 
scale $\mu_h$ modify the tree-level expression for $F_0(0)$ by $\sim 10\%$. 
With $\mu_0=1$\,GeV, the corrections arising from matching at the intermediate 
scale $\mu_i$ amount to a change of $\sim 30\%$. Our calculation also confirms 
the convergence of the hard-scattering convolution integral in a non-trivial 
case, as required by the factorization theorem (\ref{eq:soft_plus_hard}), and 
as proved on general grounds in \cite{Beneke:2003pa,Lange:2003pk}.

\section{Asymptotic behavior of Sudakov logarithms for\\
heavy-to-light form factors}
\label{sec:Sudakov}

An interesting question to ask is whether one of the soft-overlap or
hard-scattering contributions to the form factors is suppressed relative to 
the other in the formal limit $E\to\infty$ with fixed ratio $E/m_b$. As 
already noted in \cite{Lange:2003pk,Becher:2003kh}, this issue cannot be 
entirely addressed in perturbation theory, since the matrix elements for the 
soft-overlap contribution depend on $E$ in a non-perturbative way. Still, it 
is interesting to investigate the relative suppression of the Wilson 
coefficients multiplying the hadronic matrix elements at some low 
renormalization scale $\mu_0$, assuming that the matrix elements at a low 
scale have only a mild energy dependence. In that way, one takes into account
all short-distance logarithms arising from RG evolution between the scales
$\mu_h\sim 2E$ and $\mu_0$.

\begin{figure}
\begin{center}
\begin{tabular}{cc}
\psfrag{x}[]{$2E$~[GeV]}
\psfrag{y}[b]{$\exp[2S(\mu_i,\mu_0)]$}
\includegraphics[width=0.55\textwidth]{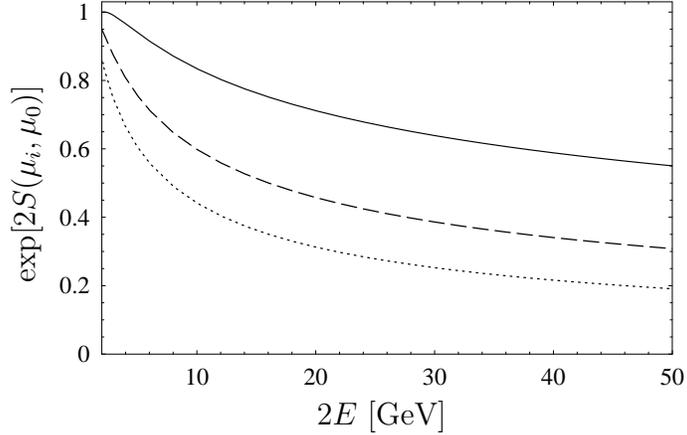}
\end{tabular}
\end{center}
\vspace{-0.5cm}
\caption{\label{fig:sudFactor}
Asymptotic Sudakov suppression of the Wilson coefficient of the soft-overlap 
contribution to a heavy-to-light form factors relative to the coefficient 
of the hard-scattering contribution. The three curves correspond to different
choices of the low scale $\mu_0$, such that $\alpha_s(\mu_0)=0.5$ (solid), 
0.75 (dashed), and 1.0 (dotted). The intermediate matching scale is taken to 
be $\mu_i=\sqrt{2E\Lambda_h}$ with $\Lambda_h=0.5$\,GeV.}
\end{figure}

In order to isolate the dominant Sudakov factor, we concentrate on only 
those terms in the RG evolution equations involving $\Gamma_{\rm cusp}$. They 
are sufficient to resum the leading Sudakov double logarithms to all orders 
in perturbation theory. In SCET$_{\rm I}$, the evolution of the $A$-type and 
$B$-type operators is then identical,
\bea\label{SCET1run}
   {d\over d\ln\mu}\,C_i^A(E,\mu)
   &\simeq& \left[ \Gamma_{\rm cusp}(\alpha_s)\,\ln{2E\over\mu} \right]
    C_i^A(E,\mu) \,, \nl
   {d\over d\ln\mu}\,C_j^B(E,u,\mu)
   &\simeq& \left[ \Gamma_{\rm cusp}(\alpha_s)\,\ln{2E \over \mu} \right]
    C_j^B(E,u,\mu) \,,
\eea
so that the leading Sudakov suppression factor from SCET$_{\rm I}$ running is 
the same for the two types of currents. Our new result (\ref{eq:decompW}) for 
the cusp contribution to the anomalous dimensions of the $B$-type currents is
a crucial ingredient to this conclusion. In SCET$_{\rm II}$, the 
leading-order heavy-light currents contributing to the soft-overlap terms 
obey the same evolution equation as in (\ref{SCET1run}) \cite{Lange:2003pk}; 
however, the hard-scattering contributions now derive from four-quark 
operators, whose coefficient functions obey
\bea
   {d\over d\ln\mu}\,D_i(E,\omega,u,\mu)
   &\simeq& \left[ \Gamma_{\rm cusp}(\alpha_s)\,\ln{\mu\over\omega}
    \right] D_i(E,\omega,u,\mu) \nl
   &=& \left[ \Gamma_{\rm cusp}(\alpha_s)\,\ln{2E\over\mu}
    - \Gamma_{\rm cusp}(\alpha_s)\,\ln{2E\omega\over\mu^2} 
    \right] D_i(E,\omega,u,\mu) \,.
\eea
If not for the term involving the intermediate scale 
$2E\omega\sim E\Lambda_{\rm QCD}\sim\mu_i^2$, the running would again be the 
same for both contributions. For a low value 
$\mu\sim\mu_0\sim\Lambda_{\rm QCD}$, the effect of this term is to reduce the
Sudakov suppression of the hard-scattering contributions relative to that of
the soft-overlap terms. It follows that, at a low scale $\mu_0\ll\mu_i$, the 
soft-overlap contribution to a heavy-to-light form factor is suppressed 
{\em relative\/} to the hard-scattering contribution by a factor 
\be
   \exp\left[ - 2\int_{\alpha_s(\mu_i)}^{\alpha_s(\mu_0)}
   {d\alpha\over\beta(\alpha)}\,\Gamma_{\rm cusp}(\alpha)
   \int_{\alpha_s(\mu_i)}^{\alpha} {d\alpha'\over\beta(\alpha')} \right]
   = e^{2S(\mu_i,\mu_0)} \,,
\ee
where the leading-order expression for $S(\mu_i,\mu_0)$ is obtained from 
(\ref{VHres}) by replacing $\mu_h\to\mu_i$ and $\mu\to\mu_0$. For realistic 
values of parameters this suppression is very mild. This is illustrated in 
Figure~\ref{fig:sudFactor}, where we show the energy dependence of 
$e^{2S(\mu_i,\mu_0)}$ for $\mu_i=\sqrt{2E\Lambda_h}$ and different choices of 
$\mu_0$.

\section{Discussion and conclusions}
\label{sec:summary}

Our analysis completes the leading-order description of large-recoil 
heavy-to-light form factors in SCET, including Sudakov resummation effects. 
The main focus was on the structure of SCET$_{\rm I}$ currents, which divide 
naturally into two classes (denoted $A$- and $B$-type), contributing 
respectively to the soft-overlap (spin-symmetry preserving) and 
hard-scattering (spin-symmetry breaking) terms.

Two interesting complications arise in the RG analysis in SCET$_{\rm I}$. The 
first, common to both types of operators, is the explicit scale dependence of 
the anomalous dimensions. The non-trivial result that the scale dependence 
appears to all orders in $\alpha_s$ only as a single logarithm times the 
universal cusp anomalous dimension follows from the cusp structure of the
Wilson loops appearing in the effective-theory operators. The resulting 
Sudakov suppression is the (formally) dominant effect of RG running. The 
second complication, specific to the $B$-type operators, is the mixing of 
operators having different hard-collinear momentum fractions. The structure 
of this mixing is constrained at leading order by a conformal symmetry in the 
hard-collinear sector of the effective theory. The corresponding 
(multiplicatively renormalized) operator eigenfunctions are expressed in 
terms of Jacobi polynomials. The heavy quark breaks the conformal 
symmetry, introducing an additional local term in the evolution equation. An 
approximate, but arbitrarily precise solution to the RG equation may be 
obtained by matrix methods with a truncated basis of conformal eigenfunctions.

To complete the form-factor analysis, we have matched the intermediate 
effective theory onto SCET$_{\rm II}$, where the $B$-type currents are 
represented in terms of non-local four-quark operators. The projection 
properties of effective-theory spinor fields severely restrict the possible 
Dirac structures appearing in the four-quark operators. As an immediate 
consequence of expressing the original QCD currents in the SCET$_{\rm II}$ 
representation, we found that for two cases the spin-symmetry relations 
holding between the soft-overlap contributions to different form factors are 
not broken at any order in perturbation theory by hard-scattering corrections. 
In the remaining cases, when matching corrections at the high scale 
$\mu_h\sim m_b$ are neglected, the resulting hard-scattering contributions can 
be related to a single factor $H_M$, the same for all $B\to M$ transitions
involving the same light final-state meson $M$. This universality implies that 
certain form-factor relations, e.g.\ those between the $B\to P$ form factors 
$F_+$, $F_0$, and $F_T$, are exact up to corrections of order $\alpha_s(m_b)$ 
and $\Lambda_{\rm QCD}/m_b$. We have presented results for the perturbative 
expansion of the quantities $H_M$ through one-loop order.    

Because of the factorized form, the RG analysis of the hard-scattering terms 
in SCET$_{\rm II}$ reduces to the renormalization of the $B$-meson and 
light-meson LCDAs. Using previous results for the relevant evolution 
equations, we have presented a complete leading-order resummation of Sudakov
logarithms. At this order, after resumming all single and double logarithms, 
the result is described by a universal RG factor, identical for all form 
factors describing the same light final-state meson. In order to investigate 
the size of matching corrections, we have performed the one-loop matching 
necessary to analyze the vector form factors $F_+$ and $F_0$ at maximum 
recoil. Contributions from matching at the high scale $\mu_h\sim 2E\sim m_b$ 
give $\sim 10\%$ corrections to the tree-level expressions, while matching 
contributions at the intermediate scale $\mu_i\sim\sqrt{2E\Lambda_{\rm QCD}}$
yield modifications of $\sim 30\%$. A complete next-to-leading order solution, 
which controls scale dependence through $\order(\alpha_s)$, will have to await 
the extension of our anomalous-dimension calculations to the two-loop order. 
Such an analysis would also require the three-loop coefficient of the cusp 
anomalous dimension $\Gamma_{\rm cusp}(\alpha_s)$, which has been calculated 
very recently \cite{Moch:2004pa}.

One of the dominant uncertainties in the phenomenological analysis of 
heavy-to-light form factors results from our ignorance about the functional 
form of the $B$-meson LCDA $\phi_B(\omega)$. The scale of the hard-scattering 
contribution to form factors is set by the quantities 
$H_M\approx(4\pi/3)\,\alpha_s f_B f_M/m_B\lambda_B\approx 0.02$--0.03 (for 
$\alpha_s\approx 0.3$ and $\lambda_B\approx 0.3$\,GeV). Since 
$\phi_B(\omega)$ and, in particular, its first inverse moment $\lambda_B$ are 
ubiquitous in the effective-theory description of exclusive $B$ decays, there 
is hope that experiment can provide significant constraints, to be checked 
against the present knowledge based on QCD sum rules 
\cite{Grozin:1996pq,Ball:2003fq,Braun:2003wx}. 

The effective theory does not predict the relative size of the soft-overlap 
and hard-scattering contributions to the form factors, other than saying that 
they are of the same order in $\Lambda_{\rm QCD}/E$ (neglecting Sudakov 
logarithms), namely of order $(\Lambda_{\rm QCD}/E)^{3/2}$. To explore this 
question further, we have investigated the asymptotic forms of the Wilson 
coefficient functions in the formal limit $E\to\infty$ with fixed $E/m_b$. We 
have found that the soft-overlap coefficients are suppressed relative to the
hard-scattering coefficients in the asymptotic large-energy limit, but that 
this suppression is ineffective for realistic energies attainable in $B$ 
decays. This conclusion is, however, tempered by the fact that the 
non-perturbative soft-overlap matrix elements have a long-distance 
sensitivity to the scale $E$, which cannot be controlled using short-distance 
methods. Therefore, the behavior of the Wilson coefficients alone does not 
necessarily give a reliable prediction for the asymptotic behavior of the 
form factors themselves. Phenomenologically, it appears that the soft-overlap 
contributions to $B\to M$ form factors are significantly larger than the 
hard-scattering terms.

The results presented in this paper will be relevant to more complicated 
decay processes.  QCD factorization formulae relate the decay 
amplitudes for rare exclusive processes such as $B\to\pi\pi$ or 
$B\to K^*\gamma$ to the sum of a $B\to M$ form-factor term plus a 
hard-scattering contribution \cite{Beneke:1999br,Beneke:2001at,Bosch:2001gv}, 
both of which are described in the effective theory by operators already 
present in the form-factor analysis.  For example, in \cite{Bauer:2004tj} it 
is argued that the jet functions appearing in $B\to\pi\pi$ are the same as those
appearing in the $B\to\pi$ form factor.  The universality of the jet functions 
discussed in the present work may have interesting implications when applied 
to these cases. Finally, using the SCET formalism it 
should also be possible to analyze heavy-to-light form factors beyond the 
leading order in the large-energy expansion.

\paragraph{Note added:}
While this paper was in writing the work \cite{Beneke:2004rc} appeared, in 
which the one-loop hard matching corrections to the subleading SCET$_{\rm I}$ 
currents are computed. Our operator basis is more convenient for studying RG 
evolution than the one adopted in that paper, since with our choice the 
two-particle ($A$-type) and three-particle ($B$-type) currents do not mix 
under renormalization. Our matching coefficient $C_S^B$ in (\ref{eq:CSBoneloop}) 
agrees with the corresponding result in \cite{Beneke:2004rc}. 

\paragraph{Acknowledgments:}
We thank M.~Beneke and D.~s.~Yang for alerting us to the scheme
dependence of the one-loop jet-functions. The work of T.B.\ and
R.J.H.\ is supported by the Department of Energy under Grant
DE-AC02-76SF00515. The research of S.J.L.\ and M.N.\ is supported by
the National Science Foundation under Grant PHY-0098631.

\end{document}